\documentclass[aps,prb,superscriptaddress,twocolumn,floatfix]{revtex4-1}
\usepackage[abs]{overpic}
\usepackage{graphicx,graphics}
\usepackage{dcolumn}
\usepackage{amsmath,amssymb,amsfonts}
\usepackage{wasysym}
\usepackage{latexsym,verbatim}
\usepackage{bm}
\usepackage{color}
\usepackage{ulem}
\usepackage[framemethod=default]{mdframed}
\usepackage[breaklinks=true,colorlinks,citecolor=blue,linkcolor=blue,urlcolor=blue]{hyperref}
\usepackage[makeroom]{cancel}
\usepackage{fancybox}

\newcommand{\be}{\begin{eqnarray}}
\newcommand{\ee}{\end{eqnarray}}
\newcommand{\nn}{\nonumber\\}

\begin{document}

\author{Matteo Carrega}
\affiliation{NEST, Istituto Nanoscienze-CNR and Scuola Normale Superiore, Piazza San Silvestro 12, 56127 Pisa, Italy}
\author{Ivan J. Vera-Marun}
\affiliation{Department of Physics and Astronomy, University of Manchester, Oxford Road, M13 9PL, Manchester, UK}
\affiliation{National Graphene Institute, University of Manchester, Oxford Road, M13 9PL, Manchester, UK}
\author{Alessandro Principi}
\affiliation{Department of Physics and Astronomy, University of Manchester, Oxford Road, M13 9PL, Manchester, UK}

\title{Tunneling spectroscopy as a probe of fractionalization in 2D magnetic heterostructures}
\begin{abstract}
In this paper we develop the theory for 2D-to-2D tunneling spectroscopy aided by magnetic or quantum-order excitations, and apply it to the description of van-der-Waals heterostructures of graphene/ultrathin $\alpha-{\rm RuCl}_3$. We study the behavior of both the differential conductance and the inelastic electron tunneling spectrum (IETS) of these heterostructures. The IETS in particular exhibits features, such as the gap of continuum spinon excitations and Majorana bound states, whose energies scale {\it cubicly} with the applied magnetic field. Such scaling, which exists for a relatively wide range of fields, is at odds with the linear one exhibited by conventional magnons and can be used to prove the existence of Kitaev quantum spin liquids.
\end{abstract}
\maketitle

\section{Introduction}
\label{sect:intro}
The quest for quantum spin liquids~\cite{Savary_rpp_2017,Zhou_rmp_2017,Wen_book} has a long history. Anderson~\cite{Anderson_mrb_1973} was the first to predict, in 1973, that quantum fluctuations of the spin degree of freedom in certain frustrated magnets could lead to the destruction of any magnetic order therein. Naively, such systems are able to avoid the symmetry breaking phenomenon usually associated with a phase transition and, even at the lowest temperatures, their spins remain disordered, hence the name of ``liquids''~\cite{Savary_rpp_2017,Zhou_rmp_2017,Wen_book}. In spite of their resemblance to paramagnets, quantum spin liquids are a fundamentally distinct class of systems. It is not temperature, but the large degree of entanglement between the spins combined with frustration, that leads to disorder in them~\cite{Savary_rpp_2017,Zhou_rmp_2017,Wen_book}. As a result, their excitations can have non-bosonic statistics~\cite{Wen_book}.

The discovery that certain quantum spin liquids can host anyons~\cite{Wen_book,Kitaev_2006}, excitations that behave neither as bosons nor as fermions, has revived the interest in such novel states of matter. A definitive proof of anyonic statistics would in fact be a major success for fundamental science. Furthermore, encoding information non-locally thanks to the large degree of entanglement and operating on the states by braiding (non-Abelian) anyons is a way to construct fault-tolerant quantum computing algorithms~\cite{Nayak_rmp_2008, adi_2008, blasi_2012}. Understanding how to manipulate anyons of quantum spin liquids could therefore constitute a major step forward towards the realization of quantum computers.

One of the models which are known to support a quantum spin liquid whose excitations can be both Abelian and non-Abelian anyons is the so-called Kitaev model~\cite{Kitaev_2006}. Thanks to a carefully engineered frustrated interaction between spin-$1/2$ magnetic moments, the model becomes solvable not just at the mean-field level~\cite{Wen_book} but exactly. The fundamental theory has been laid out by Kitaev in a seminal paper~\cite{Kitaev_2006}, where the solution has been constructed by {\it fractionalizing} each spin in terms of four Majorana particles. One of these describes the mobile excitations (the {\it spinons}), while the other three are hybridized with their counterparts from neighboring sites and give rise to a fictitious magnetic field on top of which spinons propagate~\cite{Kitaev_2006}. When a weak (real) magnetic field is turned on, novel excitations appear: Majorana particles can propagate at the edges or bound to fluxes of the {\it fictitious} magnetic field.~\cite{Kitaev_2006,Knolle_thesis}. 

Soon after Kitaev's proposal, Jackeli and Khaliullin~\cite{Jackeli_prl_2008,Jackeli_prl_2010} proposed that a similar phenomenology could be realized in Mott insulators featuring a strong spin-orbit coupling. By virtue of the complex interplay between crystal symmetry, strong interactions, spin-orbit coupling, and interference between superexchange paths~\cite{Jackeli_prl_2008,Jackeli_prl_2010,Rau_prl_2014,Winter_prb_2016}, the symmetric (Heisenberg) coupling among such spins can be made to vanish and leave the way to a frustrated interaction of the Kitaev type.  Such theoretical prediction has sparked an intense search for materials that exhibit such peculiar cocktail of features~\cite{Canals_prl_1998,Coldea_prl_2001,Itou_prb_2008,Kimchi_prb_2011,Singh_prl_2012,Knolle_prl_2014_Iridates,Yamaji_prl_2014,Kim_prb_2016,Takayama_prl_2015,Yamaji_prb_2016,Li_prb_2017,Slagle_prb_2018}, and which could therefore host emergent fractionalised quasiparticles with Abelian or non-Abelian statistics~\cite{Kitaev_2006}. The evidence for quantum spin liquid states in certain candidate materials is rapidly mounting. Among these, $\alpha-{\rm RuCl}_3$ (hereafter referred to simply as ${\rm RuCl}_3$) has been recently gaining a significant amount of attention~\cite{Plumb_prb_2014,Kim_prb_2015,Yadav_scirep_2016,Zhou_prb_2016,Sandilands_prb_2016,Banerjee_natmat_2016,Sears_prb_2017,Banerjee_science_2017,Do_natphys_2017,Leahy_prl_2017,Baek_prl_2017,Wolter_prb_2017,Ran_prl_2017,Yu_prl_2018,Shi_prb_2018,Winter_prl_2018,Cookmeyer_prb_2018,Kasahara_nature_2018,Kasahara_prl_2018,Hentrich_prb_2019,Zhou_jpcs_2019,Balz_prb_2019}.

\begin{figure}[t]
\begin{center}
\begin{tabular}{c}
\begin{overpic}[width=0.99\columnwidth]{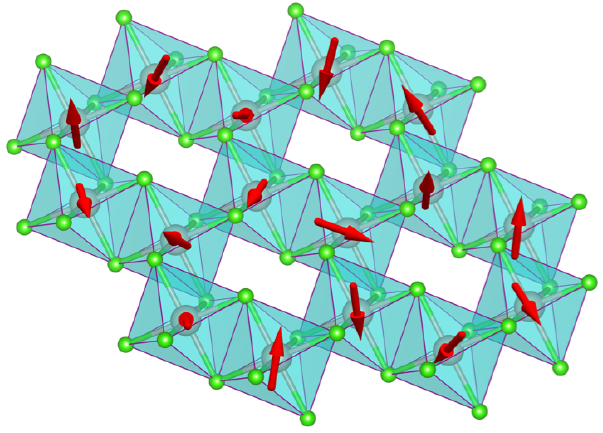}
\put(0,0){(a)}
\end{overpic}
\\
\begin{overpic}[width=0.99\columnwidth]{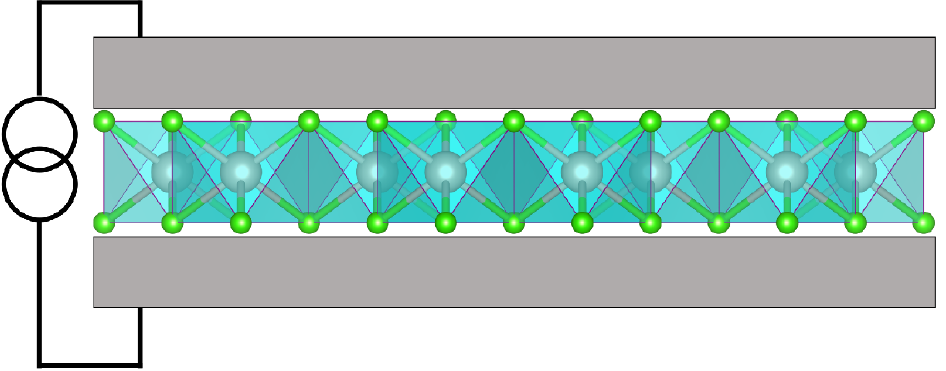}
\put(0,-7){(b)}
\end{overpic}
\end{tabular}
\end{center}
\caption{(Color online) 
Panel (a) The crystal structure of a monolayer transition metal trihalide. Transition metal atoms (large spheres), hosting localized magnetic moments (disordered arrows in the picture), are encaged in halogen (small spheres) octahedra and are coupled to each other via superexchange processes involving such atoms.
Panel (b) A schematic view of the tunneling device. The magnetic insulator is encapsulated within two graphene/thin graphite or metallic electrodes which are connected to an external voltage. 
Tunneling between them occurs via the emission of magnetic excitations in the insulator.
\label{fig:one}
}
\end{figure}

${\rm RuCl}_3$ belongs to a family of materials, the layered transition-metal trihalides~\cite{Gibertini_naturenano_2019}, whose magnetic order survives down to monolayer thicknesses~\cite{Gong_nature_2017,Huang_nature_2017} and which are at the same time widely tunable when embedded in van-der-Waals heterostructures~\cite{Huang_naturenano_2018,Song_science_2018,Klein_science_2018,Wang_natcomm_2018,Seyler_nanolett_2018,Ghazaryan_natureel_2018,Jiang_naturenano_2018,Sivadas_nanolett_2018,Chen_prx_2018,Guo_jpcm_2018,Thiel_science_2019,Cai_nanolett_2019,Zhang_nanolett_2019}. Structurally~\cite{Wang_jpcm_2011,McGuire_chemmater_2015}, transition-metal halides such as ${\rm CrI}_3$, ${\rm CrBr}_3$, ${\rm CrCl}_3$ or ${\rm RuCl}_3$, feature transition-metal atoms (${\rm Cr}$, ${\rm Ru}$) encaged in halogen (${\rm Cl}$, ${\rm Br}$, ${\rm I}$) octahedra [see Fig.~\ref{fig:one}(a)], in turn arranged to form a hexagonal lattice. The magnetic moments, localized at the transition metals as in Fig.~\ref{fig:one}(a), are coupled via superexchange processes involving the non-magnetic halogens~\cite{Wang_jpcm_2011,McGuire_chemmater_2015} [green in Fig.~\ref{fig:one}(a)]. The distances between them are of the order of $\sim 5-7$~\AA, depending on the material and its equilibrium structure. All such materials are magnetic insulators: their low-energy physics is described by effective spin Hamiltonians, whose precise form depends on the microscopic characteristics of the system under consideration.

${\rm RuCl}_3$ is a special case in this family, since it is the only one that the available evidence suggests to be a truly {\it quantum} magnet~\cite{Plumb_prb_2014,Kim_prb_2015,Yadav_scirep_2016,Zhou_prb_2016,Sandilands_prb_2016,Banerjee_natmat_2016,Sears_prb_2017,Banerjee_science_2017,Do_natphys_2017,Leahy_prl_2017,Baek_prl_2017,Wolter_prb_2017,Ran_prl_2017,Yu_prl_2018,Shi_prb_2018,Winter_prl_2018,Cookmeyer_prb_2018,Kasahara_nature_2018,Kasahara_prl_2018,Hentrich_prb_2019,Zhou_jpcs_2019,Balz_prb_2019}.
The material is exfoliable in $\mu$m-size sheets down to monolayer thicknesses, due to the weak electrostatic interlayer interactions, and it is stable at ambient condition~\cite{Zhou_jpcs_2019,Zhou_prb_2019}. 
According to the Jackeli and Khaliullin's mechanism~\cite{Jackeli_prl_2008}, the ${\rm Ru}$ $t_{2g}$-multiplets are split into effective spin-$1/2$ magnetic moments. The coupling between Ru atoms is mediated by the encaging chlorines: the interference of superexchange paths at $\pm 90^\circ$ from ${\rm Ru}-{\rm Ru}$ bonds results in a direction-dependent magnetic interaction of the Kitaev type~\cite{Jackeli_prl_2008}. Although dominant, the Kitaev coupling is not the only interaction present in ${\rm RuCl}_3$~\cite{Rau_prl_2014,Winter_prb_2016}. Its phenomenology goes beyond the ``simple'' Kitaev model and necessitates symmetric interactions to emerge~\cite{Rau_prl_2014,Kim_prb_2015,Winter_prb_2016,Sears_prb_2017}. In fact, below the critical temperature $T_{\rm c} \approx 7~{\rm K}$, the system has been found to be in a zigzag ordered phase~\cite{Banerjee_science_2017,Kim_prb_2015,Sears_prb_2017}. Above $T_{\rm c}$, the properties of bulk ${\rm RuCl}_3$ are however consistent with the formation of a Kitaev quantum spin liquid: magnon peaks, observed in neutron scattering below $T_{\rm c}$, disappear~\cite{Banerjee_science_2017}, leaving the way to a continuum of excitations as expected for a quantum spin liquid. Similarly, measurements performed above $T_{\rm c}$ report a nearly half-quantized thermal Hall conductivity~\cite{Cookmeyer_prb_2018,Kasahara_nature_2018,Kasahara_prl_2018,Hentrich_prb_2019}. In this regime, the material is well described by a Kitaev Hamiltonian~\cite{Kitaev_2006}. Although most of these observations have been performed in bulk samples, they are expected to remain valid, or be even enhanced, in thin ones~\cite{Du_2dmater_2018}. 

To address the physics of thin transition-metal halides, it is necessary to employ techniques that are specific to 2D layered systems. One of such techniques is the 2D-to-2D tunneling spectroscopy~\cite{Britnell_science_2012,Britnell_nanolett_2012,Vdovin_prl_2016,Guerrero_prb_2016,Ghazaryan_natureel_2018}, which has recently been used to study magnetic excitations of van-der-Waals magnets~\cite{Ghazaryan_natureel_2018} such as ${\rm CrBr}_3$.
In experiments, micron-sized devices are built by encapsulating the magnetic material within thin graphite electrodes~\cite{Britnell_science_2012,Britnell_nanolett_2012,Vdovin_prl_2016,Guerrero_prb_2016,Ghazaryan_natureel_2018}, as schematically depicted in Fig.~\ref{fig:one}(b). Encapsulation preserves the characteristics of the magnetic material, which interacts only weakly (with an interaction of the van-der-Waals type) with the graphite electrodes. These offer therefore a non-invasive way of probing the magnetic properties of the insulator. The interfaces between graphite and insulator are in fact atomically flat and clean, while orbitals of different materials do not hybridize. Furthermore, it has the added benefit of preserving the inner layer from contamination. 

By applying a bias voltage across the device, electrons can be made to tunnel from one electrode to the other by either elastic or inelastic processes~\cite{Britnell_science_2012,Britnell_nanolett_2012,Vdovin_prl_2016,Guerrero_prb_2016,Ghazaryan_natureel_2018}. The former conserve the energy of the tunneling particle. On the contrary, inelastic tunneling occurs via the simultaneous excitation of quasiparticles of the insulating layer and therefore electrons lose part of their energy during the process. In an idealized situation, the tunneling current exhibits jumps whenever a new channel is opened~\cite{Ghazaryan_natureel_2018,Asshoff_nanolett_2018}, {\it i.e.} when the applied bias voltage is large enough to generate excitations in the insulating layer. Tracking such steps (or, better, the peaks obtained by taking the derivative of the signal~\cite{Asshoff_nanolett_2018}) it is possible to determine the characteristic energy of excitations. In the case of magnetic systems, these can have magnetic and non-magnetic nature. Among the latter, phonons are certainly the most common and can be distinguished from magnetic ones by tracking their non-dispersive behavior under an applied magnetic field~\cite{Ghazaryan_natureel_2018,Asshoff_nanolett_2018}. 

In this paper we develop the theory of electrical tunneling involving magnetic excitations in van-der-Waals heterostructures. As the main application of the theory, we focus on the signatures of spinons and bound states in tunneling characteristics. This paper is organized as follows. 
In Sect.~\ref{sect:tunnelling} we develop the general theory of 2D-to-2D tunneling in van-der-Waals heterostructures of magnetic insulators, which constitutes the first novel aspect of our paper. We show that the tunneling characteristics can be described in terms of the spin structure factor of the insulating material. 
In Sect.~\ref{sect:probing_excitations} we specialize our result to the study of ${\rm RuCl}_3$, where we report the calculation of its spin structure factor~\cite{Baskaran_prl_2007,Knolle_thesis,Knolle_prl_2014,Knolle_prb_2015}. 
In Sect.~\ref{sect:results} we specialize the general theory to the case of doped electrodes. This allows us to concentrate on the properties of ${\rm RuCl}_3$, rather than on the physics of the electrodes themselves. By studying the tunneling conductance and its derivative (the IETS) as a function of applied magnetic field, we show which features of tunneling spectra can prove the existence of a quantum-spin-liquid phase in thin ${\rm RuCl}_3$.
Quite generally, such features are difficult to be accessed by conventional techniques such as magnetometry. Our studies therefore confirm tunneling as one of the prime tools to address the properties of atomically-thin magnetic van-der-Waals materials.
Finally, in Sect.~\ref{eq:conclusions} we draw conclusions and delineate further outlooks and applications of our theory. The appendices contain several details of the calculation.

\section{Tunneling aided by magnetic excitations}
\label{sect:tunnelling}
In this section we develop the theory of 2D-to-2D tunneling spectroscopy~\cite{Mahan_book} aided by magnetic excitations. The total Hamiltonian of the heterostructure is
\be \label{eq:H_def}
{\cal H} = {\cal H}_{0} + {\cal H}_{\rm tun}~,
\ee
where 
\be
{\cal H}_{0} = {\cal H}_{\rm t} + {\cal H}_{\rm b} + {\cal H}_{\rm m}~,
\ee
 describes the isolated top and bottom electrodes (${\cal H}_{\rm t}$ and ${\cal H}_{\rm b}$, respectively), as well as the magnetic insulator they encapsulate (${\cal H}_{\rm m}$). In Eq.~(\ref{eq:H_def}), ${\cal H}_{\rm tun}$ accounts for the tunneling between them. We now describe their features in detail.

We begin by discussing the {\it general} features of ${\cal H}_{\rm m}$. The details, {\it i.e.} the specific form it acquires for a Kitaev model, will be made explicit in Sect.~\ref{sect:probing_excitations}. We assume ${\cal H}_{\rm m}$ to be an effective low-energy spin Hamiltonian that describes the interaction between magnetic moments ${\bm s}_{{\bm r}_m}$ located at the lattice points ${\bm r}_m$. Here, $s^\gamma_{{\bm r}_m}$ ($\gamma=x,y,z$) is an operator representing the magnetic moments of the insulator. For later convenience, its magnitude is chosen to be {\it twice} that of the corresponding magnetic moment (in the case of spin-$1/2$, $s^\gamma_{{\bm r}_m}$ is therefore a Pauli operator). In what follows we will refer to it as a ``spin operator''. Spin operators are neither bosonic nor fermionic: their commutation relations are those of an angular momentum. This complicates the application of many-body techniques to derive physical quantities such as spin structure factors. (For bosons and fermions, such quantity can be calculated by means of Feyman diagrammatics, which heavily relies on Wick's theorem~\cite{Fetter_Walecka,Giuliani_and_Vignale}; such theorem does not hold for spin operators.) To overcome this obstacle, we will assume that, through one of the well-known transformations~\cite{Holstein_pr_1940,Jordan_zfp_1928,Kitaev_2006}, ${\cal H}_{\rm m}$ can be written in terms of bosonic, fermionic or Majorana operators. 

The choice of the mapping is usually dictated by the request of ``simplicity'' for the final Hamiltonian. For example, in ferromagnetic systems, spin operators are normally mapped into bosons via a Holstein-Primakoff transformation~\cite{Holstein_pr_1940}. Interactions between them can then be neglected in the limit of large magnetic moments (which usually applies to ferromagnets). The final Hamiltonian describes therefore non-interacting excitations ({\it i.e.} the magnons). In the case of the Kitaev Hamiltonian, the mapping between spins and non-interacting Majorana particles is exact, {\it i.e.} no further approximation is involved~\cite{Kitaev_2006}. Excitations are obtained by combining two of the Majorana particles and have therefore fermionic statistics~\cite{Knolle_thesis}. For the purposes of this section, it is sufficient to know the statistics of excitations (obtained via one of the mappings above) and the magnitude of the spin-spin coupling. We stress that, for the derivation of the theory of spin-assisted tunneling in van-der-Waals heterostructures, we do not need to require that ${\cal H}_{\rm m}$ is a non-interacting Hamiltonian. All the information about interactions between quasiparticles is accounted for by the spin structure factor~\cite{Knolle_thesis}.

${\cal H}_{\rm t}$ and ${\cal H}_{\rm b}$ describe two reservoirs of free electrons. As seen in Fig.~\ref{fig:one}(b), in experimental studies performed on micron-scale van-der-Waals devices the magnetic layer is usually encapsulated within thin-graphite slabs~\cite{Britnell_science_2012,Britnell_nanolett_2012,Vdovin_prl_2016,Guerrero_prb_2016,Ghazaryan_natureel_2018}, which are themselves connected to the external circuit. The carrier density in the slabs can be widely tuned. This fact enables the observation in tunneling currents of features due to (among others) the graphite's band structure, moir\`e superlattices~\cite{Bistritzer_pnas_2011} that form at the interface, and particle-particle interactions. To simplify the model, we will describe the two graphite slabs as two graphene sheets~\cite{Guerrero_prb_2016}. We point out that such approximation captures the main physics of the tunneling in the van-der-Waals heterostructure. Such approach is in fact equivalent to describing tunneling between the last atomic layer of the source electrode and the first one of the drain one~\cite{Guerrero_prb_2016,Asshoff_2Dmat_2017}.

\begin{figure}[t]
\begin{center}
\begin{tabular}{c}
\begin{overpic}[width=0.99\columnwidth]{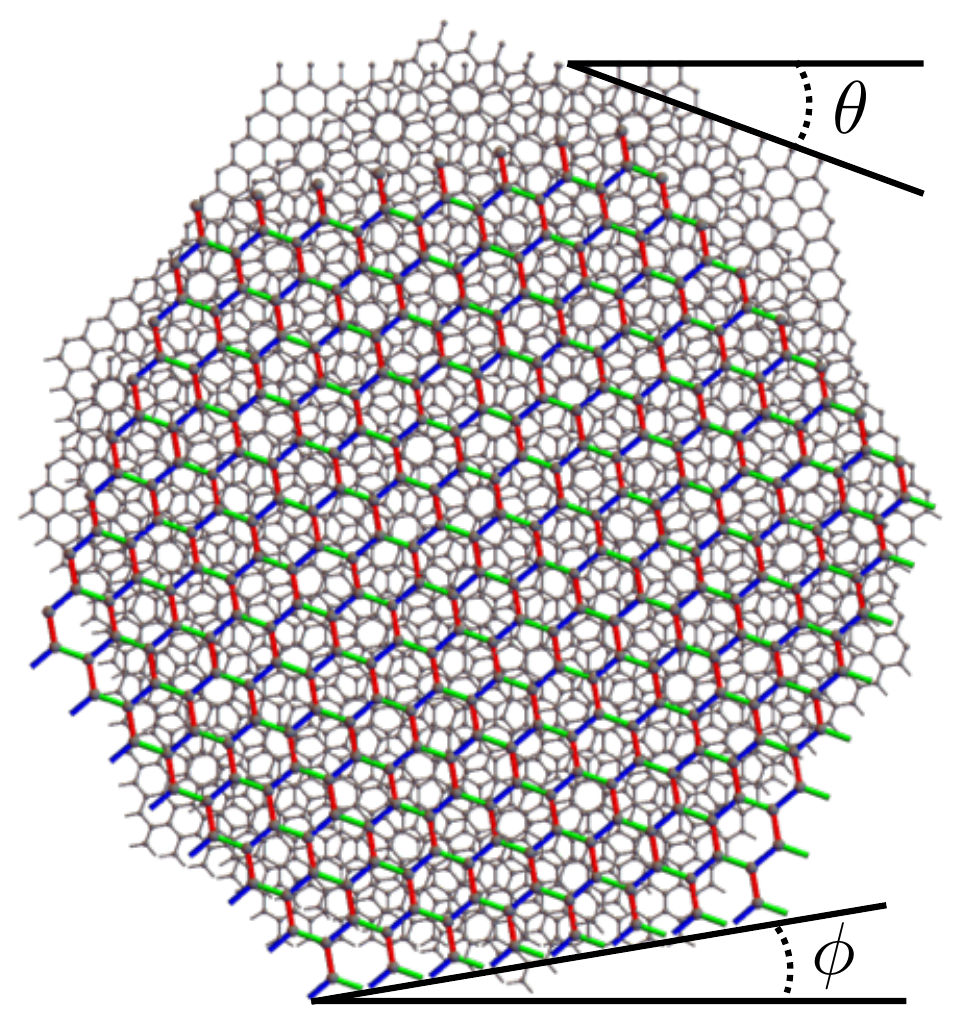}
\end{overpic}
\end{tabular}
\end{center}
\caption{(Color online) A pictorial view of the crystal structure of the heterostructure. The two graphene sheets (grey) encapsulate the magnetic insulator (colored). All materials have a hexagonal crystal structure, although with different lattice sizes (for example, the ${\rm RuCl}_3$ lattice constant is $\approx 3$ times larger than that of graphene). The upper graphene sheet and the magnetic insulator are rotated by $\theta$ and $\phi$, respectively. Rotation angles are measured with respect to the bottom layer.
\label{fig:two}
}
\end{figure}

The top and bottom graphene sheets are, in general, rotated by an arbitrary angle $\theta$ with respect to each other (see Fig.~\ref{fig:two}). In experiments, the misalignment between the layers is inherent to the stacking process and is largely unintentional. Each graphene sheet is considered to be a 2D flat hexagonal lattice with $N$ unit cells. Periodic boundary conditions are assumed to hold. Each unit cell contains two atoms, conventionally labeled $A$ and $B$. The atoms of graphene $\ell={\rm t},{\rm b}$ are located at positions ${\bm r}_\ell + {\bm \tau}_{\alpha,{\ell}}$, where ${\bm r}_\ell = n_+ {\bm a}_{+,\ell} + n_- {\bm a}_{-,\ell}$ is the position of the unit cell and ${\bm \tau}_{\alpha,\ell}$ is the position of the atom $\alpha=A,B$ within it. Here, $n_\pm$ are integer numbers and ${\bm a}_{\pm,\ell}$ are two vectors generating the triangular Bravais lattice. Since the two lattices are rotated by an angle $\theta$ with respect to each other~\cite{Bistritzer_pnas_2011}, ${\bm a}_{\pm,{\rm t}} = {\cal R}(\theta) {\bm a}_{\pm,{\rm b}}$ and ${\bm \tau}_{\alpha,{\rm t}} = {\cal R}(\theta) {\bm \tau}_{\alpha,{\rm b}}$. Here ${\cal R}(\theta)$ is the matrix of rotation by an angle $\theta$ around the axis perpendicular to the graphene planes.
Similarly, the reciprocal lattices of the bottom and top layers are also mutually rotated and are generated by 
${\bm g}_{\pm,{\rm b}}$ and ${\bm g}_{\pm,{\rm t}}\equiv{\cal R}(\theta){\bm g}_{\pm,{\rm b}}$, respectively.

We will approximate each graphene layer as a gas of massless Dirac fermions~\cite{Bistritzer_pnas_2011}. We will consider only electrons with momenta around one of the two inequivalent valleys of the Brillouin zone, for example the ${\bm K}$ point of each electrode Brillouin zone. Electrons in the other (${\bm K}'$) valley can be easily taken into account, in the absence of intervalley scattering, by an extra valley degeneracy $g_v = 2$. With these assumptions, the Hamiltonians for the graphene electrodes are~\cite{Bistritzer_pnas_2011} (from now on we set $\hbar=1$)
\be \label{eq:ham_graphene_def}
{\cal H}_{\ell} = v_{\rm F} \sum_{{\bm k},\alpha,\beta,s} c_{{\bm k},\alpha,s,\ell}^\dagger {\bm \sigma}_{\alpha\beta}\cdot({\bm k} + {\bm K}_\ell) c_{{\bm k},\beta,s,\ell} 
~.
\ee
Here the graphene Fermi velocity $v_{\rm F} = \sqrt{3} a_{\rm G} t_{\rm G}/2 = 10^6~{\rm m/s}$, $t_{\rm G}\simeq 3~{\rm eV}$ is the nearest-neighbor hopping, $a_{\rm G} = 0.246~{\rm nm}$ is the lattice constant, while ${\bm \sigma}$ is a vector of Pauli matrices acting on the sublattice degree of freedom of each layer. Finally, $c_{{\bm k},\alpha,s,\ell}^\dagger$ ($c_{{\bm k},\alpha,s,\ell}$) creates (destroys) an electron with momentum ${\bm k}$ and spin $s$ in the sublattice $\alpha,\beta =A,B$ of layer $\ell={\rm t}, {\rm b}$. In each layer, momenta are measured from the Dirac point ${\bm K}_\ell$.
Note that, since the reciprocal lattices of the two sheets are mutually rotated, the Dirac points ${\bm K}_{\rm t}$ and ${\bm K}_{\rm b}$ do not coincide.

As such, electrons must vary their in-plane momenta in the tunneling process to overcome the mismatch.
This is made possible by the fact that the tunneling amplitude $\Lambda(\delta{\bm r})$ is a periodic function~\cite{Guerrero_prb_2016,Bistritzer_pnas_2011} and depends on $\delta{\bm r} = ({\bm r}_{\rm t}+{\bm \tau}_{\alpha,{\rm t}}) - ({\bm r}_{\rm b}+{\bm \tau}_{\alpha',{\rm b}})$, where ${\bm r}_{\rm t}+{\bm \tau}_{\alpha,{\rm t}}$ and ${\bm r}_{\rm b}+{\bm \tau}_{\alpha',{\rm b}}$ are the initial and final electron positions, respectively. Tunneling is enabled or suppressed depending on how the two layers are locally stacked. This breaks momentum conservation and enables tunneling~\cite{Guerrero_prb_2016,Bistritzer_pnas_2011} between the valleys ${\bm K}_{\rm t}$ and ${\bm K}_{\rm b}$. 

Since this paper focuses on the signatures of magnetic excitations in the tunneling current, ${\cal H}_{\rm tun}$ will be taken to describe only spin-dependent processes. We will assume that a spin excitation is created/annihilated in the tunneling process at the mid-point position ${\bm r}_m = ({\bm r}_{\rm t}+{\bm \tau}_{\alpha,{\rm t}} + {\bm r}_{\rm b}+{\bm \tau}_{\alpha',{\rm b}})/2$ inside the magnetic insulator, such that the product of tunneling amplitudes from ${\bm r}_{\rm t}+{\bm \tau}_{\alpha,{\rm t}}$ to ${\bm r}_m$  and from ${\bm r}_m$ to ${\bm r}_{\rm b}+{\bm \tau}_{\alpha',{\rm b}}$ is maximized~\cite{Asshoff_2Dmat_2017}. Therefore, we postulate the following tunneling Hamiltonian~\cite{Guerrero_prb_2016,Bistritzer_pnas_2011}:
\begin{equation} \label{eq:H_tun_def_real}
{\cal H}_{\rm tun} = \sum_{\substack{{\bm r}_{\rm t},{\bm r}_{\rm b}\\ \alpha, \alpha',s,s'}} \frac{\Lambda(\delta{\bm r}) }{\sqrt{N}}
c_{{\bm r}_{\rm t},\alpha,s,{\rm t}}^\dagger \big[ {\bm \Sigma}_{ss'}\cdot{\bm s}_{{\bm r}_m} \big] c_{{\bm r}_{\rm b},\alpha',s',{\rm b}} 
+{\rm h.c.}
~,
\end{equation}
where $c^\dagger_{{\bm r}_\ell,\alpha,s,\ell}$ ($c_{{\bm r}_\ell,\alpha,s,\ell}$) is the Fourier transform of $c_{{\bm k},\alpha,s,\ell}^\dagger$ ($c_{{\bm k},\alpha,s,\ell}$) and creates (destroys) an electron of spin $s$ at position ${\bm r}_\ell+{\bm \tau}_{\alpha,\ell}$. Here ${\bm \Sigma}$ is a vector of Pauli matrices acting on the electron-spin degree of freedom. In reciprocal space, Eq.~(\ref{eq:H_tun_def_real}) reads
\be \label{eq:H_tun_def_rec}
{\cal H}_{\rm tun} &=& 
\sum_{{\bm k},{\bm k}'} \sum_{{\bm g}_{\rm t},{\bm g}_{\rm b}} \sum_{\alpha, \alpha',s,s'} \frac{\Lambda_{\tilde {\bm q}}}{\sqrt{N}}
e^{i({\bm g}_{\rm t} \cdot{\bm \tau}_{\alpha,{\rm t}} -{\bm g}_{\rm b}\cdot {\bm \tau}_{\alpha',{\rm b}})}
\nn
&\times&
c_{{\bm k},\alpha,s,{\rm t}}^\dagger \big[ {\bm \Sigma}_{ss'}\cdot{\bm s}_{{\bm q}} \big] c_{{\bm k}',\alpha',s',{\rm b}} 
+{\rm h.c.}
~,
\ee
where ${\bm q} = {\bm k} -{\bm k}' +\Delta{\bm K} + {\bm g}_{\rm t} - {\bm g}_{\rm b}$, $\Delta{\bm K}={\bm K}_{\rm t}-{\bm K}_{\rm b}$, and ${\tilde {\bm q}} = ({\bm K}_{\rm t} + {\bm K}_{\rm b} + {\bm k} + {\bm k}' + {\bm g}_{\rm t} + {\bm g}_{\rm b})/2$. Here, $\Lambda_{\tilde {\bm q}}$ and ${\bm s}_{{\bm q}}$ are, respectively, the Fourier transforms of the tunneling amplitude and of the spin operator of the magnetic insulator. To simplify this expression, we use the fact that $\Lambda_{\tilde {\bm q}}$ is a rapidly decreasing function of $|{\tilde {\bm q}}|$,~\cite{Guerrero_prb_2016,Bistritzer_pnas_2011} and therefore we can restrict ourselves to consider the tunneling amplitudes corresponding to the shortest momenta. Since ${\bm k}$ and ${\bm k}'$ are small, the shortest vectors ${\tilde {\bm q}} \simeq ({\bm K}_{\rm t} + {\bm K}_{\rm b} + {\bm g}_{\rm t} + {\bm g}_{\rm b})/2$ satisfy $|{\tilde {\bm q}}| = |{\bm K}_{\rm t} + {\bm K}_{\rm b}|/2$.~\cite{Bistritzer_pnas_2011} This implies that ${\bm g}_{\rm t} = {\cal R}(\theta){\bm g}_{\rm b}$, {\it i.e.} ${\bm g}_{\rm t}$ and ${\bm g}_{\rm b}$ are linear combinations of the generators of the reciprocal lattices (${\bm g}_{\pm,{\rm t}}$ and ${\bm g}_{\pm,{\rm b}}$, respectively) with identical coefficients. The reciprocal-lattice vectors of the bottom layer that satisfy such requirements are ${\bm g}_{\rm b} = {\bm 0}, {\bm g}_{+,{\rm b}}, -{\bm g}_{-,{\rm b}}$.~\cite{Bistritzer_pnas_2011} Calling $\Lambda_0$ the tunneling amplitude corresponding to the shortest vectors ${\tilde {\bm q}}$, and changing the summation over ${\bm k}'$ to one over ${\bm q}$, Eq.~(\ref{eq:H_tun_def_rec}) now becomes~\cite{Guerrero_prb_2016,Bistritzer_pnas_2011}
\be \label{eq:H_T_final_1}
{\cal H}_{\rm tun} &=& 
\frac{\Lambda_0}{\sqrt{N}} \sum_{{\bm k},{\bm q}} \sum_{\alpha, \alpha',s,s'} \sum_{n=0}^2
\big[T^{(n)}_{\alpha\alpha'} {\bm \Sigma}_{ss'}\cdot{\bm s}_{{\bm q}_n} \big] 
\nn
&\times&
\big( c_{{\bm k},\alpha,s,{\rm t}}^\dagger c_{{\bm k}',\alpha',s',{\rm b}} + c_{{\bm k}',\alpha,s,{\rm b}}^\dagger c_{{\bm k},\alpha',s',{\rm t}} \big)
~,
\ee
where now ${\bm k}'={\bm k}-{\bm q}$.
Here we used the fact that ${\bm g}_{\rm t} \cdot{\bm \tau}_{\alpha,{\rm t}} -{\bm g}_{\rm b}\cdot {\bm \tau}_{\alpha',{\rm b}} = {\bm g}_{\rm t} \cdot({\bm \tau}_{\alpha,{\rm t}} - {\bm \tau}_{\alpha',{\rm t}})$, and we defined ${\bm q}_n \equiv {\bm q} + {\bm G}_n$, ${\bm G}_0 = \Delta{\bm K}$, ${\bm G}_1 = {\cal R}(2\pi/3)\Delta{\bm K}$, ${\bm G}_2 = {\cal R}(4\pi/3)\Delta{\bm K}$, and~\cite{Guerrero_prb_2016,Bistritzer_pnas_2011}
\be
T^{(n)}_{\alpha\alpha'} = 
\left(
\begin{array}{cc}
1 & e^{i 2\pi n/3}
\vspace{0.2cm}\\
e^{-i 2\pi n/3} & 1
\end{array}
\right)_{\alpha\alpha'} 
~.
\ee

We now derive an expression for the tunneling current as a function of the bias voltage. 
We start from the general expression for the average current between the top and bottom layer, which reads~\cite{Mahan_book}
\be \label{eq:current_def}
I &=& 
- e \sum_n P_n \langle \psi_n(t) | I_{\rm tb} | \psi_n(t) \rangle
~,
\ee
where $-e$ is the electronic charge, $P_n$ is the occupation factor of a given eigenstate $| \psi_n \rangle$ and $| \psi_n(t) \rangle = e^{-i{\cal H}_0 t}| \psi_n \rangle$. Owing to particle conservation,~\cite{Mahan_book} the tunneling-current operator~\cite{Mahan_book} $I_{\rm tb}$ is obtained by taking the derivative with respect to time of either $N_{\rm t}$ or $-N_{\rm b}$, where the operator $N_{\ell} = \sum_{{\bm k},\alpha,s} c_{{\bm k},\alpha,s,\ell}^\dagger c_{{\bm k},\alpha,s,\ell}$ represents the total number of particles in layer $\ell$:
\be
I_{\rm tb} &=& i [{\cal H}_{\rm tun}, N_{\rm t}]
=
- \frac{i\Lambda_0}{\sqrt{N}} \sum_{{\bm k},{\bm q}} \sum_{\alpha, \alpha',s,s'} \sum_{n=0}^2 
T^{(n)}_{\alpha\alpha'} {\bm \Sigma}_{ss'}\cdot{\bm s}_{{\bm q}_n} 
\nn
&\times&
\big( 
c_{{\bm k},\alpha,s,{\rm t}}^\dagger c_{{\bm k}',\alpha',s',{\rm b}} - 
c_{{\bm k}',\alpha,s,{\rm b}}^\dagger c_{{\bm k},\alpha',s',{\rm t}}
\big)
~.
\ee
With conventional manipulations~\cite{Mahan_book} (see also App.~\ref{sect:eq:current_final_expression}), to lowest order in the tunneling amplitude, the average current~(\ref{eq:current_def}) reads~\cite{Mahan_book}
\be \label{eq:current_final_expression}
I =
-2 e \Im m\big[\chi_{AA}(\mu_{\rm t} -\mu_{\rm b}) \big]
~,
\ee
where the difference between the top and bottom chemical potentials in Eq.~(\ref{eq:current_final_expression}) ($\mu_{\rm t}$ and $\mu_{\rm b}$, respectively) is proportional to the bias voltage $-V$ applied across the junction, {\it i.e.} $e V = \mu_{\rm t} -\mu_{\rm b}$, and 
\be \label{eq:chi_AA_def}
\chi_{AA}(\omega) = -i \lim_{\eta \to 0}\int_0^{\infty} dt \langle [ A(t), A^\dagger ] \rangle e^{i(\omega+i\eta) t}
~.
\ee
Here,
\be \label{eq:A_def_main}
A = \frac{\Lambda_0}{\sqrt{N}} \sum_{{\bm k},{\bm q}} \sum_{\alpha, \alpha',s,s'} \sum_{n=0}^2 
T^{(n)}_{\alpha\alpha'} {\bm \Sigma}_{ss'}\cdot {\bm s}_{{\bm q}_n} 
c_{{\bm k},\alpha,s,{\rm t}}^\dagger c_{{\bm k}',\alpha',s',{\rm b}}
~.
\nn
\ee
The time evolution of the operator $A(t)$ in Eq.~(\ref{eq:chi_AA_def}) is generated by the ``grand-canonical'' Hamiltonian~\cite{Mahan_book,Giuliani_and_Vignale} ${\cal K}_0\equiv {\cal H}_0 - \sum_{\ell={\rm t},{\rm b}} \mu_\ell N_\ell$.
Further manipulations shown in App.~\ref{sect:eq:chi_AA_res_final} allow us to rewrite Eq.~(\ref{eq:chi_AA_def}) as
\be \label{eq:chi_AA_res_final}
&& 
\Im m\chi_{AA}(eV) =
\mp
\int\frac{d\omega'}{\pi} \big[n_{\rm B/F}(\omega') - n_{\rm B/F}(\omega'+eV)\big]
\nn
&&
\times
\sum_{n=0}^2
\sum_{{\bm q}}
\Im m Q({\bm q}_n,eV + \omega') 
\Im m \chi_{\rm tb}({\bm q},\omega',n)
~.
\ee
Here $n_{\rm B/F}(\omega) = \big[e^{\omega/(k_{\rm B} T)} \pm 1\big]^{-1}$ are the Bose-Einstein and Fermi-Dirac distribution, respectively. In Eq.~(\ref{eq:chi_AA_res_final}) the ``$\mp$'' sign and the choice of the distribution function depends on the statistics of spin excitations described by the {\it spin structure factor} $Q({\bm q}_n,\omega')$. This function contains the information about the spectrum of excitation of the insulator. For a given momentum ${\bm q}_n$, $Q({\bm q}_n,\omega')$ is in fact peaked at the frequencies $\omega'$ corresponding to the energy of magnetic excitations.
The derivation of the spin structure factor for a Kitaev model is shown in Sect.~\ref{sect:probing_excitations}. Finally, in Eq.~(\ref{eq:chi_AA_res_final}) we defined
\be
\label{eq:chi_tb_res_final_2}
&&
\Im m \chi_{\rm tb}({\bm q},\omega',n) \!\! = \!\!
\frac{\pi \Lambda_0^2}{N} \!\!
\sum_{{\bm k},\lambda,\lambda'}
\int_{-\infty}^{\infty}\frac{d\varepsilon}{\pi} 
\big[n_{\rm F}(\varepsilon) - n_{\rm F}(\varepsilon-\omega')\big]
\nn
&&
\times
\left|\rho_{{\bm k},\lambda;{\bm k}',\lambda'} + \cos\left(\phi_n\right) \sigma^x_{{\bm k},\lambda;{\bm k}',\lambda'} - \sin\left(\phi_n\right) \sigma^y_{{\bm k},\lambda;{\bm k}',\lambda'} \right|^2
\nn
&&
\times
\Im m G_{{\rm t},\lambda}({\bm k},\varepsilon)
\Im m G_{{\rm b},\lambda'}({\bm k'},\varepsilon-\omega')
~.
\ee
Here $\lambda, \lambda'=\pm$ denote the conduction and valence bands of the graphene layers ($\varepsilon_{{\bm k},\lambda} = \lambda v_{\rm F} |{\bm k}|$ is the band energy), $\phi_n = 2\pi n/3$ and, $G_{\ell, \lambda}({\bm k},\varepsilon)$ is the retarded Green's function of electrons in layer $\ell$, band $\lambda$, with momentum ${\bm k}$ and energy $\varepsilon$.
In Eq.~(\ref{eq:chi_tb_res_final_2}), $\rho_{{\bm k},\lambda;{\bm k}',\lambda'}$, $\sigma^x_{{\bm k},\lambda;{\bm k}',\lambda'}$ and $\sigma^y_{{\bm k},\lambda;{\bm k}',\lambda'}$ are the matrix elements of the density and pseudospin operators between graphene eigenstates with momenta ${\bm k}$ and ${\bm k}'$ in bands $\lambda$ and $\lambda'$, respectively. The function $\Im m \chi_{\rm tb}({\bm q},\omega',n)$ describes the spectrum and DOS of the electrodes, as well as the tunneling probability between them due to the overlap of the electron wavefunctions.

Eqs.~(\ref{eq:chi_AA_res_final})-(\ref{eq:chi_tb_res_final_2}) describe electrons of momentum ${\bm k}$ and energy $\varepsilon$ tunneling from the top to the bottom layer. In the final state, their momentum is ${\bm k'}$ and the energy $\varepsilon-\omega'$. In the process, excitations of the insulator are emitted. The probability of the latter process is described by $Q({\bm q}_n,\omega')$, where ${\bm q}_n$ and $\omega'$ are the momentum and energy of the emitted excitation. Therefore, $I\equiv I(V)$ in Eq.~(\ref{eq:current_final_expression}) corresponds to the contribution to the tunneling current from channels opened by inelastic spin-non-conserving processes. We remind the reader that there are two other sources of tunneling currents: elastic and inelastic spin-conserving processes. To get rid of the first contribution is sufficient, at low enough temperatures (much smaller than chemical potential and exchange parameters), to measure the ``{\it inelastic electron tunneling spectrum}''~\cite{Asshoff_nanolett_2018} (IETS) $dG/dV \equiv d^2 I(V)/dV^2$ (where $G = dI/dV$ is the differential conductance). Such quantity contains only the information regarding inelastic (spin-conserving and spin-non-conserving) processes. Non-magnetic (spin-conserving) processes can be filtered in the IETS by noting that they do not depend on an applied magnetic field, whereas inelastic tunneling aided by magnetic excitations in general does~\cite{Asshoff_nanolett_2018}. 
In conventional (anti)ferromagnets the IETS is dominated by magnon peaks at small momenta. Their energy increases linearly with the magnetic field. At odds with magnons, the energy of excitations of a Kitaev model scales {\it cubicly} with the applied magnetic field~\cite{Knolle_thesis,Kitaev_2006}. As we show in Sect.~\ref{sect:results}, this peculiar behavior can be used to distinguish fractionalized excitations from usual magnons.

\section{The spin structure factor of the Kitaev model}
\label{sect:probing_excitations}
We now specialize to the tunneling of electrons in a graphene-${\rm RuCl}_3$-graphene heterostructure, with the aim of describing signatures of excitation fractionalization in the IETS. Each layer of ${\rm RuCl}_3$ is assumed to contribute independently to the tunneling, and therefore will be treated as an independent Kitaev model~\cite{Kitaev_2006}. In fact, to simplify the treatment we will consider the tunneling through a single ${\rm RuCl}_3$ sheet, and assume that the total spin structure factor is just the number of ${\rm RuCl}_3$ layers times the spin structure factor of a single sheet. The number of insulating layer is therefore a parameter that can be reabsorbed in the definition of the tunneling amplitude. This approximation amounts to assuming that spin excitations can be created in any layer and that the interlayer magnetic coupling is weak. ${\rm RuCl}_3$ satisfies to a large degree such assumptions~\cite{Banerjee_science_2017}. Note also that this approximation takes also care of the possible transfer of charge from the graphene electrodes to the ${\rm RuCl}_3$ stack~\cite{Zhou_prb_2019}, which is an important issue for transport experiments, but much less severe for tunneling ones. In fact, since charges tends to accumulate into external layers~\cite{Zhou_prb_2019}, the result is that the insulating slab becomes effectively thinner. Thus, the metallic leads get effectively closer. Charge transfer can therefore be accounted for by reducing the number of insulating layers participating in the tunneling. Note that such reduction is compensated by a concomitant enhancement of the tunneling amplitude, which depends exponentially on the insulator thickness. We start by briefly describing the Kitaev model of ${\rm RuCl}_3$ and the calculation of its structure factor. The details of the derivations are given in Apps.~\ref{sect:Kitaev_1}-\ref{sect:QSL_spin_spin_momentum_frequency}.

\subsection{Kitaev model of ${\rm RuCl}_3$}
\label{sect:KitaevRuCl3}
A good approximation for ${\rm RuCl}_3$, above~\cite{Banerjee_natmat_2016,Banerjee_science_2017} $T_{\rm c} = 7~{\rm K}$, is the effective Kitaev spin Hamiltonian~\cite{Kitaev_2006} in applied magnetic field, which reads [we denote ${\bm s}_i \equiv {\bm s}_{{\bm r}_i}$]
\be \label{eq:Ham_m_def}
{\cal H}_{\rm m} = -J \sum_{\langle i,j\rangle_\gamma} s^\gamma_i  s^\gamma_j - \sum_i {\bm h}\cdot {\bm s}_i
~.
\ee
This model describes the behavior of magnetic moments located at the ${\rm Ru}$ sites~\cite{Jackeli_prl_2008} (for~\cite{Banerjee_natmat_2016,Banerjee_science_2017} ${\rm RuCl}_3$, $J\approx 1.3~{\rm meV}$). Hence, spins are organized in a hexagonal lattice. Each unit cell, whose position is a linear combinations of the lattice vectors ${\bm a}_{\pm}$ ($|{\bm a}_\pm|\approx 7$~\AA) with integer coefficients, contains two identical sites, $A$ and $B$. Their positions within the cell are determined by the vectors ${\bm \tau}_{\alpha}$ ($\alpha=A,B$). The lattice is rotated by an arbitrary angle $\phi$ with respect to the bottom graphene electrode. 

\begin{figure}[t]
\begin{center}
\begin{tabular}{c}
\begin{overpic}[width=0.99\columnwidth]{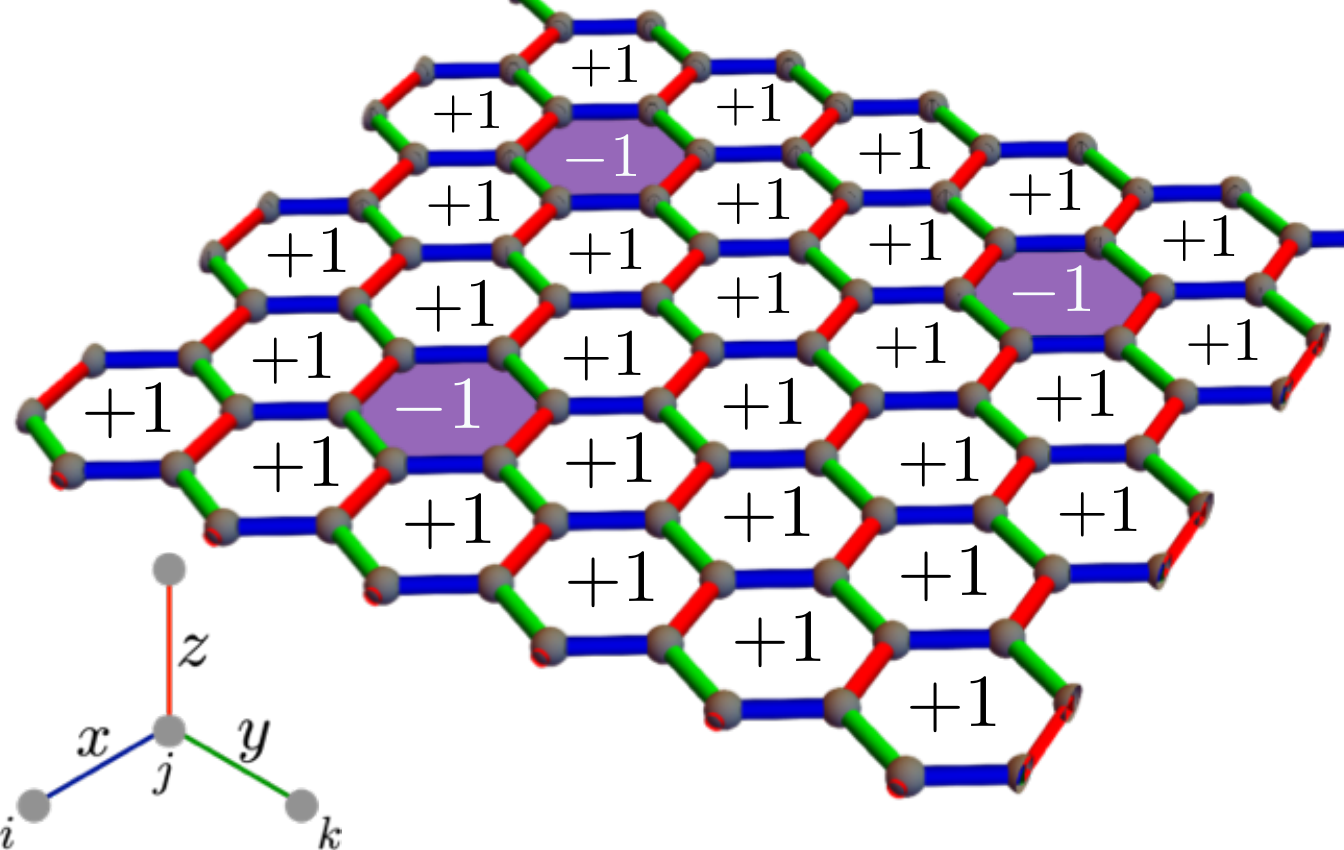}
\put(0,100){(a)}
\put(0,40){(b)}
\end{overpic}
\\
\begin{overpic}[width=0.99\columnwidth]{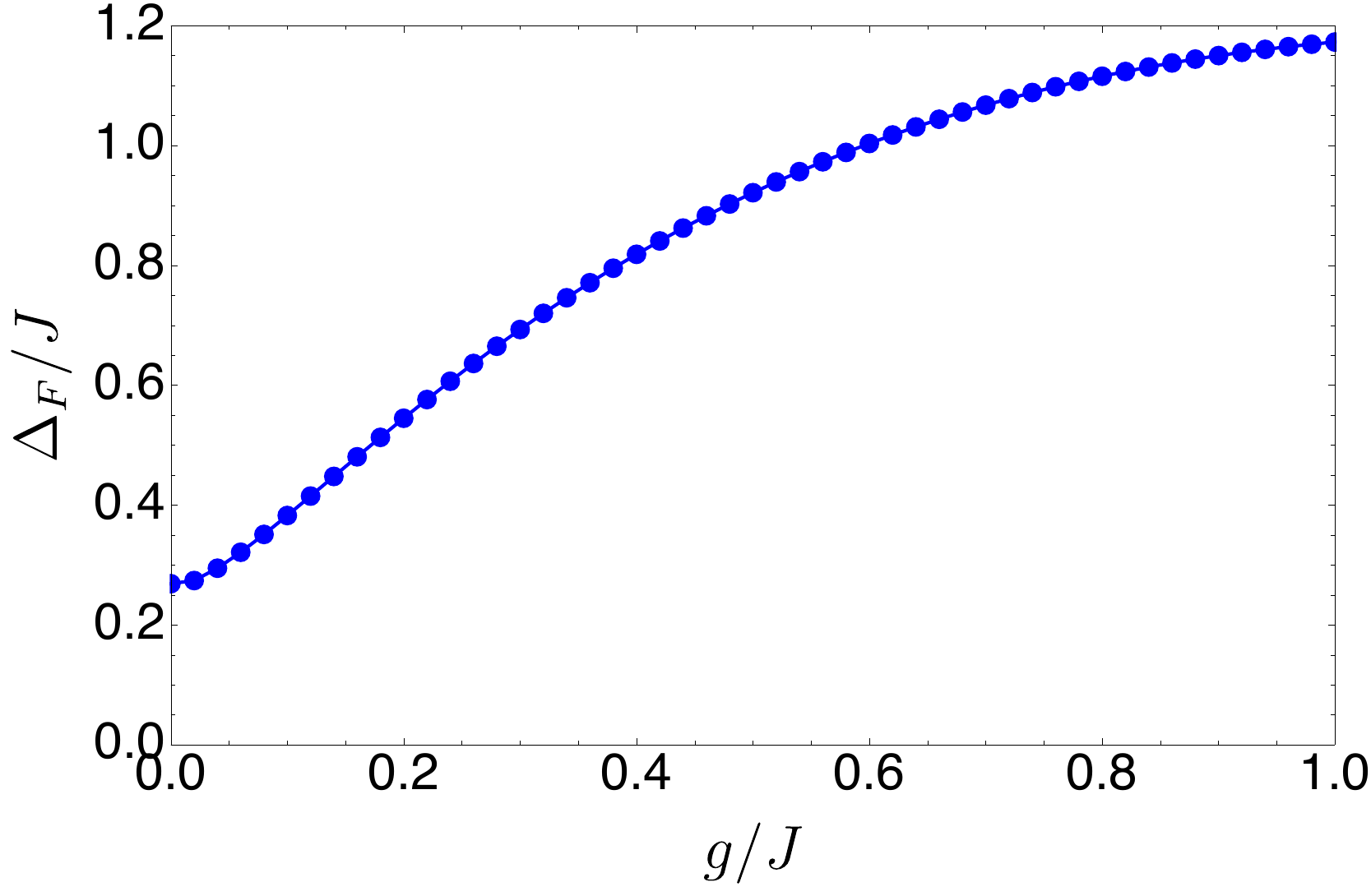}
\put(0,0){(c)}
\end{overpic}
\end{tabular}
\end{center}
\caption{(Color online) 
Panel (a) A pictorial view of the Kitaev model. The hexagonal lattice is composed by magnetic sites (grey). Bonds are colored accordingly to the spin component that is coupled along each of them. The coupling is of the type $s^z_i s_j^z$ along red bonds,  $s^y_i s_j^y$ along green ones, and $s^x_i s_j^x$ along blue ones. Note that all parallel bonds have the same type of coupling. The figure also shows a generic flux configuration, whereby $\Phi_p = \pm 1$, depending on the plaquette $p$. 
Panel (b) A zoom in on a particular site and its neighbors. The color code is the same as in panel (a). 
Panel (c) The two-flux gap $\Delta_F$, obtained by exactly diagonalizing the Hamiltonian~(\ref{eq:Hamil_QSL_Majorana_eff_main}), corresponding to the energy required to create two fluxes in neighboring plaquettes (and, therefore, to the minimum energy of spin excitations), in units of the Kitaev coupling $J$ and as a function of $g/J$ ({\it i.e.} of the magnetic field).
\label{fig:three}
}
\end{figure}

The first sum in Eq.~(\ref{eq:Ham_m_def}) runs over all pairs of nearest-neighbor sites. There, $\langle i,j\rangle_\gamma$ denotes the bond between the sites $i$ and $j$. Note that, depending on the direction of the bond, only one component ($\gamma \in \{x,y,z\}$) of the spins ${\bm s}_i$ and ${\bm s}_j$ is coupled~\cite{Kitaev_2006}. The coupling involves the {\it same} spin component if two bonds are parallel. There are, clearly, three different types of bonds starting at each site $i$ at $120^\circ$ from each other, and therefore a different component of the spin ${\bm s}_i$ is coupled to each of its neighbors. Fig.~\ref{fig:three}(a) and~(b) offer a pictorial view of the model Hamiltonian. 
Lattice sites are denoted with latin indices $i$ and $j$. The second sum in Eq.~(\ref{eq:Ham_m_def}) runs over all lattice sites. To keep the presentation simple, we do not include other interactions between the spins that are present in the real material and can induce ordering at low-temperatures in the material~\cite{Rau_prl_2014,Kim_prb_2015,Winter_prb_2016,Sears_prb_2017}. The focus will therefore be on the description of the quantum-spin-liquid phase which in the real material emerges above $\sim 7~{\rm K}$.

The model described by Eq.~(\ref{eq:Ham_m_def}) is exactly soluble~\cite{Kitaev_2006} only for ${\bm h}={\bm 0}$. For small applied magnetic fields, {\it i.e.} for $|{\bm h}|\ll J$ (note that ${\bm h}$ has the unit of an energy), one can resort to the same approximation used in Ref.~\onlinecite{Kitaev_2006}. It is then possible to derive an {\it effective} Hamiltonian ${\cal H}_{\rm m}^{\rm eff}$ that describes the dynamics of excitations in the low-energy sector (see App.~\ref{sect:Kitaev_1} for details). Crucially, terms that have a nontrivial impact on the spin structure factor are at least of third order in the magnetic field~\cite{Kitaev_2006}:
\be \label{eq:Hamil_QSL_eff_main}
{\cal H}_{\rm m}^{\rm eff}  &=& -J \sum_{\langle i,j\rangle_\gamma} s^\gamma_i s^\gamma_j
- g \sum_{\langle \langle i, k\rangle \rangle} 
s^{\gamma}_i s^{\gamma'}_j s^{\gamma''}_k
~,
\ee
where $g =  h_x h_y h_z/\Delta_F^2$, with $\Delta_F$ the minimum energy of spin excitations~\cite{Kitaev_2006,Knolle_thesis}.
For $g$ to be non-zero, the magnetic field ${\bm h}$ should {\it not} be aligned with any of the spin quantization axis.
As shown in Fig.~\ref{fig:three}(c), $\Delta_F$ is itself a function of $g$. However, since the numerical results we will present are obtained for small $g/J$ [see, e.g., Fig.~\ref{fig:nine}(b) below], we will henceforth use its zero-field value~\cite{Kitaev_2006,Knolle_thesis} in the definition of $g$, {\it i.e.} $\Delta_F(h=0) = 0.26 J$.
In Eq.~(\ref{eq:Hamil_QSL_eff_main}), $\langle \langle i, k\rangle \rangle$ is a pair of next-nearest neighbors, while $j$ is the only site connected to both $i$ and $k$. Furthermore, $\gamma$ and $\gamma''$ are the types of the two bonds $\langle i,j\rangle_{\gamma}$ and $\langle j,k\rangle_{\gamma''}$ (which are clearly different from each other, since the bonds point in different directions from $j$). Finally, $\gamma'=x,y,z$ is the only index left that is different from both $\gamma$ and $\gamma''$. The indices $\gamma\neq \gamma'\neq \gamma''$, as well as $j$, are therefore uniquely determined for the next-nearest-neighbor pair $\langle \langle i, k\rangle \rangle$.
An example is shown in Fig.~\ref{fig:three}(b).

Note that, once energies are scaled with the Kitaev exchange coupling $J$, the only parameter left in the theory is $g/J$. Therefore, all physical observables (e.g., the energy of excitation) scale to lowest order linearly with this parameter. This in turn implies that they scale at least {\it cubicly} with the magnetic field. 

The approximate model of Eq.~(\ref{eq:Hamil_QSL_eff_main}) can be solved exactly, {\it i.e.} the spectrum and statistics of spin excitations can be completely determined. Following Kitaev~\cite{Kitaev_2006}, we introduce four Majorana operators per lattice site, $c_i$ and $b_i^\gamma$, where $\gamma = x,y,z$, such that
\be \label{eq:spin_majorana_main}
s^\gamma_i = i b_i^\gamma c_i
~.
\ee
The Majorana operators satisfy anticommutation relations $\{ b_i^\gamma, b_j^\eta \} = 2\delta_{ij} \delta_{\gamma\eta} \equiv 2 \delta_{ij}^{\gamma\eta}$, $\{c_i,c_j\} = 2\delta_{ij}$ and $\{ b_i^\gamma, c_j \} = 0$. The Hamiltonian~(\ref{eq:Hamil_QSL_eff_main}), written in terms of Majorana operators, now becomes
\begin{equation} \label{eq:Hamil_QSL_Majorana_eff_main}
{\cal H}_{\rm m}^{\rm eff}  = iJ \sum_{\langle i,j\rangle_\gamma} u_{ij}^\gamma c_{i} c_{j} + i g \sum_{\langle \langle i, k\rangle \rangle} \varepsilon_{\gamma,\gamma',\gamma''}
u_{ij}^{\gamma} D_{j} u_{jk}^{\gamma''} c_{i} c_{k}
~,
\end{equation}
where $u_{ij}^\gamma = i b_i^\gamma b_j^\gamma$, $\varepsilon_{\gamma,\gamma',\gamma''}$ is the Levi-Civita tensor and~\cite{Kitaev_2006} $D_j \equiv -i s^x_j s^y_j s^z_j = b_j^x b_j^y b_j^z c_j$. 
By definition, physical states satisfy $D_j = 1$. It can be shown~\cite{Kitaev_2006} that all the $u_{ij}^\gamma$ commute with the Hamiltonian~(\ref{eq:Hamil_QSL_Majorana_eff_main}) and are, therefore, constants of motion. Since all eigenstates of the Hamiltonian are simultaneously eigenstates of all of $u_{ij}^\gamma$, such operators can be replaced by their eigenvalues in Eq.~(\ref{eq:Hamil_QSL_Majorana_eff_main}). [Given that each $u_{ij}^\gamma$ is the product of two Majorana operators, its eigenvalues are, by construction, $\pm 1$.]

Once such eigenvalue ``pattern'' is fixed, it specifies a Hilbert subspace in which  Eq.~(\ref{eq:Hamil_QSL_Majorana_eff_main}) reduces to the Hamiltonian of {\it free} Majorana particles $c_i$ (the spinons) propagating on top of the $\mathbb{Z}_2$ vector potential generated by the $u_{ij}^\gamma$. ``$\mathbb{Z}_2$'' here stands for the fact that, as explained, along each bond the vector potential can only acquire two values, $\pm 1$.
The vector potential produces a $\mathbb{Z}_2$ ``magnetic field'': for each hexagonal plaquette $p$, we can define the $\mathbb{Z}_2$ ``magnetic flux'' threading it as $\Phi_p = \prod_{\langle i,j\rangle_\gamma\in p} u_{ij}^\gamma$, where the notation ``$\langle i,j\rangle_\gamma\in p$'' means that we take the product of all $u_{ij}^\gamma$ such that the bond $\langle i,j\rangle_\gamma$ is an edge of the hexagonal plaquette $p$. Then, $\Phi_p = -1$ is interpreted as having a $\mathbb{Z}_2$ flux threading the hexagonal plaquette $p$. Conversely, $\Phi_p = +1$ signifies that no flux is present. Lieb's theorem~\cite{Lieb_prl_1989} constrains the ground state to have zero total flux, {\it i.e.} $\Phi_p = +1$ for all plaquettes. Introducing fluxes into the system by flipping the sign of one or more bond eigenvalues $u_{ij}^\gamma$ costs a finite amount energy: the zero-flux sector is therefore separated in energy from all other Hilbert subspaces~\cite{Kitaev_2006}. Flipping the sign of one of the $u_{ij}^\gamma$ introduces a pair of fluxes in the two neighboring plaquettes sharing the bond $\langle i,j\rangle_\gamma$. This has an energy cost $\Delta_F$, which is shown in Fig.~\ref{fig:three}(c). At low enough temperatures, therefore, the system does not present any $\mathbb{Z}_2$ flux~\cite{Kitaev_2006}. 

One possible vector-potential configuration compatible with such constraint is obtained by setting all $u_{ij}^\gamma = -1$, ($u_{ij}^\gamma = 1$) with $i$ and $j$ sites of type $A$ and $B$ ($B$ and $A$), respectively. We observe that the flux operator can be represented in term of the original spin operators~\cite{Kitaev_2006}, and that $\Phi_p$ (and not $u_{ij}^\gamma$) is conserved by the original spin Hamiltonian. The conservation of all $u_{ij}^\gamma$ is, in a sense, ``spurious'' to the Majorana representation, whose introduction has enlarged the Hilbert space by adding unphysical states~\cite{Kitaev_2006} for which $D_j \neq 1$. Both issues are solved as follows: once the calculation has been performed for a given choice of the $u_{ij}^\gamma$, and a state $|\psi\rangle$ has been obtained, the physical one is constructed by projecting it onto the physical subspace. The physical state is therefore obtained as~\cite{Kitaev_2006} $\prod_j (1+D_j)|\psi\rangle/2$.

\begin{figure}[t]
\begin{center}
\begin{tabular}{cc}
\begin{overpic}[width=0.49\columnwidth]{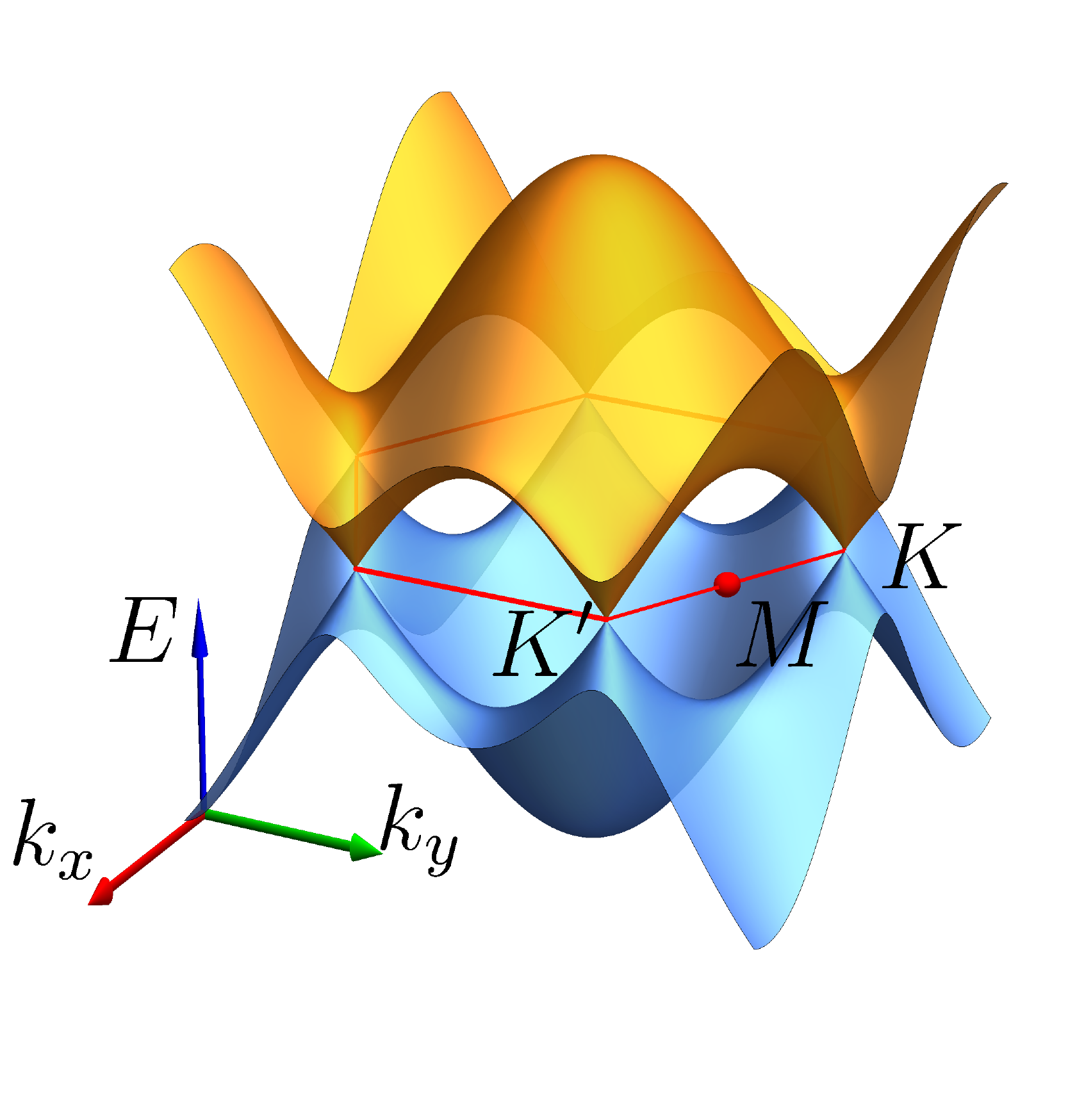}
\put(4,10){(a)}
\end{overpic}
&
\begin{overpic}[width=0.49\columnwidth]{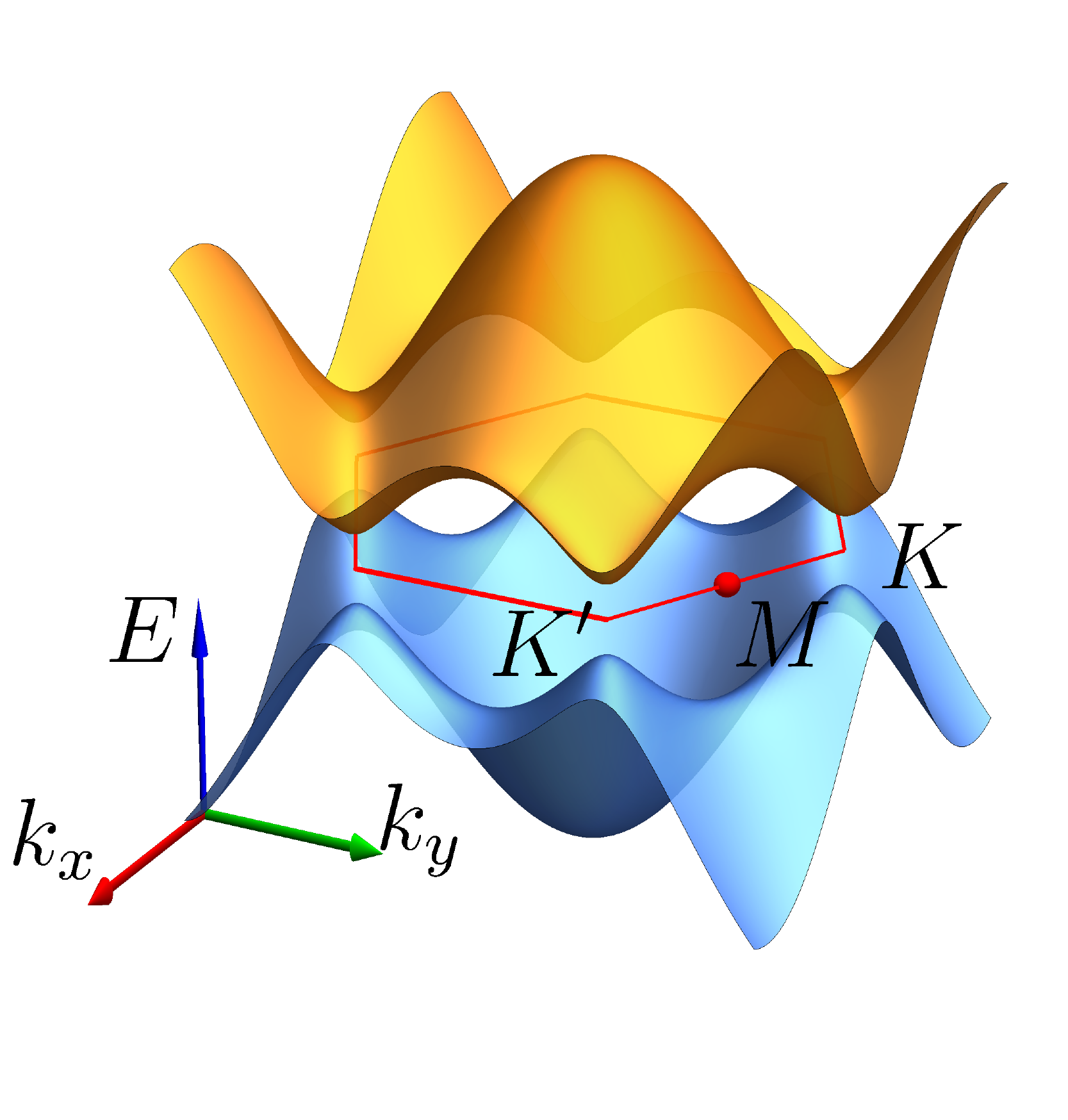}
\put(4,10){(b)}
\end{overpic}
\end{tabular}
\end{center}
\caption{(Color online) The energy dispersion of Majorana particles, clearly symmetrical with respect to the mid-point. 
Panel (a) $g/J = 0$. Note the conical intersection at the corners of the Brillouin zone ({\it i.e.} at the inequivalent points ${\bm K}$ and ${\bm K}'$).
Panel (b) $g/J = 0.06$. A gap opens and the two bands are separated in energy. Creating a fermionic excitation costs a finite amount of energy proportional to $g$.
In both cases, a saddle point (van-Hove singularity) is present at the ${\bm M}$ point of the Brillouin zone ({\it i.e.} at the mid-point of its side).
\label{fig:four}
}
\end{figure}

The Hamiltonian~(\ref{eq:Hamil_QSL_Majorana_eff_main}) is quadratic in the Majorana (spinon) operators $c_i$ and, given that in the zero-flux sector all $u_{ij}^\gamma$ and $D_{j}$ have the same value, it can be easily diagonalized~\cite{Kitaev_2006}. Details are given in App.~\ref{app:Majorana_flux_free}. In Fig.~\ref{fig:four} we plot the energy dispersion of the Majorana particles for two values of the magnetic field, namely $g=0$ [panel (a)] and $g=0.05J$ [panel (b)]. We obtain two bands symmetric around $E=0$. 
For $g=0$, similarly to graphene, the band structure exhibits Dirac cones at the two inequivalent points ${\bm K}$ and ${\bm K}'$ of the hexagonal Brillouin zone~\cite{Kitaev_2006}. This in turn implies that spinons can be created at no cost at zero magnetic field, {\it i.e.} their dispersion is gapless. This does {\it not} imply that {\it spin} excitations are gapless. In fact, such excitations are complex objects born of the fusion of spinons and fluxes (the $c$ and $b$ Majorana operators), and require the creation of both Majorana particles and flux pairs in neighboring plaquettes~\cite{Kitaev_2006,Knolle_thesis}. The latter cost a finite amount of energy $\Delta_F$ which is in fact the minimum energy of spin excitations. Note also that, as graphene~\cite{Castro_Neto_rmp_2009}, the band structure of the Kitaev model for $g=0$ exhibits van-Hove singularities at the ${\bm M}$ points of the Brillouin zone. As we shall see in Sect.~\ref{sect:results}, the spinon density-of-states diverges logarithmically at the energy corresponding to such singularities.

A gap opens at the points ${\bm K}$ and ${\bm K}'$ of the spinon Brillouin zone when the magnetic field ${\bm h}$ is turned on~\cite{Kitaev_2006} ({\it i.e.} for $g\neq 0$). Its effect is highly nontrivial: the second term on the right-hand side of Eq.~(\ref{eq:Hamil_QSL_Majorana_eff_main}) plays a role analogous to the Haldane term for electrons in a hexagonal lattice~\cite{Haldane_prl_1988}. In fact, one can show that such term breaks time-reversal symmetry~\cite{Kitaev_2006} and opens a gap in the spinons band dispersion (see Fig.~\ref{fig:four}). The gap has opposite signs in the two valleys and introduces a nontrivial spinon topology. 
This term stabilizes (i) edge spinons in finite systems and (ii) Majorana bound states at vortex cores~\cite{Knolle_thesis}.

\subsection{The spin structure factor}
\label{sect:KitaevQ}
We now describe the derivation of the spin structure factor, which crucially determines the tunneling current aided by magnetic excitations.
At any given temperature $T$, the spin structure factor $Q({\bm q},\omega)$ that appears in Eq.~(\ref{eq:chi_AA_res_final}) can be derived from the correlation function~\cite{Baskaran_prl_2007,Knolle_thesis} for the spin polarization $\gamma$ defined on the imaginary-time axis~\cite{Fetter_Walecka,Mahan_book,Giuliani_and_Vignale}:
\be \label{eq:spin_spin_corr_summed_main}
Q^\gamma({\bm r}_i,{\bm r}_{i'},\tau) &=& 
- \langle {\cal T} s_i^\gamma(\tau) s_{i'}^\gamma \rangle
~.
\ee
Here the imaginary-time evolution of the spin operator ${\bm s}_i(\tau)$ is generated by the effective Hamiltonian ${\cal H}_{\rm m}^{\rm eff}$ of Eq.~(\ref{eq:Hamil_QSL_Majorana_eff_main}), $\langle \ldots \rangle$ represents the average over a thermal state, while ${\cal T}$ is the imaginary-time ordering~\cite{Giuliani_and_Vignale}. From Eq.~(\ref{eq:spin_spin_corr_summed_main}) we obtain the spin structure factor for a given spin polarization $\gamma$, $Q^\gamma({\bm q},\omega)$, by first taking its Fourier transform in both space and imaginary time [the latter is restricted to the finite interval~\cite{Giuliani_and_Vignale} $-(k_{\rm B}T)^{-1}\leq \tau \leq (k_{\rm B}T)^{-1}$] and by then analytically continuing the result to real frequencies~\cite{Fetter_Walecka,Mahan_book,Giuliani_and_Vignale}. In App.~\ref{app:QSL_spin_spin} we prove that $Q^\gamma({\bm r}_i,{\bm r}_{i'},\tau)$ depends only on ${\bm r}_i-{\bm r}_{i'}$, and therefore its Fourier transform depends only on one momentum variable, ${\bm q}$. Finally, the spin structure factor needed in Eq.~(\ref{eq:chi_AA_res_final}) is obtained by summing over all spin polarizations: $\Im m Q({\bm q}, \omega) = \sum_{\gamma} \Im m Q^\gamma({\bm q}, \omega)$. 

We now summarize the calculation~\cite{Baskaran_prl_2007,Knolle_thesis,Knolle_prl_2014,Knolle_prb_2015} of $\Im m Q^\gamma ({\bm q},\omega)$. More details are given in App~\ref{app:QSL_spin_spin}.
Starting from Eq.~(\ref{eq:spin_spin_corr_summed_main}), we rewrite it as
\be \label{eq:QSL_T_ord_corr_1_main}
Q^\gamma({\bm r}_i,{\bm r}_{i'},\tau) &=& \langle {\cal T} b_i^\gamma(\tau) c_{i}(\tau) b_{i'}^\gamma c_{{i'}} \rangle
\nn
&=&
\langle {\cal T} c_{i}(\tau) S_{i,\gamma}(\tau) c_{i'} (ib_{i}^\gamma b_{i'}^\gamma ) \rangle
~.
\ee
In the second line of Eq.~(\ref{eq:QSL_T_ord_corr_1_main}) we commuted the $b_i$ operator with the time evolution, in order to isolate the product $ib_i^\gamma b_{i'}^\gamma$, and as a result obtained
\be \label{eq:S_igamma_tau}
S_{i,\gamma}(\tau) &\equiv& e^{{\cal H}_{\rm m}^{\rm eff} \tau}e^{-({\cal H}_{\rm m}^{\rm eff} + V^{(1)}_{i,\gamma} + V^{(2)}_{i,\gamma}) \tau} 
\nn
&=&
{\cal T} \exp\left(- \int_0^\tau d\tau' \big[V^{(1)}_{i,\gamma}(\tau') + V^{(2)}_{i,\gamma}(\tau')\big] \right)
~.
\ee
Eq.~(\ref{eq:QSL_T_ord_corr_1_main}) strongly resembles the calculation of a time-dependent correlation function following a quench. Such analogy has been noted in Refs.~\onlinecite{Baskaran_prl_2007,Knolle_thesis,Knolle_prl_2014,Knolle_prb_2015}, which have used it to evaluate $Q^\gamma({\bm r}_i,{\bm r}_{i'},\tau)$. In this context, $S_{i,\gamma}(\tau)$ plays the role of the $S$-matrix usually encountered in many-body problems~\cite{Fetter_Walecka,Mahan_book,Giuliani_and_Vignale}, which stems from the potential $V^{(1)}_{i,\gamma}(\tau') + V^{(2)}_{i,\gamma}(\tau')$ (in the interaction picture) being turned on between the time $\tau'=0$ and $\tau'=\tau$. The specific form of $V^{(1)}_{i,\gamma}$ and $V^{(2)}_{i,\gamma}$ is rather complicated and is given in App.~\ref{app:QSL_spin_spin}. We briefly comment on their origin. The operator $b_i^\gamma$ does {\it not} commute with the generator of the time evolution, the effective Hamiltonian ${\cal H}_{\rm m}^{\rm eff}$ in Eq.~(\ref{eq:Hamil_QSL_Majorana_eff_main}). In particular, its action on the Hamiltonian is to flip the sign of all operators $u_{\ell \ell'}^\alpha$ such that $\alpha = \gamma$ and either $\ell$ or $\ell'$ is equal to $i$. The two terms $V^{(1)}_{i,\gamma}$ and $V^{(2)}_{i,\gamma}$ emerge when the operator $b_i^\gamma$ is commuted with the first and second terms of ${\cal H}_{\rm m}^{\rm eff}$ defined in Eq.~(\ref{eq:Hamil_QSL_Majorana_eff_main}), respectively. By changing the sign of bond operators, $b_i^\gamma$ introduces fluxes in two of the three plaquettes that contain the site $i$ (depending on the bond-type $\gamma$). Therefore, Eqs.~(\ref{eq:QSL_T_ord_corr_1_main})-(\ref{eq:S_igamma_tau}) describe the response to a quantum quench consisting in the introduction of fluxes in the otherwise flux-free state~\cite{Baskaran_prl_2007,Knolle_thesis,Knolle_prl_2014,Knolle_prb_2015}.

We now observe that, since all $u_{\ell \ell'}^\alpha$ commute with the Kitaev Hamiltonian, the density matrix used to take the average $\langle\ldots\rangle$ factorizes into a product of two density matrices, one for spinons and one for fluxes~\cite{Baskaran_prl_2007}. In line with the choice of working with the effective Hamiltonian ${\cal H}_{\rm m}^{\rm eff}$, we will assume that the density matrix for the flux sector represents a pure state with no fluxes~\cite{Kitaev_2006}. The average of $ib_i^\gamma b_{i'}^\gamma$ is then nonzero and equal to $i$, if $i={i'}$, or to $u_{ij}^\gamma = \pm 1$, if $i$ and ${i'}=j$ are nearest-neighbors connected by a bond of type $\gamma$. The sign here depends on whether the site $i$ is of type $A$ or $B$. To account for all this explicitly and in a convenient way, we now choose the unit cell of the model to include the site $i$ and its nearest neighbor $j$ along the bond of type $\gamma$. The site $j$ coincides with ${i'}$ if ${i'}\neq i$. We then introduce the fermionic operators $f_{\bm r} = (c_{i} +i c_{j})/2$ and $f_{\bm r}^\dagger = (c_{i} - i c_{j})/2$, if $i$ and $j$ are sites of type $A$ and $B$, respectively. In the opposite case ($i$ is of type $B$ and $j$ of type $A$), the two sites are swapped in the definitions of $f_{\bm r}$ and $f_{\bm r}^\dagger$. Here ${\bm r}$ is the position of the unit cell containing both sites. Eq.~(\ref{eq:QSL_T_ord_corr_1_main}) then becomes
\begin{equation} \label{eq:QSL_T_ord_corr_6_rew_main}
Q^\gamma({\bm r}_i,{\bm r}_{i'},\tau) =
\varsigma_{\alpha\alpha'}
\langle {\cal T} \big[f_{\bm r}(\tau) + \eta_{\alpha} f_{\bm r}^\dagger(\tau)\big] S_{{\bm r}}(\tau) (f_{\bm r} + \eta_{\alpha'} f_{\bm r}^\dagger) \rangle
,
\end{equation}
where $\varsigma_{\alpha\alpha'} = - (\sigma^z_{\alpha\alpha'} + i \sigma^y_{\alpha\alpha'})$, $\sigma^y$ and $\sigma^z$ are two Pauli matrices, while $\alpha$ and $\alpha'$ are, respectively the types of sites $i$ and ${i'}$ ($A$ or $B$). In this equation we introduced $\eta_\alpha$ such that $\eta_A = 1$ and $\eta_B = -1$. The operator $S_{{\bm r}}(\tau)$ is obtained from that in Eq.~(\ref{eq:S_igamma_tau}) by going to the fermionic basis. To simplify the following calculation, we neglect the term $V^{(2)}_{i,\gamma}(\tau')$ in Eq.~(\ref{eq:S_igamma_tau}). In fact, such term is proportional to $g$ and, in the limit of $g\ll J$ ({\it i.e.} for magnetic fields $|{\bm h}|\lesssim 10~{\rm T}$ -- see results below) is negligible compared to $V^{(1)}_{i,\gamma}(\tau')$. Retaining only the latter term, we get
\be \label{eq:S_r_tau}
S_{{\bm r}}(\tau) \equiv
{\cal T} \exp\left(2 J \int_0^\tau d\tau' \big[2 f_{\bm r}^\dagger(\tau') f_{\bm r}(\tau') - 1\big] \right) 
~.
\ee
Notably, $S_{{\bm r}}(\tau)$, which describes the quench following the introduction of fluxes in an otherwise flux-free state, acquires now the same form of the S-matrix due to the interaction of $f$-fermions with a localized impurity potential~\cite{Bruus_Flensberg}. This observation makes the problem exactly soluble, once the following further approximation is introduced.

\begin{figure}[t]
\begin{center}
\begin{tabular}{c}
\begin{overpic}[width=0.99\columnwidth]{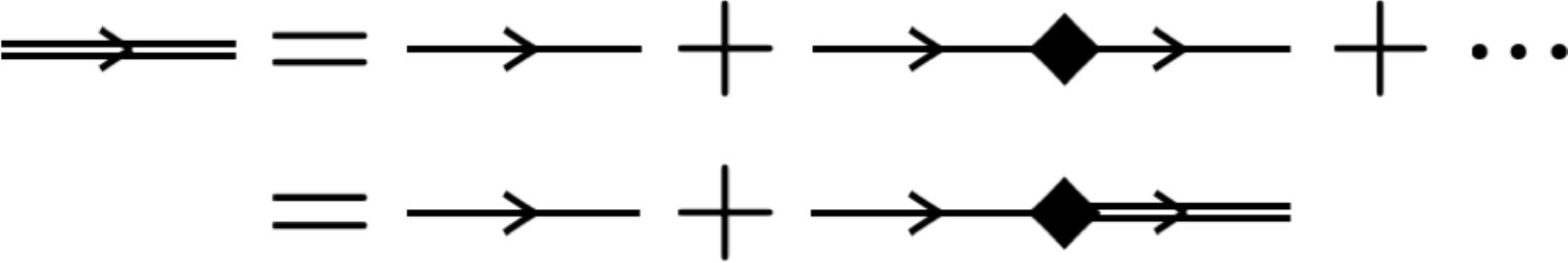}
\end{overpic}
\end{tabular}
\end{center}
\caption{(Color online) The series of Feynman diagrams needed to determine the spin correlation function of the Kitaev model. Single (double) lines denote bare (dressed) Green's functions, whereas diamonds stand for insertions of $V_{\rm imp}({\bm r})$. Note that the series is identical to that needed to determine the Green's function of electrons in graphene in the presence of a single impurity. 
\label{fig:five}
}
\end{figure}

Instead of solving the full quantum-quench problem, that can only be tackled numerically, we utilize the ``adiabatic approximation'' introduced in Refs.~\onlinecite{Knolle_thesis,Knolle_prl_2014,Knolle_prb_2015}. A detailed comparison between such approximation and numerical simulations has been performed in, e.g., Ref.~\onlinecite{Knolle_thesis}. There, the adiabatic approximation has been shown to well reproduce the features of the spin structure factor of the Kitaev model. This assumes that the potential in the $S$-matrix~(\ref{eq:S_r_tau}) is {\it adiabatically} turned on for the whole duration of the imaginary-time evolution~\cite{Giuliani_and_Vignale}, {\it i.e.} until $\tau=\beta$.
We therefore multiply and divide Eq.~(\ref{eq:QSL_T_ord_corr_6_rew_main}) by the same quantity, $\langle {\cal T} S_{{\bm r}}(\tau) \rangle$, and then set $\tau=\beta= (k_{\rm B} T)^{-1}$ in $S_{{\bm r}}(\tau)$ in the ratio $\langle {\cal T} \big[f_{\bm r}(\tau) + \eta_{\alpha} f_{\bm r}^\dagger(\tau)\big] S_{{\bm r}}(\tau) (f_{\bm r} + \eta_{\alpha'} f_{\bm r}^\dagger) \rangle/\langle {\cal T} S_{{\bm r}}(\tau) \rangle$. 
One then recognizes such ratio as the Green's function of graphene-like electrons in the presence of a single impurity located at a given lattice site. Such Green's function can be calculated {\it exactly} by resumming an infinite (geometric) series of Feynman diagrams,~\cite{Giuliani_and_Vignale,Bruus_Flensberg} see Fig.~\ref{fig:five}. There, crosses correspond to the impurity potential $V_{\rm imp}({\bm r}) = 4 J f_{\bm r}^\dagger f_{\bm r}$.

One can also prove that $\langle {\cal T} f_{\bm r}(\tau) f_{\bm r} \rangle = \langle {\cal T} f^\dagger_{\bm r}(\tau) f^\dagger_{\bm r} \rangle =  0$ because of the sublattice symmetry of the model~\cite{Knolle_thesis}. Since the S-matrix does not feature anomalous couplings between fermions (it represents a simple scalar potential), the {\it dressed} anomalous Green's functions obtained by resumming the corresponding Feynman diagrams are zero. Therefore, Eq.~(\ref{eq:QSL_T_ord_corr_6_rew_main}) now becomes
\begin{equation}
\label{eq:QSL_T_ord_corr_7b_main}
Q^\gamma({\bm r}_i,{\bm r}_{i'},\tau) = 
\big[Q_{{\rm c}}^\gamma({\bm r},\tau,0) - \eta_\alpha \eta_{\alpha'} Q_{{\rm c}}^\gamma({\bm r},0,\tau)\big]
\langle {\cal T} S_{{\bm r}}(\tau) \rangle 
~,
\end{equation}
where
\be \label{eq:Q_c_Q_L_def_main}
&&
Q_{{\rm c}}^\gamma({\bm r},\tau,\tau') \equiv - \frac{ \langle {\cal T} f_{\bm r}(\tau) f_{\bm r}^\dagger(\tau') S_{{\bm r}}(\beta) \rangle }{ \langle {\cal T} S_{{\bm r}}(\beta) \rangle } 
~,
\ee
is the dressed Green's function. Its Fourier transform reads~\cite{Knolle_thesis,Knolle_prl_2014,Knolle_prb_2015}
\be \label{eq:Q_c_result_main}
Q_{{\rm c}}^\gamma({\bm r},\omega) = \frac{Q_{{\rm c},0}^\gamma({\bm r},\omega)}{1 + 4 J Q_{{\rm c},0}^\gamma({\bm r},\omega)}
~,
\ee
where $Q_{{\rm c},0}^\gamma({\bm r},\tau,\tau')$, the anti-Fourier transform of $Q_{{\rm c},0}^\gamma({\bm r},\omega)$, is obtained from Eq.~(\ref{eq:Q_c_Q_L_def_main}) by replacing $S_{{\bm r}}(\beta) \to 1$ ({\it i.e.} it is the non-interacting Green's function of the $f$-fermions). As shown in App.~\ref{eq:non_int_Q}, in the flux-free sector $Q_{{\rm c},0}^\gamma({\bm r},\omega)$ and $Q_{{\rm c}}^\gamma({\bm r},\tau,\tau')$ are actually independent of the position ${\bm r}$. Their dependence on such variable will therefore be neglected in what follows.

\begin{figure}[t]
\begin{center}
\begin{tabular}{c}
\begin{overpic}[width=0.99\columnwidth]{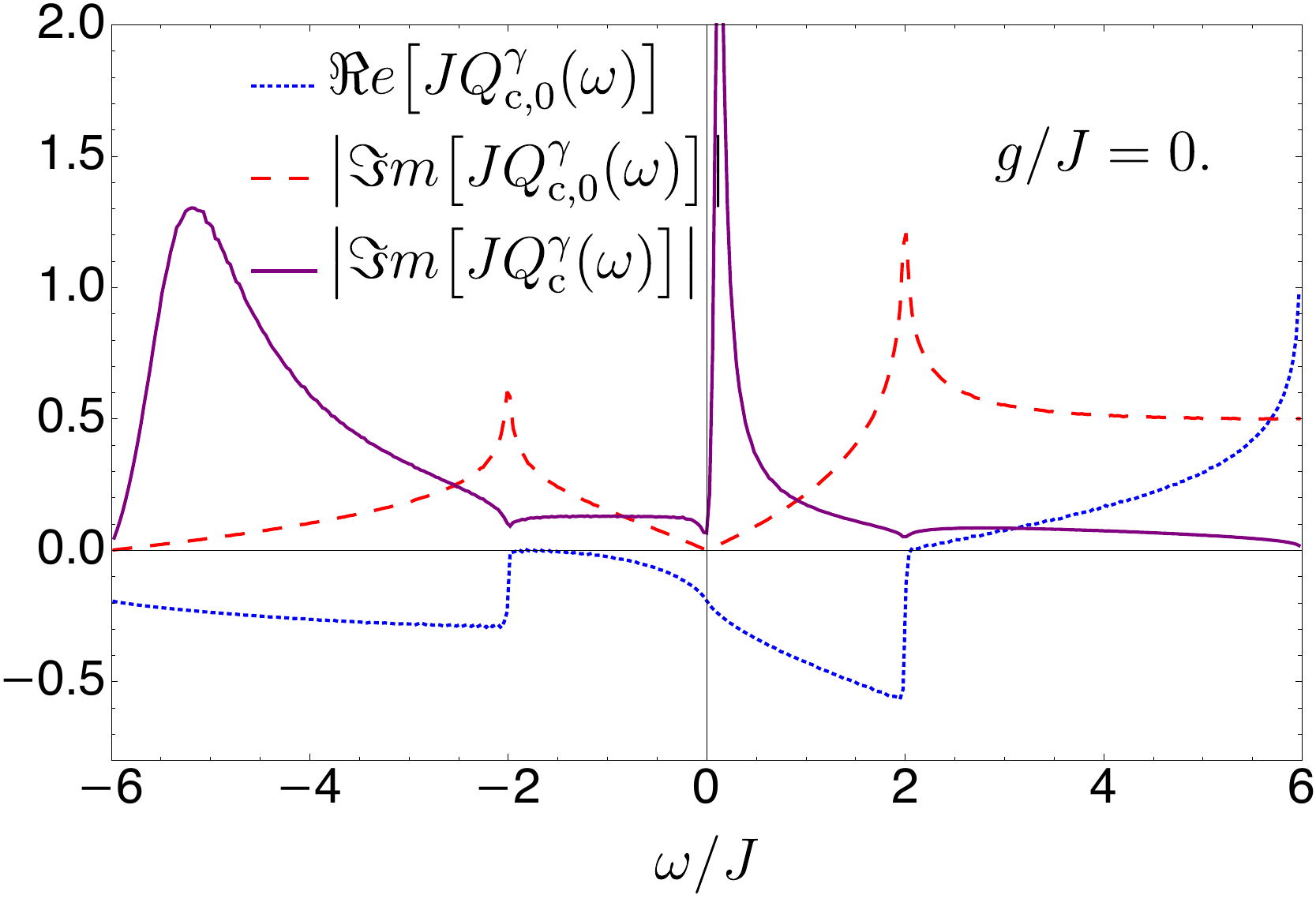}
\put(25,35){(a)}
\end{overpic}
\\
\begin{overpic}[width=0.99\columnwidth]{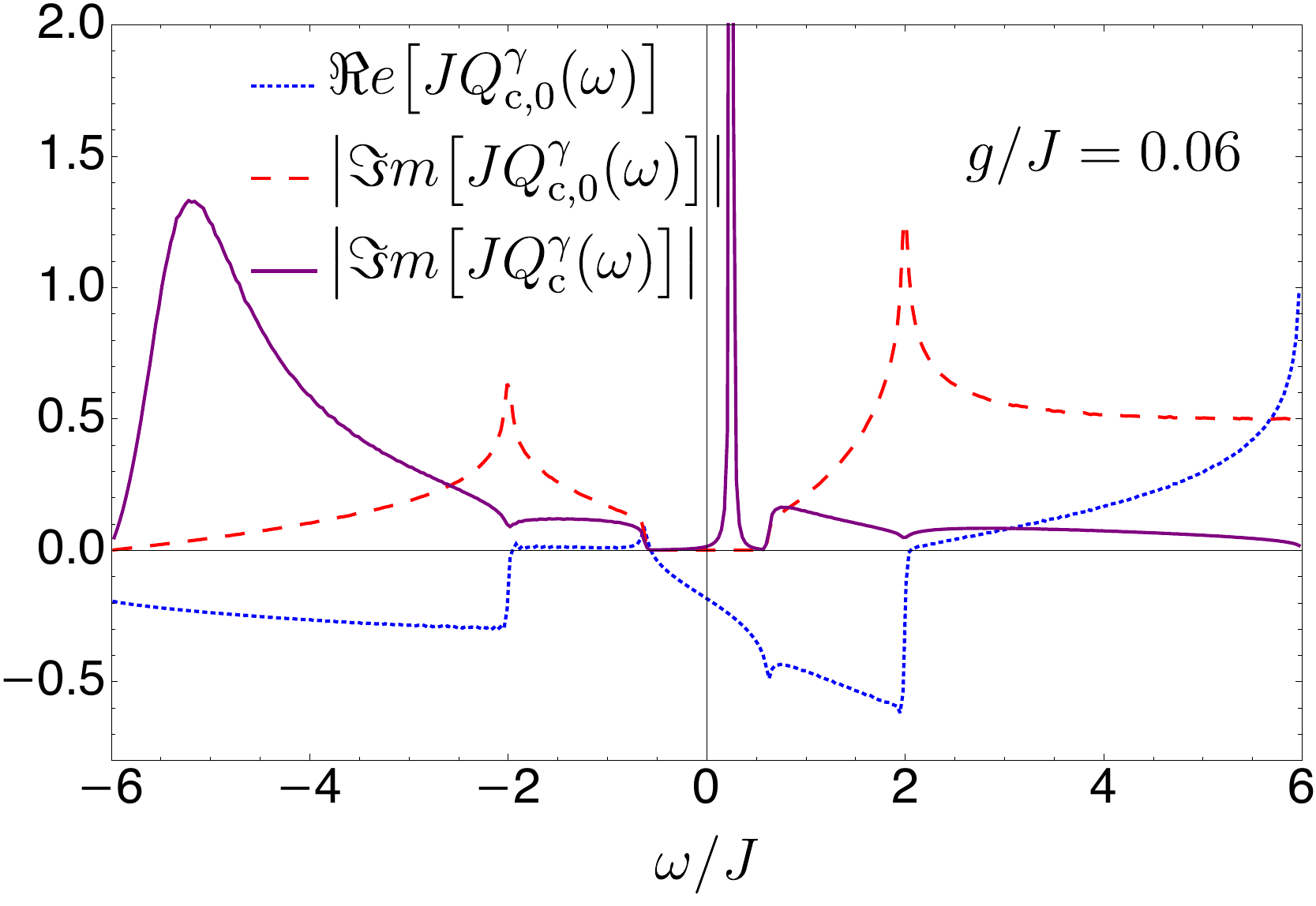}
\put(25,35){(b)}
\end{overpic}
\end{tabular}
\end{center}
\caption{(Color online) The Majorana Green's functions as a function of the energy $\omega$, in units of $J^{-1}$.
Panel (a) $g/J = 0$. 
Panel (b) $g/J=0.06$. A broadening has been added by shifting $\Im m\big[Q_{{\rm c},0}^\gamma(\omega)\big] \to \Im m\big[Q_{{\rm c},0}^\gamma(\omega)\big] + 10^{-3} J^{-1}$  to evidence the below-the-gap bound state.
\label{fig:six}
}
\end{figure}

In Fig.~\ref{fig:six}, we show plots of $\Re e\big[Q_{{\rm c},0}^\gamma(\omega)\big]$, $\Im m\big[Q_{{\rm c},0}^\gamma(\omega)\big]$, as well as of $\Im m\big[Q_{{\rm c}}^\gamma(\omega)\big]$ for two values of the parameter $g$ ({\it i.e.} of the magnetic field ${\bm h}$). These results agree with the ones reported in Refs.~\onlinecite{Knolle_thesis,Knolle_prl_2014,Knolle_prb_2015}. At $g=0$, $\Im m\big[Q_{{\rm c},0}^\gamma(\omega)\big]$ exhibits a continuum of spinon excitations that grows linearly for small $|\omega|$ and diverges at the energy corresponding to the van-Hove singularity. This behavior strongly resembles that of the DOS of graphene treated within the nearest-neighbor tight-binding approximation~\cite{Castro_Neto_rmp_2009}. Note that the singularity in $\Im m\big[Q_{{\rm c},0}^\gamma(\omega)\big]$ translates in a dip of $\Im m\big[Q_{{\rm c}}^\gamma(\omega)\big]$. At finite $g$, a gap opens in both $\Im m\big[Q_{{\rm c},0}^\gamma(\omega)\big]$ and $\Im m\big[Q_{{\rm c}}^\gamma(\omega)\big]$ at low energies $|\omega|$. Such gap refers to the spinons and does not incorporate the energy cost of the two-flux insertion $\Delta_F$, hence it is absent at zero magnetic field. A fundamental feature of $\Im m\big[Q_{{\rm c}}^\gamma(\omega)\big]$ for $g=0$ is the peak at low positive energies, which evolves in a true below-the-gap resonance at finite $g$, as shown in panel (b) of Fig.~\ref{fig:six}. Such feature corresponds to Majorana particles bound to fluxes~\cite{Knolle_thesis}. 

As we show below, the Majorana bound states appear as sharp peaks in the spin structure factor~\cite{Knolle_thesis}, whereas spinons contribute a continuum of excitations at energies above $\Delta_F$. The energies of both types of excitations grow with the magnetic field. In fact, at finite magnetic field, both the creation of Majorana particles and the insertion of fluxes cost a finite amount of energy. The extra energy required to add Majorana particles grows linearly in $g$ (and therefore cubicly in $|{\bm h}|$). It is such peculiar dependence of the energy of excitations with magnetic field that can constitute a proof of a quantum-spin-liquid phase in ${\rm RuCl}_3$.

Finally, as explained in App.~\ref{app:QSL_spin_spin}, the term $\langle {\cal T} S_{{\bm r}}(\tau) \rangle$ in Eq.~(\ref{eq:QSL_T_ord_corr_7b_main}) is manipulated to give 
$\langle {\cal T} S_{{\bm r}}(\tau,0) \rangle  \simeq e^{-\Delta_F\tau}$. Such exponential factor has a fundamental physical effect. At low temperatures it suppresses the response below the two-flux excitation energy and is therefore responsible for the gap in spin excitations even at zero magnetic field~\cite{Knolle_thesis,Knolle_prl_2014,Knolle_prb_2015}.
To show this, we take the Fourier transform of Eq.~(\ref{eq:QSL_T_ord_corr_7b_main}) with respect to both time and space. Since $Q^\gamma({\bm r}_i,{\bm r}_{i'},\tau)$ is expressed in terms of the imaginary time, when we take its Fourier transform with respect to such variable, we obtain the coefficients $Q^\gamma({\bm r}_i,{\bm r}_{i'},i\omega_n)$ of the corresponding Matsubara sum. Here $\omega_n =\pi  k_{\rm B} T (2n+1)$ is the fermionic Matsubara frequency. To get the spin structure factor, we then have to analytically continue these coefficients to the real-frequency axis, {\it i.e.} we must take $i\omega_n\to \omega + i0^+$ (for $\omega>0$). All details of the calculation are given in App.~\ref{sect:QSL_spin_spin_momentum_frequency}. Here we quote the final result, which in the limit of zero temperature reads
\be\label{eq:ImQ_semifinal_main}
\Im m\big[Q^\gamma({\bm r}_i,{\bm r}_{i'},\omega)\big] &=& 
\Theta(\omega-\Delta_F)
\Big\{
\Im m\big[Q_{{\rm c}}^\gamma(\omega - \Delta_F)\big] 
\nn
&+&
\eta_\alpha \eta_{\alpha'} \Im m\big[Q_{{\rm c}}^\gamma(\Delta_F-\omega)\big]
\Big\}
~.
\ee
It is clear that the obtained function vanishes for $0<\omega<\Delta_F$ and therefore a gap appears in the spectrum of spin excitations. 
Each process described by $\Im m\big[Q^\gamma({\bm r}_i,{\bm r}_{i'},\omega)\big]$ corresponds to the simultaneous insertion of fluxes in neighboring plaquettes and creation/annihilation of spinons. Each spin excitation is in fact, by virtue of the fractionalization introduced in Eq.~(\ref{eq:spin_majorana_main}), the combination of a spinon creation and flux insertion and the two are inextricably connected. While, for $g=0$, spinon excitations can be created at no cost, inserting fluxes in the system costs a finite amount of energy $\Delta_F$ and therefore spin excitations are, as a whole, gapped. At finite magnetic field, both flux and spinon creation are gapped, with the cost of both increasing approximately linearly with $g\propto |{\bm h}|^3$. Therefore, the overall gap increases.

Next, we take the Fourier transform over space variables in Eq.~(\ref{eq:ImQ_semifinal_main}). We first recall that ${\bm r}_i$ and ${\bm r}_{i'}$ either coincide or are two nearest neighbors~\cite{Knolle_thesis,Knolle_prl_2014,Knolle_prb_2015}. Furthermore, in App.~\ref{eq:non_int_Q} we show that $Q_{{\rm c}}^\gamma({\bm r},\omega)\equiv Q^\gamma_{{\rm c}}(\omega)$ is independent of ${\bm r}$. Using these two facts and the expression reported in Eq.~(\ref{eq:Q_c_result_main}) we find (see also App.~\ref{sect:QSL_spin_spin_momentum_frequency})
\begin{eqnarray}
\label{eq:ImQ_final_FT_main}
\Im m\big[Q^{\gamma}({\bm q},\omega)\big] &=& 
2\Theta(\omega-\Delta_F)
\Big\{
\Im m\big[Q_{{\rm c}}^\gamma(\omega - \Delta_F)\big] 
f_{{\bm q},+}^\gamma
\nn
&+&
\Im m\big[Q_{{\rm c}}^\gamma(\Delta_F,\omega)\big]
f_{{\bm q},-}^\gamma
\Big\}
~,
\end{eqnarray}
where $f_{{\bm q},\pm}^\gamma \equiv 1 \pm \cos({\bm q}\cdot {\bm \delta}_{\rm K}^\gamma)$.
We recall that ${\bm \delta}_{\rm K}^\gamma$ are the positions of the three nearest neighbor of an atom of type $A$ in the direction $\gamma = x,y,z$.

\section{Signatures of fractionalized excitations}
\label{sect:results}
The goal of this section is to highlight signatures in tunneling current, differential conductance $G(V)$ and, in particular, IETS~\cite{Ghazaryan_natureel_2018} ($\equiv dG/dV$) that can be unequivocally attributed to the excitation of fractionalized quasiparticles. 
To distinguish spin-conserving from spin-non-conserving tunneling events, one can study the behavior of $dG/dV$ as a function of the applied magnetic field~\cite{Ghazaryan_natureel_2018} ${\bm h}$. Tunneling processes involving the emission of, e.g., phonons or other non-magnetic quasiparticles are in fact not susceptible to variations of ${\bm h}$. 

Resonances in the IETS occur at the energy of the magnetic quasiparticles excited in the tunneling process, as we proceed to show. Since the focus of this paper is on the properties of the magnetic insulator rather than on the graphite itself, we will from now on assume the two graphene slabs to be doped with typical electron concentrations $\sim 10^{12}~{\rm cm}^{-2}$. This will allow us to neglect most of the features of the graphene sheets themselves, for example the reconstruction of the band structure due to the formation of the superlattice, to concentrate on the physics of the magnetic insulator. At such densities, the chemical potentials of the two graphene sheets $\mu_{\rm t}$ and $\mu_{\rm b}$ are of the order of $\sim 100~{\rm meV}$, much larger than the typical temperatures. Note that the massless Dirac fermion approximation for graphene electrons is valid up to Fermi energies of several hundreds of ${\rm meV}$~\cite{Castro_Neto_rmp_2009} (densities up to $\sim 7-10\times 10^{12}~{\rm cm}^{-2}$), and therefore is applicable to the current problem.

As shown in Eq.~(\ref{eq:ImQ_final_FT_main}), the spin structure factor of the Kitaev model $Q({\bm q}+{\bm G}_n,\omega)$ has two contributions. One is momentum independent, while the other depends on momentum as $\sum_\gamma \cos\big[({\bm q}+{\bm G}_n)\cdot {\bm \delta}_K^\gamma\big]$. For the materials we will consider~\cite{Sears_prb_2015}, $|{\bm \delta}_{\rm K}| \lesssim 0.7~{\rm nm}$. At low temperatures, when electron tunneling occurs between Fermi-surface states, the exchanged momenta are constrained by $|k_{{\rm F},{\rm t}}-k_{{\rm F},{\rm b}}|\leq |{\bm q}|\leq k_{{\rm F},{\rm t}}+k_{{\rm F},{\rm b}}$, where $k_{{\rm F},{\rm t}}$ and $k_{{\rm F},{\rm b}}$ are the Fermi momenta of the top and bottom graphene layers, respectively. For typical doping concentrations ($n\sim 10^{12}~{\rm cm}^{-2}$), $k_{{\rm F},{\rm t}} \sim k_{{\rm F},{\rm b}} \sim 0.1~{\rm nm}^{-1}$. Therefore, the product ${\bm q}\cdot {\bm \delta}_K \lesssim 0.14$ can be taken to be zero. Note that $Q({\bm q} + {\bm G}_n,\omega) \simeq Q({\bm G}_n,\omega)$ still depends on the twist angle $\theta$ between the two graphene sheets, as well as on the angle between these and the insulator, via its dependence on ${\bm G}_n$. We remind the reader that generic formulas for the tunneling current are given in Eqs.~(\ref{eq:chi_AA_res_final})-(\ref{eq:chi_tb_res_final_2}), while the dependence of the spin structure factor has been made explicit in Eq.~(\ref{eq:ImQ_final_FT_main}).

\begin{figure}[t]
\begin{center}
\begin{tabular}{c}
\begin{overpic}[width=0.7\columnwidth]{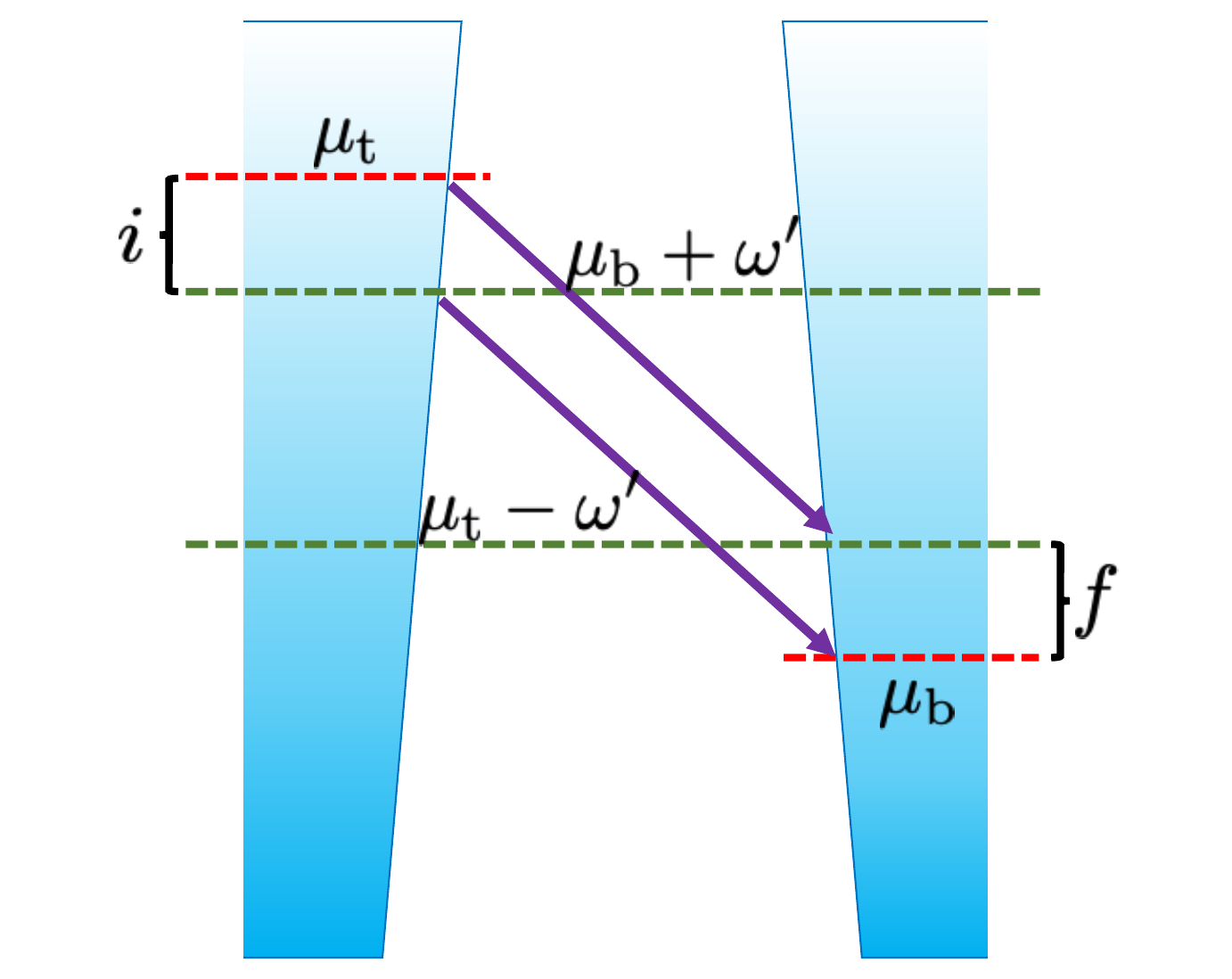}
\end{overpic}
\end{tabular}
\end{center}
\caption{(Color online) A schematics of the tunneling processes described by Eq.~(\ref{eq:chi_AA_simplification2}) in the limit of zero temperature. The shaded trapezoids represent the graphene DOS. Quantum-spin-liquid excitations (arrows) with energies $0\leq \omega'\leq \mu_{\rm t}-\mu_{\rm b}$ enable the tunneling of electrons with energies $\mu_{\rm b}+\omega'\leq \varepsilon \leq \mu_{\rm t}$ (marked as ``$i$'' in the figure). After tunneling, these fill states with energies $\mu_{\rm b}\leq \varepsilon\leq \mu_{\rm t}-\omega'$ (``$f$'' in the figure).
\label{fig:seven}
}
\end{figure}

Since the spin structure factor of the magnetic layer can be taken to be independent of ${\bm q}$, it becomes now possible to perform the sum over such momentum variable in Eq.~(\ref{eq:chi_AA_res_final}). After few manipulations, the tunneling current in Eq.~(\ref{eq:current_final_expression}) finally reads
\be \label{eq:chi_AA_simplification2}
I &=& 
\mp 
\alpha_0
\int_{-\infty}^\infty d\omega' 
\big[n_{{\rm F}/{\rm B}}(\omega'-e V) - n_{{\rm F}/{\rm B}}(\omega')\big]
\nn
&\times&
\Im m {\bar Q}(\omega')
N_{\rm tb}(\omega',eV)
~.
\ee
where we introduced $\Im m {\bar Q}(\omega) \equiv \sum_{n=0}^2 \Im m Q({\bm G}_n, \omega)$, $\alpha_0 = 2 \pi \Lambda_0^2 (V_{\rm uc}^{({\rm G})})^2 N e$, the area of the graphene unit cell $V_{\rm uc}^{({\rm G})} = \sqrt{3} a_{\rm G}^2/2$ and the joint density of states (DOS)
\be
\label{eq:joint_DOS}
N_{\rm tb}(\omega',\omega) &=&
\int_{-\infty}^\infty d\varepsilon
\big[n_{\rm F}(\varepsilon) - n_{\rm F}(\varepsilon+\omega-\omega')\big]
\nn
&\times&
N_{\rm t}(\varepsilon + \mu_{\rm t})
N_{\rm b}(\varepsilon-\omega'+\mu_{\rm t})
~.
\ee
Here,
\be 
N_\ell(\varepsilon + \mu_\ell) = -\frac{1}{\pi N V_{\rm uc}} \sum_{{\bm k},\lambda} \Im m G_{\ell,\lambda}({\bm k},\varepsilon)
~,
\ee
is the DOS of the individual graphene sheet in the massless-Dirac-fermion approximation ($\ell={\rm t},{\rm b}$ is the layer index). We assume the magnetic field to be low enough to allow the graphene electrodes to remain in the semiclassical regime, in which the linear-in-energy expression for their DOS applies, {\it i.e.} $N_\ell(\varepsilon) = g_v|\mu_{\ell} + \varepsilon|/(2\pi v_{\rm F}^2)$. Since both graphene sheets are n-doped, $\mu_{\rm t},\mu_{\rm b} > 0$. A schematic representation of zero-temperature tunneling processes is shown in Fig.~\ref{fig:seven}. There, we show that magnetic excitations with energies $0\leq \omega'\leq \mu_{\rm t}-\mu_{\rm b}$ aid the tunneling of electrons with energies $\mu_{\rm b}+\omega'\leq \varepsilon \leq \mu_{\rm t}$ towards states in the range $\mu_{\rm b}\leq \varepsilon\leq \mu_{\rm t}-\omega'$

We consider the regime of low temperatures, {\it i.e.} $k_{\rm B}T \ll \mu_{\rm t}, \mu_{\rm b}, J$. By taking $e V \sim J \ll \mu_{\rm t}, \mu_{\rm b}$, we focus on the contribution of magnetic excitations to the interlayer tunneling. Under these approximations, the chemical potentials of the top and bottom layers can be taken to be nearly identical, {\it i.e.} $\mu_{\rm t} \simeq \mu_{\rm b}\equiv \mu$. We also note that, thanks to the Fermi and Bose distributions in Eq.~(\ref{eq:chi_AA_simplification2})-(\ref{eq:joint_DOS}), $\omega'\simeq \varepsilon \simeq e V$. Therefore, Eq.~(\ref{eq:chi_AA_simplification2}) can be approximated as
\be \label{eq:chi_AA_simplification4}
I(V) &=& 
\frac{e}{\hbar} \frac{4 g_v^2 N}{3 \pi} \left(\frac{\Lambda_0}{t_{\rm G}}\right)^2 \left(\frac{\mu}{t_{\rm G}}\right)^2 {\cal I}(V)
~,
\nn
\ee
where we defined
\be \label{eq:cal_I_def}
{\cal I}(V) &=& -\int_0^{eV}  d\omega' (eV-\omega')
\Im m {\bar Q}(\omega')
~.
\ee
Taking the second derivative with respect to $V$ of Eq.~(\ref{eq:cal_I_def}) we obtain the IETS
\be \label{eq:dGdV_ImQ}
\frac{dG}{dV} = -\frac{e}{\hbar} \frac{4 g_v^2 N}{3 \pi} \left(\frac{\Lambda_0}{t_{\rm G}}\right)^2 \left(\frac{\mu}{t_{\rm G}}\right)^2 \Im m {\bar Q}(eV)
~,
\ee
which is clearly proportional to the averaged spin structure factor $\Im m {\bar Q}(eV) \equiv \sum_{n=0}^2 \sum_\gamma \Im m Q^\gamma({\bm G}_n, \omega)$.
To keep the presentation concise, in what follows we will show results obtained by fixing $\theta = 3^\circ$ and $\phi = 5^\circ$. This choice is dictated by the purpose of avoiding small twist angles that give rise to large moir\'e periodicities. In this case, the superlattice has a size of $\approx 4~{\rm nm}$, well below the tens or even hundreds of nanometers of aligned graphene/hBN~\cite{Woods_natphys_2014} or magic-angle twisted bilayer graphene~\cite{Cao_nature_2018_1,Cao_nature_2018_2}. We note that samples of relatively large twist angles are much more common that nearly-aligned ones. Their production requires in fact advanced tear-and-stack techniques that have been developed only recently~\cite{Kim_nanolett_2016}. Therefore, our results are applicable to a large variety of devices. 
We have checked that results remain qualitatively similar in a range of $\theta$ and $\phi$ about $10^\circ$ around the chosen ones, as long as $\theta, \phi \gtrsim 2^\circ$. Therefore, the ones we discuss hereafter are good representative choices. For such twist angles, the product ${\bm G}_n\cdot {\bm \delta}_K\simeq 1$ and therefore the dependence of $\Im m Q^\gamma({\bm G}_n, \omega)$ on ${\bm G}_n$ cannot be neglected.

\begin{figure}[t]
\begin{center}
\begin{tabular}{c}
\begin{overpic}[width=0.99\columnwidth]{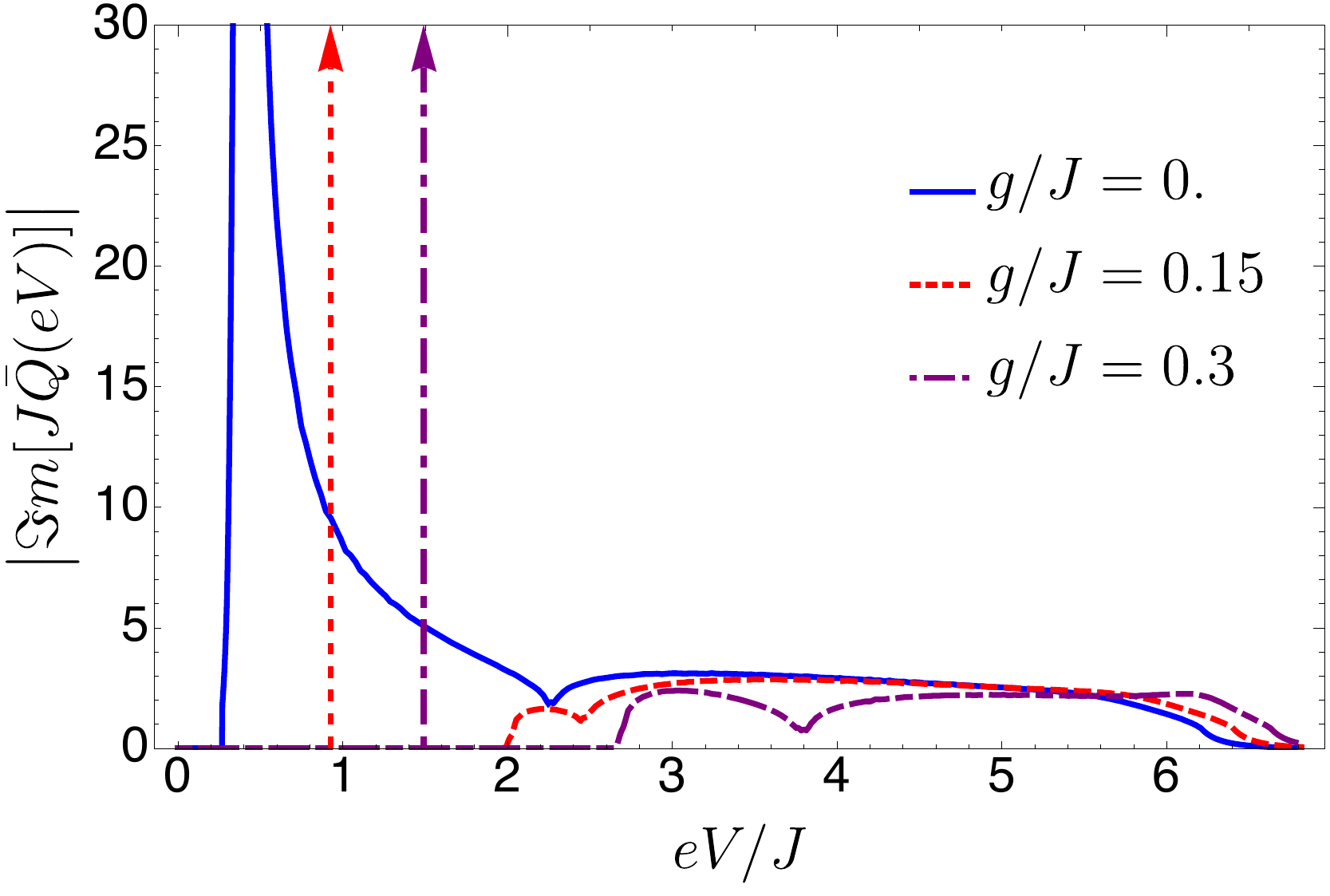}
\put(10,10){(a)}
\end{overpic}
\\
\begin{overpic}[width=0.99\columnwidth]{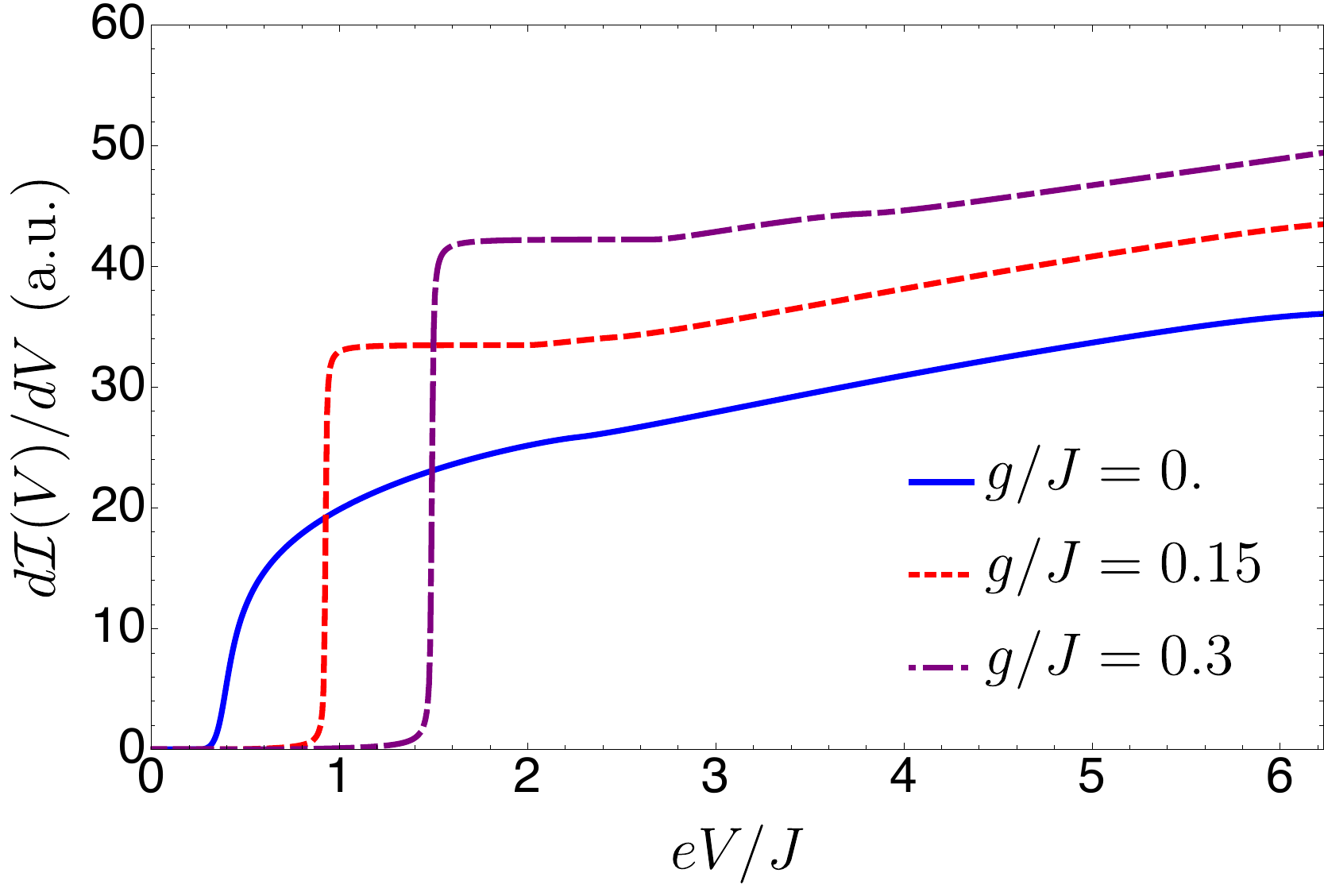}
\put(10,10){(b)}
\end{overpic}
\end{tabular}
\end{center}
\caption{(Color online) 
Panel (a) The spin structure factor of the Kitaev model $\Im m\big[J {\bar Q}(eV)\big]$, proportional to the IETS, plotted as a function of the bias potential $eV$ and for three values of the coupling $g$ (in units of $J$). In this plot the below-the-gap peaks have been marked with vertical arrows.
Panel (b) The (dimensionless) differential conductance plotted as a function of the bias potential $eV$ and for the same three values of the coupling $g$ (in units of $J$) used in Panel (a).
\label{fig:eight}
}
\end{figure}

In Fig.~\ref{fig:eight}(a) we show the spin structure factor, proportional to the IETS, as a function of the bias potential $\omega=eV$ and for three values of the parameter $g$. 
We notice several important features in the plots of Fig.~\ref{fig:eight}(a). First, a low frequency gap is present in the spectrum of the spin structure factor for all values of $g$. Its size equals the sum of the minimum energies of spinon and flux excitations. As shown in Sect.~\ref{sect:probing_excitations}, they both grow approximately linearly with $g$, and therefore cubicly with the magnetic field. 
Notably, a peak appears just above the gap at $g=0$. When the magnetic field increases, such peak develops in a well-defined quasiparticle excitation {\it below} the full spin-excitation gap. As shown in Ref.~\onlinecite{Knolle_thesis}, its energy corresponds to the (dynamical) creation of spinons localized at the cores of the two neighboring fluxes, introduced by the (dynamical) flipping of the sign of a bond eigenvalue. Finally, the dip visible at $eV/J \simeq 2.2$ for $g=0$ and that evolves with $g$ is a signature of the van-Hove singularity at the ${\bm M}$ point of the spinon Brillouin zone.

In Fig.~\ref{fig:eight}(b) we plot the dimensionless conductance ${\bar G} \equiv d{\cal I}(V)/dV$ [obtained by differentiating Eq.~(\ref{eq:cal_I_def})] as a function of bias voltage and magnetic field. 
For any given value of $g/J$, the function ${\bar G}$ appears to be ``gapped'': it remains exactly zero until the inelastic channel involving the excitations of Majorana bound states is opened. At the value of the potential for which this happens, ${\bar G}$ quickly rises to a finite value. The energy for which this happens matches that of the sharp peak observed in the IETS of Fig.~\ref{fig:eight}(a). Following the development of such gap in the tunneling conductance represents an alternative way to detect fractionalized excitations. In fact, the size of the gap scales cubicly in the magnetic field (as expected, since it follows the position of the below-the-gap resonance), in sharp contrast to what is expected for conventional magnetic excitations, for which it would scale linearly~\cite{Ghazaryan_natureel_2018}.

\begin{figure}[t!]
\begin{center}
\begin{tabular}{c}
\begin{overpic}[width=0.99\columnwidth]{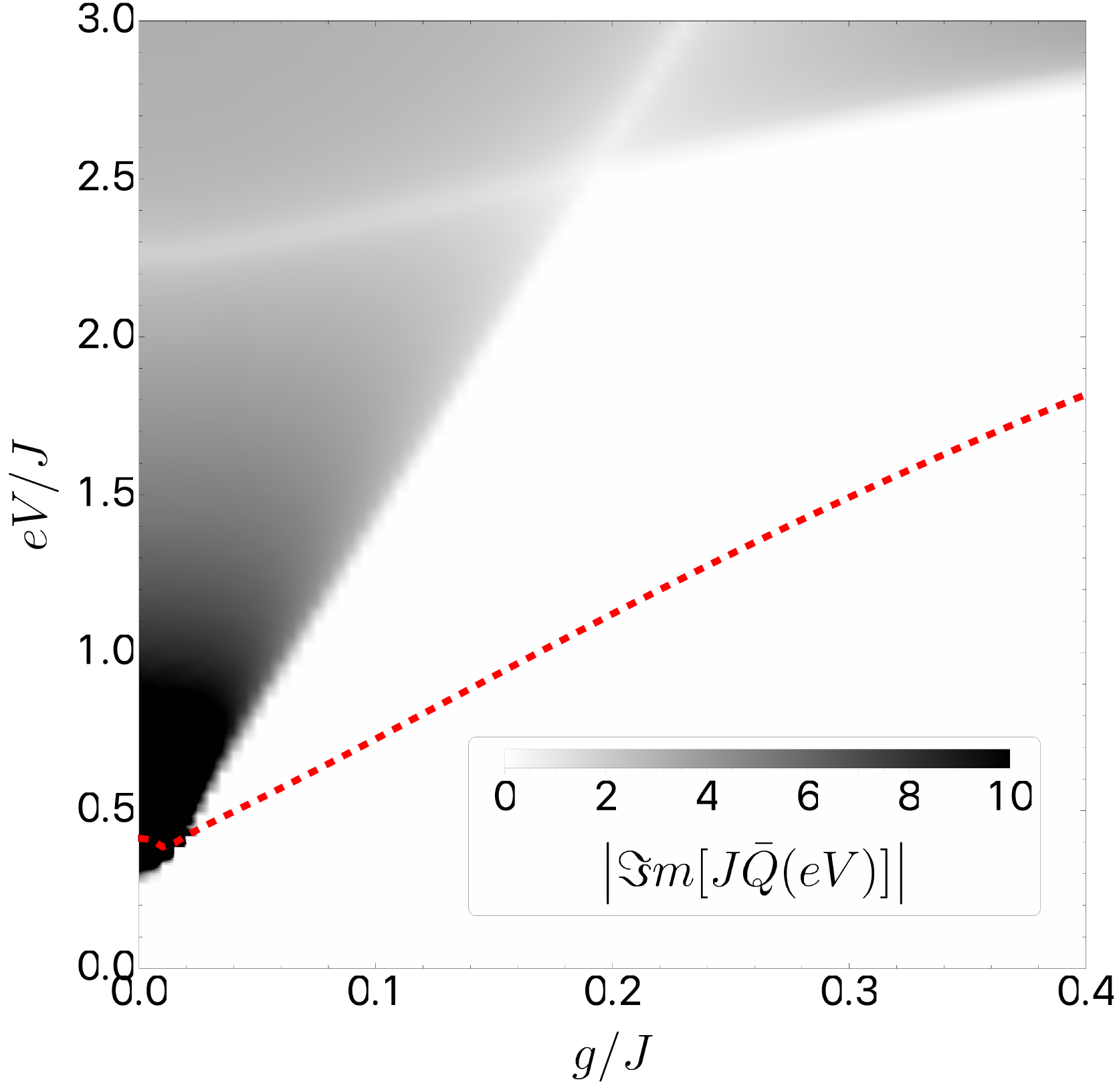}
\put(10,10){(a)}
\end{overpic}
\\
\begin{overpic}[width=0.99\columnwidth]{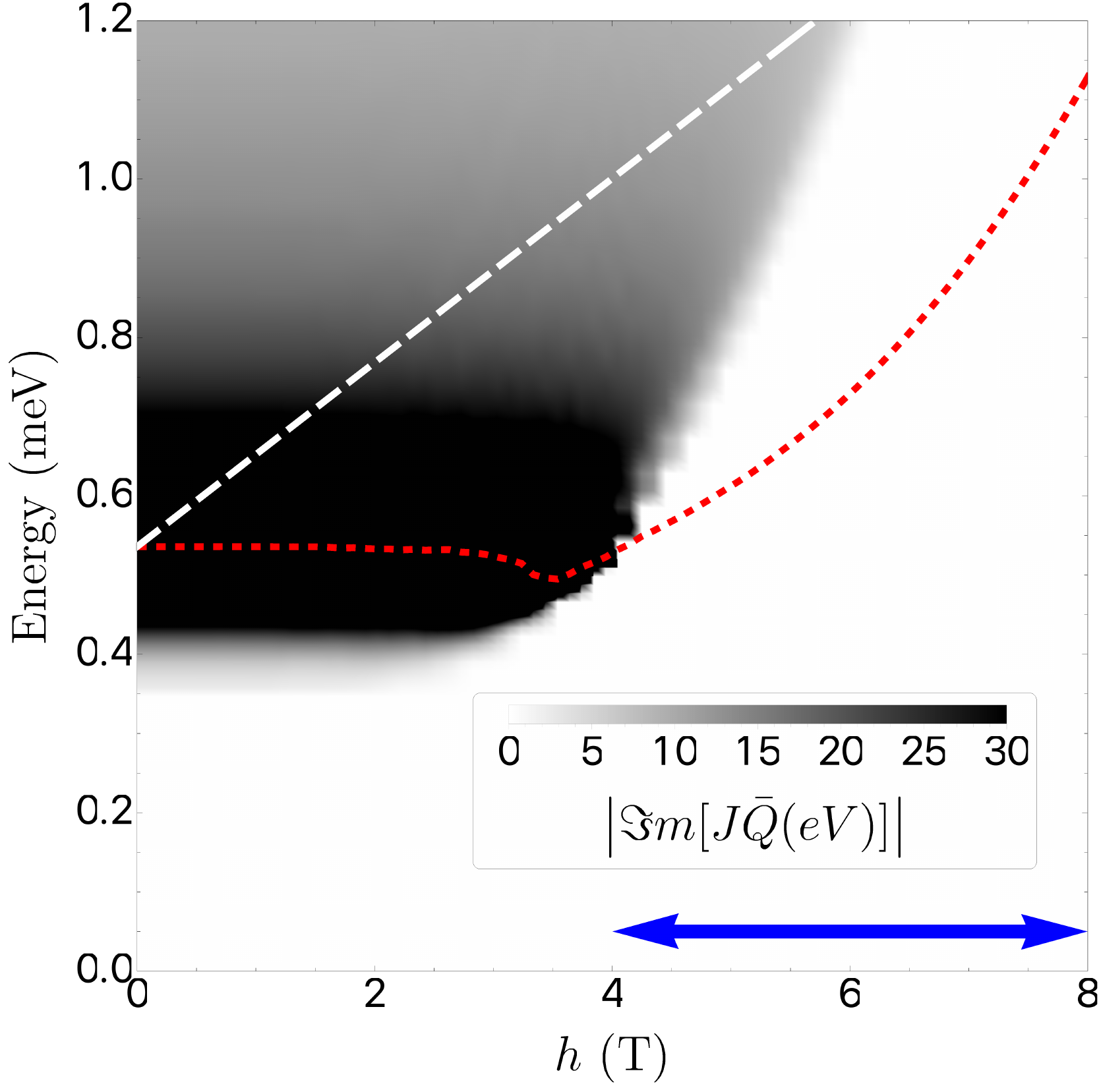}
\put(10,0){(b)}
\end{overpic}
\end{tabular}
\end{center}
\caption{(Color online) 
Panel (a) The spin structure factor, proportional to the IETS, as a function of both the bias potential $eV/J$ and coupling $g/J$. Dark regions correspond to the continuum of energies for which spin excitations can be generated. The red dotted line denotes the position of the resonance peak, which evolves in a below-the-gap bound state. 
Panel (b) Same as in Panel (a), but now the function is plotted against magnetic field and energy. 
The double arrow highlights the range of magnetic fields for which it is possible to track the bound-state energy (red dotted line) and validate the Kitaev description of ${\rm RuCl}_3$.
For comparison, the dashed line represents the energy dispersion of a conventional magnon linear in the magnetic field~\cite{Ghazaryan_natureel_2018}.
\label{fig:nine}
}
\end{figure}

Finally, in Fig.~\ref{fig:nine}(a) we show a 2D plot of the dimensionless spin structure factor $\Im m\big[J{\bar Q}(eV)\big]$ as a function of both the bias potential $eV$ and coupling $g$, both expressed in units of the exchange parameter $J$. We see that the gap of the continuum of excitations rapidly grows, linearly as a function of $g$, and that it changes slope at about $g/J\simeq 0.2$ when it meets the dip due to the van-Hove singularity. We also plot the position of the quasiparticle resonance, whose energy also increases (approximately) linearly with $g$. The fact that all these features of Fig.~\ref{fig:nine}(a) exhibit qualitatively the same dependence in $g$ is not surprising: as explained after Eq.~(\ref{eq:Hamil_QSL_eff_main}), the parameter $g/J$ is the only parameter controlling the theory. Such common behavior is very informative of the nature of the ground state of ${\rm RuCl}_3$, {\it i.e.} of the realization of a quantum spin liquid, and can be used to extract the model parameters ({\it i.e.} the Kitaev exchange $J$).

We warn the reader that the portion of Fig.~\ref{fig:nine}(a) at relatively large values of $g/J$ should be taken with caution. In fact, our results have been obtained by (i) perturbatively accounting for the effect of the magnetic field, and (ii) by neglecting $V^{(2)}_{i,\gamma}(\tau')$ in Eq.~(\ref{eq:S_igamma_tau}). Both these approximations are justified for small $g/J$ and become less reliable as the magnetic field is increased. As we proceed to show, however, magnetic fields corresponding to the unreliable zone of Fig.~\ref{fig:nine}(a) are above those needed to induce phase transitions to other quantum-spin-liquid or ferromagnetically ordered states~\cite{Balz_prb_2019} ($h\gtrsim 7.5~{\rm T}$).

In Fig.~\ref{fig:nine}(b) we show a magnification of the bottom left corner of Fig.~\ref{fig:nine}(a). To obtain such plot, we have converted magnetic fields and excitations energies to Tesla and ${\rm meV}$, respectively, using the measured value of~\cite{Banerjee_natmat_2016,Banerjee_science_2017} $J = 1.3~{\rm meV}$. Since $g = h_x h_y h_z/\Delta_F^2$ has a complex behavior with both the modulus and direction of the magnetic field, in the conversion we have adopted the following strategy. On the one hand, owing to the fact that we are focusing on small values of $g/J$, we have assumed that the two-flux gap has a value approximately equal to the zero-field one, {\it i.e.} $\Delta_F = 0.26 J$. On the other hand, we have fixed the direction of ${\bm h}$ such that the product $h_x h_y h_z$ is maximum ({\it i.e.} $h_x h_y h_z \approx 0.2 |{\bm h}|^3$). As a result, in Fig.~\ref{fig:nine}(b) $h=8~{\rm T}$ corresponds to $g/J\approx 0.1$. We see that the resonance becomes a below-the-gap bound state at the experimentally accessible magnetic field of $4~{\rm T}$. After that, it shows a characteristic cubic scaling with the magnetic field. Therefore, there is a relatively wide range of magnetic fields -- between $4$ and $7.5-8~{\rm T}$, as highlighted by the double arrow in Fig.~\ref{fig:nine}(b) -- for which it is possible to track the behavior of bound-state energy and use this to recognize it as a fractionalized excitation. 
We believe that the cubic scaling of the bound-state energy would start at smaller magnetic field (ideally at $|{\bm h}| = 0$), but such behavior is masked by its interaction with the continuum of spinon excitations. As a result, at small magnetic fields ($\lesssim 4~{\rm T}$) the resonance energy is non-dispersive and exhibits a dip just before exiting the spinon continuum.
We compare the bound-state energy dispersion to that of a magnon [dashed line~\cite{Ghazaryan_natureel_2018} in Fig.~\ref{fig:nine}(b)]: we clearly see that the two have a very different magnetic-field dependence. 

The {\it cubic} growth with the applied magnetic field of the bound-state and van-Hove-dip energy, as well as of the gap of continuum excitation, are all distinctive signatures of the Kitaev model. Therefore, tracking their energy as a function of the magnetic field in tunneling experiments can be used to validate the applicability of the presented model to ${\rm RuCl}_3$, and ultimately provide a proof of the existence of a quantum-spin-liquid phase and fractionalized excitations therein.

\section{Conclusions and outlook}
\label{eq:conclusions}
In this paper we have derived the theory of tunneling assisted by magnetic excitations in vertical van-der-Waals heterostructures. A magnetic insulator is encapsulated by thin graphite slabs whose interfaces are atomically smooth and clean. Their orbitals do not hybridize, and the graphite layers serve simultaneously as a protection from the environment and as electrodes, once connected by metallic contacts to source and drain voltages~\cite{Ghazaryan_natureel_2018}. The magnetic insulator is described in terms of an effective spin Hamiltonian which, depending on the material under consideration, supports classical or quantum excitations~\cite{Gibertini_naturenano_2019}. After having derived the general theory, we have focused on the tunneling assisted by the excitations of ${\rm RuCl}_3$ in the regime in which graphite is doped. This enable us to focus on the properties of the magnetic insulator by neglecting most of the features of the graphite itself, for example the modification of its band structure due to moir\`e superlattices.

${\rm RuCl}_3$, which belongs to the family of transition-metal halides, has recently attracted a significant deal of attention~\cite{Plumb_prb_2014,Kim_prb_2015,Yadav_scirep_2016,Zhou_prb_2016,Sandilands_prb_2016,Banerjee_natmat_2016,Sears_prb_2017,Banerjee_science_2017,Do_natphys_2017,Leahy_prl_2017,Baek_prl_2017,Wolter_prb_2017,Ran_prl_2017,Yu_prl_2018,Shi_prb_2018,Winter_prl_2018,Cookmeyer_prb_2018,Kasahara_nature_2018,Kasahara_prl_2018,Hentrich_prb_2019,Zhou_jpcs_2019,Balz_prb_2019}. This material is in fact a Mott insulator characterized by a relatively large spin-orbit coupling~\cite{Jackeli_prl_2008} and, as such, exhibits a phenomenology which well approximates that of the Kitaev model~\cite{Kitaev_2006}, famous for supporting an {\it exact} quantum-spin-liquid phase.

With a minimalistic model focussed on the spin excitations of the quantum-spin-liquid phase of ${\rm RuCl}_3$, we have shown that contributions due to inelastic processes involving excitations of the magnetic insulator can be singled out in experiments and are very informative about the nature of such quasiparticles~\cite{Ghazaryan_natureel_2018}. By tracking the dependence of their energy on applied magnetic fields, one can (i) distinguish them from non-magnetic ones~\cite{Ghazaryan_natureel_2018} (e.g., phonons) and (ii) recognize them as fractionalized excitations. In fact, the peculiar (cubic) scaling of the excitation energy with the magnetic field can be used to distinguish them from conventional magnons, whose energy increases linearly with the applied field. This constitutes one of the novel aspects of our work. 

This study establishes electron tunneling as a prime tool to address the phases realized in ${\rm RuCl}_3$. Contrary to more conventional techniques such as neutron scattering~\cite{Banerjee_science_2017}, tunneling is specific for 2D systems and particularly suitable for atomically-thin van-der-Waals heterostructures~\cite{Ghazaryan_natureel_2018}. Furthermore, it requires much smaller samples and can enable the proof of Kitaev physics directly for thin devices, ideally in the monolayer limit. The Kitaev model is, in fact, purely two-dimensional, whereas current experiments address its signatures in 3D bulk systems. This allows one to automatically filter spurious effects, such as interlayer interactions or magnetic couplings emerging from stacking faults or strain fields~\cite{Banerjee_science_2017}.

We stress that the present paper represents a viability study of tunneling as a novel methodology to address quantum-ordered phases emerging in ${\rm RuCl}_3$. A quantitative comparison with experiments must account for effects due to (presently neglected) beyond-Kitaev interactions. These are responsible for the zigzag order observed~\cite{Banerjee_science_2017} below $T_{\rm c} \approx 7~{\rm K}$, and can also have a potential impact on the putative quantum-spin-liquid phases that emerge at moderate magnetic fields~\cite{Winter_prb_2016}. Furthermore, we have not yet addressed the impact of strong magnetic fields in modifying the DOS of graphene when the system enters in the quantum Hall regime. Although the emerging phenomenology is rather intriguing, accounting for such effects is beyond the scope of the present work, and will be the subject of future studies.

\acknowledgments
A.P. and M.C. acknowledge support from the Royal Society International Exchange grant IES\textbackslash R3\textbackslash 170252.
M.C. acknowledges support from the Quant-EraNet project ``Supertop".

\begin{widetext}

\appendix

\section{Derivation of Eq.~(\ref{eq:current_final_expression})}
\label{sect:eq:current_final_expression}

We start from Eq.~(\ref{eq:current_def}) and we introduce the interaction picture by defining $| \psi_n(t) \rangle = e^{-i {\cal H}_0 t} | {\tilde \psi}_n(t) \rangle$. The latter wavefunction satisfies
\be
i \partial_t | {\tilde \psi}_n(t) \rangle = {\cal H}_{\rm tun}(t) | {\tilde \psi}_n(t) \rangle
~,
\ee
where ${\cal H}_{\rm tun}(t) = e^{i {\cal H}_0 t} {\cal H}_{\rm tun} e^{-i {\cal H}_0 t}$ is the tunneling Hamiltonian in the interaction picture. The evolution operator in the interaction picture $U_{\rm tun}(t,t_0)$, such that $| {\tilde \psi}_n(t) \rangle = U_{\rm tun}(t,t_0) | {\tilde \psi}_n(t_0) \rangle$, satisfies
\be
i \partial_t U_{\rm tun}(t,t_0) = {\cal H}_{\rm tun}(t) U_{\rm tun}(t,t_0)
~.
\ee
To first order in the tunneling Hamiltonian, $U_{\rm tun}(t,t_0)$ is then given by
\be \label{eq:U_solution}
U_{\rm tun}(t,t_0) \simeq \openone -i \int_{t_0}^t dt' {\cal H}_{\rm tun}(t')
~.
\ee
The current in Eq.~(\ref{eq:current_def}) then reads
\be \label{eq:current_manip_1}
I = - e \sum_n P_n \langle {\tilde \psi}_n(t) | I_{\rm tb}(t) | {\tilde \psi}_n(t) \rangle
~,
\ee
where $I_{\rm tb}(t) \equiv e^{i {\cal H}_0 t} I_{\rm tb} e^{-i {\cal H}_0 t}$.
Using the result of Eq.~(\ref{eq:U_solution}), we get
\be \label{eq:current_manip_1b}
I &\simeq&
- e i \int_{t_0}^t dt' \sum_n P_n \langle {\tilde \psi}_n(t_0) | [ {\cal H}_{\rm tun}(t'), I_{\rm tb}(t) ] | {\tilde \psi}_n(t_0) \rangle
\nn
&=&
- e i \int_{-\infty}^{+\infty} dt' \theta(t-t') \langle [ {\cal H}_{\rm tun}(t'),  I_{\rm tb}(t) ] \rangle
~,
\ee
where in the last line we took the limit $t_0\to -\infty$ and denoted with $\langle \ldots \rangle$ the trace over the initial density matrix (at $t_0=-\infty$). 
We rewrite the two terms in the commutator on the last line of Eq.~(\ref{eq:current_manip_1b}) as
\be \label{eq:H_T_K}
{\cal H}_{\rm tun}(t') &=& 
\frac{\Lambda_0}{\sqrt{N}} \sum_{{\bm k},{\bm k}'} \sum_{n} \sum_{\alpha, \alpha',s,s'}
\big[T^{(n)}_{\alpha\alpha'} {\bm \Sigma}_{ss'}\cdot{\bm s}_{{\bm k} -{\bm k}' + {\bm G}_n}(t') \big] 
\nn
&\times&
\big( e^{-i(\mu_{\rm t} -\mu_{\rm b})t'} c_{{\bm k},\alpha,s,{\rm t}}^\dagger(t') c_{{\bm k}',\alpha',s',{\rm b}}(t') + e^{-i(\mu_{\rm b} -\mu_{\rm t})t'} c_{{\bm k}',\alpha,s,{\rm b}}^\dagger(t') c_{{\bm k},\alpha',s',{\rm t}}(t') \big)
\nn
&\equiv&
e^{-i(\mu_{\rm t} -\mu_{\rm b})t'} A(t') + e^{i(\mu_{\rm t} -\mu_{\rm b})t'} A^\dagger(t')
~,
\ee
and
\be \label{eq:DtN_T_K}
I_{\rm tb}(t) &=& 
-i \frac{\Lambda_0}{\sqrt{N}} \sum_{{\bm k},{\bm k}'} \sum_{n} \sum_{\alpha, \alpha',s,s'}
\big(T^{(n)}_{\alpha\alpha'} {\bm \Sigma}_{ss'}\cdot {\bm s}_{{\bm k} -{\bm k}' + {\bm G}_n}(t) \big)
\nn
&\times&
\big( 
e^{-i(\mu_{\rm t} -\mu_{\rm b})t} 
c_{{\bm k},\alpha,s,{\rm t}}^\dagger (t) c_{{\bm k}',\alpha',s',{\rm b}} (t) - 
e^{-i(\mu_{\rm b} -\mu_{\rm t})t} 
c_{{\bm k}',\alpha,s,{\rm b}}^\dagger (t) c_{{\bm k},\alpha',s',{\rm t}} (t)
\big)
\nn
&\equiv&
-i\big(e^{-i(\mu_{\rm t} -\mu_{\rm b})t} A(t) - e^{i(\mu_{\rm t} -\mu_{\rm b})t} A^\dagger(t)\big)
~,
\ee
where in Eqs.~(\ref{eq:H_T_K})-(\ref{eq:DtN_T_K}) the time evolution of operators is generated by the ``grand-canonical'' Hamiltonian ${\cal K}_0$ [defined after Eq.~(\ref{eq:A_def_main})], {\it i.e.} $c_{{\bm k},\alpha,s, \ell}^{(\dagger)}(t) \equiv e^{i {\cal K}_0 t} c_{{\bm k},\alpha,s, \ell}^{(\dagger)} e^{-i {\cal K}_0 t}$ and ${\bm s}_{{\bm k}}(t) \equiv e^{i {\cal K}_0 t} {\bm s}_{{\bm k}} e^{-i {\cal K}_0 t}$. The relations above can be easily proven by using that
\be
c_{{\bm k},\alpha,s,{\rm t}}^\dagger c_{{\bm k}',\alpha',s',{\rm b}} (\mu_{\rm t} N_{\rm t} + \mu_{\rm b} N_{\rm b}) = (\mu_{\rm t} N_{\rm t} + \mu_{\rm b} N_{\rm b} + \mu_{\rm t} - \mu_{\rm b}) c_{{\bm k},\alpha,s,{\rm t}}^\dagger c_{{\bm k}',\alpha',s',{\rm b}}
~.
\ee
Plugging Eqs.~(\ref{eq:H_T_K})-(\ref{eq:DtN_T_K}) into Eq.~(\ref{eq:current_manip_1}), we then obtain 
\be \label{eq:current_manip_2}
I &=& 
- e \int_{-\infty}^{+\infty} dt' \theta(t-t') \Big[
e^{i(\mu_{\rm t} -\mu_{\rm b})(t'-t)}
\langle [ A^\dagger(t'), A(t) ] \rangle
-
e^{-i(\mu_{\rm t} -\mu_{\rm b})(t'-t)}
\langle [ A(t'), A^\dagger (t) ] \rangle
\Big]
\nn
&=&
-2 e \Im m\big[\chi_{AA}(\mu_{\rm t} -\mu_{\rm b}) \big]
~,
\ee
which coincides with Eq.~(\ref{eq:current_final_expression}).

\section{Derivation of Eq.~(\ref{eq:chi_AA_res_final})}
\label{sect:eq:chi_AA_res_final}

To calculate the retarded response function at finite temperature, we start from the imaginary-time-ordered response~\cite{Giuliani_and_Vignale}  $\chi_{AA}^{({\cal T})}(\tau)\equiv -\langle {\cal T}  A(\tau) A^\dagger \rangle$, where the imaginary-time ordering ${\cal T}[\ldots]$ orders operators according to decreasing imaginary time (lower times to the left). The imaginary-time-ordered response function is periodic of period $2\beta$, where $\beta \equiv (k_{\rm B}T)^{-1}$, $k_{\rm B}$ is the Boltzmann constant and $T$ the electronic temperature. Its discrete Fourier transform reads
\be
\chi_{AA}^{({\cal T})}(i\omega_m) &=& -\int_0^{\beta} d\tau e^{i\omega_m \tau} \langle {\cal T} A(\tau) A^\dagger \rangle
\nn
&=&
-\frac{\Lambda_0^2}{N}
\sum_{\gamma,\gamma'}
\sum_{{\bm k},{\bm k}'} \sum_{n=0}^2 \sum_{\alpha, \alpha',s,s'}
\sum_{{\bm k}'',{\bm k}'''} \sum_{n'=0}^2 \sum_{\alpha'', \alpha''',s'',s'''}
\big(T^{(n)}_{\alpha\alpha'} \Sigma^{\gamma}_{ss'}  \big)
\big(T^{(n')}_{\alpha''\alpha'''} \Sigma^{\gamma'}_{s''s'''}\big)
\nn
&\times&
\int_0^\beta d\tau e^{i\omega_m \tau}
\langle {\cal T} s^\gamma_{{\bm k} -{\bm k}' + {\bm G}_n}(\tau) c_{{\bm k},\alpha,s,{\rm t}}^\dagger (\tau) c_{{\bm k}',\alpha',s',{\rm b}} (\tau) s^{\gamma'}_{{\bm k}'' -{\bm k}''' + {\bm G}_{n'}} c_{{\bm k}'',\alpha'',s'',{\rm b}}^\dagger c_{{\bm k}''',\alpha''',s''',{\rm t}} \rangle
~.
\ee
Using that
\be
\Xi &\equiv& 
\langle {\cal T} s^\gamma_{{\bm k} -{\bm k}' + {\bm G}_n}(\tau) c_{{\bm k},\alpha,s,{\rm t}}^\dagger (\tau) c_{{\bm k}',\alpha',s',{\rm b}} (\tau) s^{\gamma'}_{{\bm k}'' -{\bm k}''' + {\bm G}_{n'}} c_{{\bm k}'',\alpha'',s'',{\rm b}}^\dagger c_{{\bm k}''',\alpha''',s''',{\rm t}} \rangle
\nn
&=&
\langle {\cal T} s^\gamma_{{\bm k} -{\bm k}' + {\bm G}_n}(\tau) s^{\gamma'}_{{\bm k}'' -{\bm k}''' + {\bm G}_{n'}} \rangle
\langle {\cal T} c_{{\bm k},\alpha,s,{\rm t}}^\dagger (\tau) c_{{\bm k}''',\alpha''',s''',{\rm t}} \rangle
\langle {\cal T} c_{{\bm k}',\alpha',s',{\rm b}} (\tau) c_{{\bm k}'',\alpha'',s'',{\rm b}}^\dagger \rangle
\nn
&=&
\delta_{{\bm k}, {\bm k}'''} \delta_{s,s'''}
\delta_{{\bm k}', {\bm k}''} \delta_{s',s''}
\delta_{n,n'}
Q^{\gamma\gamma'}({\bm k} -{\bm k}' + {\bm G}_n,\tau) G_{\alpha''',\alpha}^{({\rm t})}({\bm k},-\tau) G_{\alpha'\alpha''}^{({\rm b})}({\bm k}',\tau)
~,
\ee
where we used the fact that graphene electrons are not spin polarized, we find
\be \label{eq:chi_AA_res1}
\chi_{AA}^{({\cal T})}(i\omega_m) &=& 
-\frac{\Lambda_0^2}{N}
\sum_{{\bm k},{\bm k}'} \sum_{n=0}^2 
\int_0^{\beta} d\tau e^{i\omega_m \tau}
Q({\bm k} -{\bm k}' + {\bm G}_n,\tau) {\rm Tr}\big[ G_{\rm b}({\bm k}',\tau) T^{(n)} G_{\rm t}({\bm k},-\tau) T^{(n)}\big]
~.
\ee
Here we used that ${\rm Tr}(\Sigma^{i} \Sigma^{j}) = \delta_{ij}$, and we defined the spin structure factor of the magnetic layer
\be \label{eq:spin_spin_corr_summed}
Q({\bm k} -{\bm k}' + {\bm G}_n,\tau) &=& 
-\sum_\gamma \langle {\cal T} s^\gamma_{{\bm k} -{\bm k}' + {\bm G}_n}(\tau) s^{\gamma}_{{\bm k}'' -{\bm k}''' + {\bm G}_{n'}} \rangle
~.
\ee
In Eq.~(\ref{eq:chi_AA_res1}), $G_{{\rm t},\alpha''',\alpha}({\bm k},-\tau) = - \langle {\cal T} c_{{\bm k},\alpha,s,{\rm t}}^\dagger (\tau) c_{{\bm k}''',\alpha''',s''',{\rm t}} \rangle$ and $G_{{\rm b},\alpha'\alpha''}({\bm k}',\tau) = - \langle {\cal T} c_{{\bm k}',\alpha',s',{\rm b}} (\tau) c_{{\bm k}'',\alpha'',s'',{\rm b}}^\dagger \rangle$ are the electron Green's functions in the top and bottom layer, respectively. In Eq.~(\ref{eq:chi_AA_res1}) the trace is over the indices $\alpha,\alpha',\alpha'',\alpha'''$. Using that
\be
T^{(n)} = \openone + \cos\left(\frac{2n\pi}{3}\right) \sigma^x - \sin\left(\frac{2n\pi}{3}\right) \sigma^y
~,
\ee
Eq.~(\ref{eq:chi_AA_res1}) is rewritten as
\be \label{eq:chi_AA_res2}
\chi_{AA}^{({\cal T})}(i\omega_m) &=& 
-\frac{\Lambda_0^2}{N}
\sum_{{\bm k},{\bm k}'} \sum_{\lambda,\lambda'} \sum_{n} 
\int_0^{\beta} d\tau e^{i\omega_m \tau}
Q({\bm k} -{\bm k}' + {\bm G}_n,\tau) 
G_{{\rm t},\lambda}({\bm k},-\tau)
G_{{\rm b},\lambda'}({\bm k}',\tau)
\nn
&\times&
\left|\rho_{{\bm k},\lambda;{\bm k}',\lambda'} + \cos\left(\frac{2n\pi}{3}\right) \sigma^x_{{\bm k},\lambda;{\bm k}',\lambda'} - \sin\left(\frac{2n\pi}{3}\right) \sigma^y_{{\bm k},\lambda;{\bm k}',\lambda'} \right|^2
~.
\ee
Finally, introducing
\be
&&
G_{{\rm t},\lambda}({\bm k},i\varepsilon_n) = \int_0^{\beta} d\tau e^{i\varepsilon_n \tau} G_{{\rm t},\lambda}({\bm k},\tau)
~,
\nn
&&
G_{{\rm b},\lambda'}({\bm k}',i\varepsilon_{n'}) = \int_0^{\beta} d\tau e^{i\varepsilon_{n'} \tau} G_{{\rm b},\lambda'}({\bm k}',\tau)
~,
\ee
we get
\be \label{eq:chi_AA_res3}
\chi_{AA}^{({\cal T})}(i\omega_m) &=& 
-\frac{\Lambda_0^2}{N}
\sum_{{\bm k},{\bm k}'} \sum_{\lambda,\lambda'} \sum_{n} 
\left|\rho_{{\bm k},\lambda;{\bm k}',\lambda'} + \cos\left(\frac{2n\pi}{3}\right) \sigma^x_{{\bm k},\lambda;{\bm k}',\lambda'} - \sin\left(\frac{2n\pi}{3}\right) \sigma^y_{{\bm k},\lambda;{\bm k}',\lambda'} \right|^2
\nn
&\times&
\frac{1}{\beta^2} \sum_{\varepsilon_n,\varepsilon_{n'}}
Q({\bm k} -{\bm k}' + {\bm G}_n,i\omega_m + i\varepsilon_n - i\varepsilon_{n'}) 
G_{{\rm t},\lambda}({\bm k},i\varepsilon_n)
G_{{\rm b},\lambda'}({\bm k}',i\varepsilon_{n'})
~.
\ee
Here $\varepsilon_n = (2n+1)\pi/\beta$ and $\varepsilon_{n'} = (2n'+1)\pi/\beta$, while $\omega_m$ is either a bosonic [$= 2m\pi/\beta$] or a fermionic [$= (2m+1)\pi/\beta$] Matsubara frequency. The choice is due to the statistics of spin excitations [{\it i.e.} whether $Q({\bm q},\tau)$ is symmetric or antisymmetric in the interval $\tau\in(-\beta,\beta)$]. Recall indeed that, in general, these have no preferred statistics: they can be bosonic (as in the case of magnons) or fermionic (like spinons). 

Introducing $\omega_{m'} = \varepsilon_n - \varepsilon_{n'}$ and ${\bm q} = {\bm k}-{\bm k}'$, we get
\be \label{eq:chi_AA_res4}
\chi_{AA}^{({\cal T})}(i\omega_m) &=& 
-\Lambda_0^2
\sum_{{\bm q}, n} 
\frac{1}{\beta} \sum_{\omega_{m'}}
Q({\bm q} + {\bm G}_n,i\omega_m + i\omega_{m'}) 
\chi_{\rm tb}({\bm q},i\omega_{m'})
~,
\ee
where
\be
\chi_{\rm tb}({\bm q},i\omega_{m'}) &\equiv&
\frac{1}{N}\sum_{{\bm k},\lambda,\lambda'}
\left|\rho_{{\bm k},\lambda;{\bm k}-{\bm q},\lambda'} + \cos\left(\frac{2n\pi}{3}\right) \sigma^x_{{\bm k},\lambda;{\bm k}-{\bm q},\lambda'} - \sin\left(\frac{2n\pi}{3}\right) \sigma^y_{{\bm k},\lambda;{\bm k}-{\bm q},\lambda'} \right|^2
\nn
&\times&
\frac{1}{\beta} \sum_{\varepsilon_n}
G_{{\rm t},\lambda}({\bm k},i\varepsilon_n)
G_{{\rm b},\lambda'}({\bm k}-{\bm q},i\varepsilon_{n}-i\omega_{m'})
~.
\ee
Upon analytical continuation to real frequencies, and taking the imaginary part, we finally get Eq.~(\ref{eq:chi_AA_res_final}).

\section{Diagonalization of the Kitaev Hamiltonian}
\label{sect:Kitaev_1}
%
We recall that the Kitaev Hamiltonian is [see Eq.~(\ref{eq:Ham_m_def})]
\be \label{eq:QSL_ham_def}
{\cal H}_{\rm m} = -J \sum_{\langle i,j\rangle_\gamma} s_i^\gamma s_j^\gamma - \sum_i {\bm h}\cdot {\bm s}_i
~.
\ee
where $s_i^\gamma$ represents the $\gamma$-component ($\gamma=x,y,z$) of the spin-$1/2$ magnetic moment at site $i$. The exchange term couples only one spin component ($\gamma$) along a given bond $\langle i,j\rangle_\gamma$. The same spin component is coupled along bonds in the same direction. The bond type is defined as follows: we set the three nearest-neighbor vectors starting from an $A$ site and ending in a $B$ one:
\be
{\bm d}^A_{x} = \frac{\sqrt{3} a_{\rm K}}{3} \left(-1,0\right)
~, 
{\bm d}^A_{y} = \frac{\sqrt{3} a_{\rm K}}{3} \left(\frac{1}{2},-\frac{\sqrt{3}}{2}\right)
~, 
{\bm d}^A_{z} = \frac{\sqrt{3} a_{\rm K}}{3} \left(\frac{1}{2},\frac{\sqrt{3}}{2}\right)
~.
\ee
Here $a_{\rm K}\sim 7~{\rm \AA}$ is the lattice constant corresponding to the ${\rm Ru}-{\rm Ru}$ distance in ${\rm RuCl}_3$.
The nearest neighbors from a site of type $B$ are located at ${\bm d}^B_{\gamma}=-{\bm d}^A_{\gamma}$, where $\gamma=x,y,z$. If a link is parallel to the vector ${\bm d}^A_{\gamma}$, is said to be of type $\gamma$. 
When the magnetic field ${\bm h} = {\bm 0}$, the Hamiltonian can be exactly diagonalized via the introduction of four Majorana fermions per lattice site~\cite{Kitaev_2006}, $c_i, b_i^\gamma$, where $\gamma = x,y,z$, as explained in the main text. The Kitaev Hamiltonian then becomes
\be \label{eq:Hamil_QSL_Majorana}
{\cal H}_{\rm m} = iJ \sum_{\langle i,j\rangle_\gamma} u_{ij}^\gamma c_{i} c_{j}
~,
\ee
where $u_{ij}^\gamma = ib_{i}^\gamma b_{j}^\gamma$ as defined after Eq.~(\ref{eq:Hamil_QSL_Majorana_eff_main}). Note that $u_{ij}^\gamma = - u_{ji}^\gamma$, and that such operators commute among themselves and with the Hamiltonian. As such, the Hilbert space becomes the direct sum of subspaces, each characterized by a given configuration of $u_{ij}^\gamma$. The latter play the role of a gauge field on top of which the $c$-Majorana particles propagate. Not all configurations of the gauge field are independent: actually, several of them are equivalent~\cite{Kitaev_2006}. By fixing a set of variables $\alpha_i =\pm 1$ for all sites $i$, one can define a gauge transformation ${\tilde u}_{ij}^\gamma = \alpha_i u_{ij}^\gamma \alpha_j$. The resulting gauge-field configuration ${\tilde u}_{ij}^\gamma$ yields the same physical properties as $u_{ij}^\gamma$. As explained in the main text, the conserved quantities are in fact the fluxes threading the plaquettes. One state in the zero flux sector is obtained by setting $u_{ij}^\gamma = -1$ when $i$ and $j$ are sites of type $A$ and $B$, respectively.

In the presence of a magnetic field, the Hamiltonian~(\ref{eq:QSL_ham_def}) cannot be exactly diagonalized. We will therefore derive an effective Hamiltonian, from perturbation theory, that has this properties. We start by rewriting Eq.~(\ref{eq:QSL_ham_def}) in the Majorana representation. It becomes~\cite{Kitaev_2006,Knolle_thesis}
\be \label{eq:Hamil_QSL_Majorana_h}
{\cal H}_{\rm m} = i J \sum_{\langle i,j\rangle_\gamma} u_{ij}^\gamma c_{i} c_{j} - i \sum_{i,\gamma} h_\gamma b_{i}^\gamma c_{i}
~.
\ee
To find the effective Hamiltonian in the flux-free sector of the Hilbert space~\cite{Kitaev_2006}, we consider the perturbative expansion of the energy 
\be
E = E_0 + \delta E_1 + \delta E_2 + + \delta E_3
~,
\ee
where 
\be
&& E_0 = \langle 0 | {\cal H}_{{\rm m},0} | 0 \rangle
~,
\nn
&& \delta E_1 = \langle 0 | {\cal H}_{{\rm m},1} | 0 \rangle
~,
\nn
&& \delta E_2 = \sum_{n\neq 0} \frac{\langle 0 | {\cal H}_{{\rm m},1} | n \rangle \langle n | {\cal H}_{{\rm m},1} | 0 \rangle}{E_0 - E_n}
~,
\nn
\nn
&& \delta E_3 = \sum_{n,m\neq 0} \frac{\langle 0 | {\cal H}_{{\rm m},1} | n \rangle \langle n | {\cal H}_{{\rm m},1} | m \rangle \langle m | {\cal H}_{{\rm m},1} | 0 \rangle}{(E_0 - E_n)(E_0 - E_m)}
~.
\ee
Here $|n\rangle$ and $|m\rangle$ are eigenstates of the unperturbed Hamiltonian 
\be
{\cal H}_{{\rm m},0} = i J \sum_{\langle i,j\rangle_\gamma} u_{ij}^\gamma c_{i} c_{j}
\ee
which do {\it not} belong to the flux-free sector, and whose energies are $E_n$ and $E_m$, respectively. Conversely, $|0\rangle$ denotes states belonging to the flux-free sector, whose energy is $E_0$. Finally,
\be
{\cal H}_{{\rm m},1} =  - i \sum_{i,\gamma} h_\gamma b_{i}^\gamma c_{i}
\ee
is the perturbing Hamiltonian. Note that, because of the gauge redundancy, in this equations $\langle 0 |$ is the Hermitian conjugate of a {\it gauge equivalent} state of $|0\rangle$.

To perform the calculation, it is useful to introduce the following ``bond fermion'' operators~\cite{Knolle_thesis}
\be
\chi_{\langle i,j\rangle_\gamma} = \frac{b_{i}^\gamma + i b_{j}^\gamma}{2}
~,
\ee
where $i$ and $j$ are the neighbors along the direction $\gamma$, 
such that
\be \label{u_ij_chi}
u_{ij}^\gamma = i b_{i}^\gamma b_{j}^\gamma = 2 \chi_{\langle i,j\rangle_\gamma}^\dagger \chi_{\langle i,j\rangle_\gamma} - 1
~.
\ee
Note that the site $j$ is uniquely determined by the choice of $i$ and $\gamma$.
Clearly, $\chi_{\langle i,j\rangle_\gamma}^\dagger$ and $\chi_{\langle i,j\rangle_\gamma}$ change the number of bond fermions along $\langle i,j\rangle_\gamma$ and therefore change the sign of $u_{ij}^\gamma$. As such, each of the the two operators adds a pair of fluxes in the two plaquettes sharing the bond $\langle i,j\rangle_\gamma$.
In terms of $\chi_{\langle i,j\rangle_\gamma}$ and $\chi_{\langle i,j\rangle_\gamma}^\dagger$, the spin operators read~\cite{Knolle_thesis}
\be
&&
s_{i}^\gamma = i(\chi_{\langle i,j\rangle_\gamma} + \chi_{\langle i,j\rangle_\gamma}^\dagger)c_{i}
~,
\nn
&&
s_{j}^\gamma = (\chi_{\langle i,j\rangle_\gamma} - \chi_{\langle i,j\rangle_\gamma}^\dagger)c_{j}
~.
\ee
Hence, they introduce a $c$-Majorana particle while simultaneously changing the number of bond fermions ({\it i.e.} flipping the sign of a bond operator or, equivalently, introducing a pair of fluxes in neighboring plaquettes). 
With these definitions, 
\be
&&
{\cal H}_{{\rm m},0} = iJ  \sum_{\langle i,j\rangle_\gamma} (2 \chi_{\langle i,j\rangle_\gamma}^\dagger \chi_{\langle i,j\rangle_\gamma} - 1) c_{i,A} c_{j,B}
~,
\nn
&&
{\cal H}_{{\rm m},1} = - i \sum_{i,\gamma} h_\gamma  (\chi_{\langle i,j\rangle_\gamma} + \chi_{\langle i,j\rangle_\gamma}^\dagger) c_{i}
~.
\ee
We know that the eigenstates of ${\cal H}_{{\rm m},0}$ are also eigenstates of $u_{ij}^\gamma$ and that the ground state lies in the no-flux sector of the Hilbert space. We observe that ${\cal H}_{{\rm m},1}$ changes the number of fluxes by 2 (by flipping the sign of one bond operator). Therefore, the first-order correction to the energy, $\langle 0 | {\cal H}_{{\rm m},1} | 0 \rangle$, vanishes exactly, since the initial and final states cannot be in the same flux sector (the vector ${\cal H}_{{\rm m},1}|0\rangle$ has zero overlap with $|0\rangle$). The second-order term is non-zero, but it can be shown to only renormalize the nearest-neighbor hopping amplitude of the $c$-Majorana particles. This effects is negligible, since it only makes the hopping in the three directions slightly anisotropic, but not to the point of merging two Dirac points and opening a gap~\cite{Kitaev_2006}.

The first non-trivial term, that opens a field-dependent gap in the spinon dispersion, appears at third order in perturbation theory. In this case, the states $|m\rangle$ and $|n\rangle$ must contain two fluxes each. Following Kitaev~\cite{Kitaev_2006}, we replace $E_0-E_m \simeq E_0 - E_n \simeq - \Delta_F$. We can then perform the sums over $m$ and $n$ using that $\sum_n |n\rangle \langle n |$ and $\sum_m |m\rangle \langle m |$ are both equal to the identity. Therefore, the third-order correction becomes
\be \label{eq:deltaE3_1}
\delta E_3 &=&  \frac{\langle 0 | {\cal H}_{{\rm m},1}^3 | 0 \rangle}{\Delta_F^2}  = i \sum_{i,\gamma} \sum_{i',\gamma'} \sum_{i'',\gamma''}  \frac{h_\gamma h_{\gamma'} h_{\gamma''}}{\Delta_F^2} 
\nn
&\times&
\langle 0 | 
(\chi_{\langle i,j\rangle_\gamma} + \chi_{\langle i,j\rangle_\gamma}^\dagger) c_{i} 
(\chi_{\langle i',j'\rangle_{\gamma'}} + \chi_{\langle i',j'\rangle_{\gamma'}}^\dagger) c_{i'} 
(\chi_{\langle i'',j''\rangle_{\gamma'}} + \chi_{\langle i'',j''\rangle_{\gamma''}}^\dagger) c_{i''} 
| 0 \rangle
~.
\ee
We note that, since $| 0 \rangle$ belongs to the zero-flux sector, $\chi_{\langle i,j\rangle_{\gamma}} |0\rangle = 0$ for all $\chi_{\langle i,j\rangle_{\gamma}}$. Hence, the last round bracket becomes $\chi_{\langle i'',j''\rangle_{\gamma'}} + \chi_{\langle i'',j''\rangle_{\gamma''}}^\dagger \to \chi_{\langle i'',j''\rangle_{\gamma''}}^\dagger$, which flips the sign of the bond operator $u_{i''j''}^{\gamma''}$. To go back to the initial state with no fluxes, it is necessary to flip the signs of the other two bonds connected to either the site $i''$ or to $j''$. In this way, two fluxes are first created at the two sides of the bond $\langle i'',j''\rangle_{\gamma''}$ and one of then is carried around either $i''$ or $j''$ and finally annihilated with the one that has been left behind. As a consequence, $\langle i,j\rangle_\gamma$, $\langle i',j'\rangle_{\gamma'}$ and $\langle i'',j''\rangle_{\gamma''}$ are three bonds of different types all connected to one common site (and therefore $h_\gamma h_{\gamma'} h_{\gamma''} = h_xh_yh_z$, independently of the order of the $\gamma$'s). The described process corresponds to selecting only the term containing three operators $\chi^\dagger$ in~(\ref{eq:deltaE3_1}). Going back to spin operators, we have~\cite{Kitaev_2006}
\be \label{eq:deltaE3_2}
\delta E_3 &=&  - \frac{h_x h_y h_z}{\Delta_F^2}  \sum_{i,\gamma} \sum_{i',\gamma'} \sum_{i'',\gamma''} 
\langle 0 | s_i^\gamma s_{i'}^{\gamma'} s_{i''}^{\gamma''} | 0 \rangle
~.
\ee
There are two possibilities for the three sites $i,i',i''$:
\begin{enumerate}
\item $i,i',i''$ are the three nearest-neighbours of a given site and belong to the same sublattice. In the Majorana representation~(\ref{eq:deltaE3_1}), the indices $j=j'=j''$ all correspond to the site in the middle of the triangle $i,i',i''$.
As explained in Ref.~\onlinecite{Kitaev_2006} this term does not directly contribute to the quadratic part of the Hamiltonian and will be ignored.

\item $i,i',i''$ are consecutive sites, not necessarily in this order. Without loss of generality, we can order them by commuting the spin operators in Eq.~(\ref{eq:deltaE3_2}) and relabeling them. Therefore, we assume that $i'$ is a nearest-neighbor of both $i$ and $i''$. Hence, in Eq.~(\ref{eq:deltaE3_1}) $j=j''=i'$, while $j'$ is the third nearest-neighbor of $i'$ that does not coincide with either $i$ or $i''$. Note that, once two next-nearest neighbors $i$ and $i''$ have been chosen, $i'$ is uniquely determined by the fact that it has to be the nearest neighbor of both (in this case, it is $i'$ the site that all of $\langle i,j\rangle_\gamma$, $\langle i',j'\rangle_{\gamma'}$ and $\langle i'',j''\rangle_{\gamma''}$ share). Similarly, since there is only one path connecting them, also $\gamma$ and $\gamma''$ are fixed, and coincide with the types of the bonds $\langle i,i'\rangle_\gamma$ and $\langle i',i''\rangle_{\gamma''}$, respectively. This fact follows from the constraint $j=j''=i'$. Hence, $\gamma'$, which has to be different from both $\gamma$ and $\gamma''$ (to carry one flux along a closed path), is also constrained by the choice of $i$ and $i''$. Eq.~(\ref{eq:deltaE3_2}) then reads~\cite{Kitaev_2006}
\be \label{eq:deltaE3_case2_1}
\delta E_3 &=&  - \frac{h_x h_y h_z}{\Delta_F^2}  \sum_{\langle\langle i,k\rangle \rangle} \langle 0 | s_i^\gamma s_{j}^{\gamma'} s_{k}^{\gamma''} | 0 \rangle
~,
\ee
where, as explained above, $j,\gamma,\gamma',\gamma''$ are completely determined by the choice of the next-nearest neighbors $\langle\langle i,k\rangle \rangle$. We can further manipulate Eq.~(\ref{eq:deltaE3_case2_1}), going back to the Majorana representation. We obtain
\be \label{eq:deltaE3_case2_2}
\delta E_3 &=& i \frac{h_x h_y h_z}{\Delta_F^2}  \sum_{\langle\langle i,k\rangle \rangle}
\langle 0 | b_i^\gamma c_{i} b_{j}^{\gamma'} c_{j} b_{k}^{\gamma''} c_{k} | 0 \rangle
\nn
&=&
i  \frac{h_x h_y h_z}{\Delta_F^2} \sum_{\langle\langle i,k\rangle \rangle} \varepsilon_{\gamma\gamma'\gamma''}
\langle 0 | u_{ij}^{\gamma} D_{j} u_{jk}^{\gamma''} c_{i} c_{k} | 0 \rangle
~.
\ee
Here $\varepsilon_{\gamma\gamma'\gamma''}$ is the Levi-Civita tensor, which emerges after $b_{j}^{\gamma} b_{j}^{\gamma'} b_{j}^{\gamma''} c_{j}$ has been reordered to give $D_{j}=b_{j}^{x} b_{j}^{y} b_{j}^{z} c_{j}$, such that $D_{j} = 1$ on the physical states.

\end{enumerate}
The final effective Hamiltonian is therefore~\cite{Kitaev_2006}
\be \label{eq:Hamil_QSL_Majorana_eff}
{\cal H}_{\rm m}^{\rm eff}  &=& iJ \sum_{\langle i,j\rangle_\gamma} u_{ij}^\gamma c_{i} c_{j} + i g \sum_{\langle \langle i, k\rangle \rangle} \varepsilon_{\gamma\gamma'\gamma''}
u_{ij}^{\gamma} D_{j} u_{jk}^{\gamma''} c_{i} c_{k}
~,
\ee
where $g = h_xh_yh_z/\Delta_F^2$. Since $j$ must be in between $i$ and $k$, and there is only one path that connects all three of them, $\gamma\neq \gamma'\neq \gamma''$ are uniquely determined for each pair of next-nearest neighbors $\langle \langle i, k\rangle \rangle$. 
For future purposes we can also rewrite the effective Hamiltonian as
\be \label{eq:Hamil_QSL_Majorana_eff_2}
{\cal H}_{\rm m}^{\rm eff}  &=& iJ \sum_{\langle i,j\rangle_\gamma} u_{ij}^\gamma c_{i} c_{j} + i \frac{g}{2} \sum_{\langle i, j \rangle_\gamma} \sum_{\langle j, k \rangle_{\gamma''}} \varepsilon_{\gamma\gamma'\gamma''} u_{ij}^{\gamma} D_j u_{jk}^{\gamma''} c_{i} c_{k}
~,
\ee
where the factor $1/2$ corrects for double counting.

\section{Majorana excitations of the flux-free sector}
\label{app:Majorana_flux_free}
We focus on the case of constant $u_{ij}^\gamma=-1$, where $i$ is assumed to be an $A$-site and $j$ a $B$ one (in the opposite case, $u_{ij}^\gamma=+1$). This corresponds to having no flux piercing the system ($\chi^\dagger\chi = 0$ over all links oriented from an $A$ to a $B$ site). Lieb's theorem guarantees the ground state of the zero-field Kitaev model to be in such sector~\cite{Kitaev_2006}. We rewrite Eq.~(\ref{eq:Hamil_QSL_Majorana_eff}) as
\be \label{eq:Hamil_QSL_Majorana_2}
{\cal H}_{\rm m}^{\rm eff} = -iJ \sum_{{\bm r},{\bm \delta}_1} c_{{\bm r},A} c_{{\bm r} + {\bm \delta}_1,B} - i \frac{g}{2} \sum_{{\bm r},{\bm \delta}_2,\alpha} \varsigma_\alpha({\bm \delta}_2)  c_{{\bm r},\alpha} c_{{\bm r}+{\bm \delta}_2,\alpha}
~,
\ee
where ${\bm r} = n_+ {\bm a}_{+} + n_- {\bm a}_{-}$ is the position of a unit cell [${\bm a}_\pm = a_{\rm K} (\sqrt{3}/2, \pm 1/2)$, while $n_\pm$ are integers] and ${\bm \delta}_1 \in \{{\bm 0}, {\bm a}_{+}, {\bm a}^{-}\}$ [${\bm \delta}_2 \in \{ \pm {\bm a}_{+}, \pm {\bm a}_{-}, \pm ({\bm a}_{-} - {\bm a}_{+}) \}$] are the position of the first- (second-)nearest-neighboring cells of a given unit cell. 
In the first term, we explicitly accounted for the sublattice type ($A$ or $B$) to avoid double counting the bonds $\langle i,j\rangle_\gamma$. In the second term, $\varsigma_\alpha({\bm \delta}_2) = \pm 1$ depending on the sublattice and next-nearest-neighbor vector. Given the structure of the second term on the right-hand side of Eq.~(\ref{eq:Hamil_QSL_Majorana_eff_2}), $\varsigma_B({\bm \delta}_2) = -\varsigma_A({\bm \delta}_2)$, $\varsigma_\alpha(-{\bm \delta}_2) = -\varsigma_\alpha({\bm \delta}_2)$ ($\alpha=A,B$ labels the sublattice type) and, therefore, $\varsigma_B(-{\bm \delta}_2) = \varsigma_A({\bm \delta}_2)$. These equalities stem from properties of the Levi-Civita tensor. From the structure of the second term on the right-hand side of Eq.~(\ref{eq:Hamil_QSL_Majorana_eff_2}), we find
\be
\varsigma_A({\bm a}_{+,{\rm K}}) = \varsigma_A(-{\bm a}_{-,{\rm K}}) = \varsigma_A({\bm a}_{-,{\rm K}}-{\bm a}_{+,{\rm K}}) = +1
~.
\ee
The resulting Hamiltonian is analogous to that of the Haldane model~\cite{Haldane_prl_1988}. 
To diagonalize such Hamiltonian, we introduce the fermion operators~\cite{Knolle_thesis}
\be \label{eq:f_op_def}
f_{\bm r} = \frac{c_{{\bm r},A} +i c_{{\bm r},B}}{2}
~,
\nn
f_{\bm r}^\dagger = \frac{c_{{\bm r},A} - i c_{{\bm r},B}}{2}
~,
\ee
so that
\be \label{eq:c_f_op_def}
c_{{\bm r},\alpha} = \iota_{\alpha} (f_{\bm r} + \eta_{\alpha} f_{\bm r}^\dagger)
~,
\ee
where $\iota_A = 1$, $\iota_B = i$, $\eta_A = 1$, $\eta_B = -1$ ($\eta_\alpha = \iota_\alpha^2$). 
With these definitions, the Hamiltonian~(\ref{eq:Hamil_QSL_Majorana_2}) becomes
\be \label{eq:Hamil_QSL_Majorana_3}
{\cal H}_{\rm m}^{\rm eff} = J \sum_{{\bm r},{\bm \delta}_1} (f_{\bm r} + f_{\bm r}^\dagger) (f_{{\bm r}+{\bm \delta}_1} - f_{{\bm r}+{\bm \delta}_1}^\dagger)
- i \frac{g}{2} \sum_{{\bm r},{\bm \delta}_2,\alpha} \varsigma_\alpha({\bm \delta}_2)  \iota_{\alpha} (f_{\bm r} + \eta_{\alpha} f_{\bm r}^\dagger) \iota_{\alpha} (f_{{\bm r}+{\bm \delta}_2} + \eta_{\alpha} f_{{\bm r}+{\bm \delta}_2}^\dagger)
~.
\ee
Using that $\iota_\alpha^2 = \eta_\alpha$ and $\eta_\alpha \varsigma_\alpha({\bm \delta}_2) = \varsigma_A({\bm \delta}_2)$, we can perform the sum over $\alpha$ in the second term on the right-hand side of Eq.~(\ref{eq:Hamil_QSL_Majorana_3}). We immediately see that the terms containing $f_{\bm r} f_{{\bm r}+{\bm \delta}_2}^\dagger$ and $f_{\bm r}^\dagger f_{{\bm r}+{\bm \delta}_2}$ vanish, since $\sum_\alpha \eta_{\alpha}=0$. We get~\cite{Knolle_thesis}
\be \label{eq:Hamil_QSL_Majorana_3b}
{\cal H}_{\rm m}^{\rm eff} = J \sum_{{\bm r},{\bm \delta}_1} (f_{\bm r} + f_{\bm r}^\dagger) (f_{{\bm r}+{\bm \delta}_1} - f_{{\bm r}+{\bm \delta}_1}^\dagger)
- i g \sum_{{\bm r},{\bm \delta}_2} \varsigma_A({\bm \delta}_2)  (f_{\bm r} f_{{\bm r}+{\bm \delta}_2} +  f_{\bm r}^\dagger f_{{\bm r}+{\bm \delta}_2}^\dagger)
~.
\ee

The next step consist in Fourier-transforming the operators, introducing
\be
&&
f_{\bm r} = \frac{1}{\sqrt{N_{\rm K}}} \sum_{\bm k} e^{i {\bm k}\cdot {\bm r}} f_{\bm k} 
~,
\nn
&&
f^\dagger_{\bm r} = \frac{1}{\sqrt{N_{\rm K}}} \sum_{\bm k} e^{-i {\bm k}\cdot {\bm r}} f_{\bm k}^\dagger 
~,
\ee
where $N_{\rm K}$ is the number of unit cells of the ${\rm RuCl}_3$ lattice.
Eq.~(\ref{eq:Hamil_QSL_Majorana_3b}) then becomes~\cite{Knolle_thesis}
\be \label{eq:Hamil_QSL_Majorana_5}
{\cal H}_{\rm m}^{\rm eff} &=& 
\sum_{{\bm k}} 
\left(  f_{\bm k}^\dagger,  f_{-{\bm k}} \right)
\left(
\begin{array}{cc}
\xi_{\bm k} & \kappa_{\bm k} - i \Delta_{\bm k}
\\
\kappa_{\bm k} + i \Delta_{\bm k} & -\xi_{\bm k}
\end{array}
\right)
\left(  
\begin{array}{c}
f_{\bm k}
\\
f_{-{\bm k}}^\dagger
\end{array}
\right)
~,
\ee
where 
\be
&&
\xi_{\bm k} = J \Re e\Big[ \sum_{{\bm \delta}_1} e^{i{\bm k}\cdot{\bm \delta}_1} \Big] = J \big[1 + \cos({\bm k}\cdot{\bm a}_{+,{\rm K}}) + \cos({\bm k}\cdot{\bm a}_{-,{\rm K}}) \big]
~,
\nn
&&
\Delta_{\bm k} = J \Im m\Big[ \sum_{{\bm \delta}_1} e^{i{\bm k}\cdot{\bm \delta}_1} \Big] = J \big[1 + \sin({\bm k}\cdot{\bm a}_{+,{\rm K}}) + \sin({\bm k}\cdot{\bm a}_{-,{\rm K}}) \big]
\nn
&&
\kappa_{\bm k} = - i g \sum_{{\bm \delta}_2} \varsigma_A({\bm \delta}_2) e^{i{\bm k}\cdot{\bm \delta}_2} = 2 g \big[\sin({\bm k}\cdot{\bm a}_{+,{\rm K}}) - \sin({\bm k}\cdot{\bm a}_{-,{\rm K}}) + \sin\big({\bm k}\cdot({\bm a}_{-,{\rm K}}-{\bm a}_{+,{\rm K}})\big) \big]
~.
\ee
Such Hamiltonian is diagonalized by introducing~\cite{Knolle_thesis}
\be \label{Bogoliubov_transf_h}
\left(  
\begin{array}{c}
f_{\bm k}
\\
f_{-{\bm k}}^\dagger
\end{array}
\right) 
=
\left(  
\begin{array}{cc}
u_{\bm k} & i v_{\bm k}
\vspace{0.2cm}\\
i v_{\bm k}^\star & u_{\bm k}
\end{array}
\right)
\left(  
\begin{array}{c}
a_{\bm k}
\\
a_{-{\bm k}}^\dagger
\end{array}
\right)
~,
\ee
where
\be \label{Bogoliubov_coeffs_h}
&&
u_{\bm k} = \frac{\sqrt{\varepsilon_{\bm k} + \xi_{\bm k}}}{\sqrt{2 \varepsilon_{\bm k}}}
~,
\nn
&&
v_{\bm k} = \frac{\Delta_{\bm k} + i\kappa_{\bm k}}{\sqrt{\kappa_{\bm k}^2 + \Delta_{\bm k}^2}} \frac{\sqrt{\varepsilon_{\bm k} - \xi_{\bm k}}}{\sqrt{2 \varepsilon_{\bm k}}}
~,
\ee
and $\varepsilon_{\bm k} = \sqrt{\xi_{\bm k}^2 + \Delta_{\bm k}^2+ \kappa_{\bm k}^2}$ is the energy dispersion of the upper band (the lower band is $-\varepsilon_{\bm k}$). Finally,
\be \label{eq:Hamil_QSL_Majorana_6}
{\cal H}_{\rm m}^{\rm eff} &=& 
\sum_{{\bm k}} 
\left(  a_{\bm k}^\dagger,  a_{-{\bm k}} \right)
\left(
\begin{array}{cc}
\varepsilon_{\bm k} & 0
\\
0 & -\varepsilon_{\bm k}
\end{array}
\right)
\left(  
\begin{array}{c}
a_{\bm k}
\\
a_{-{\bm k}}^\dagger
\end{array}
\right)
=\sum_{{\bm k}} 
\varepsilon_{\bm k} (2 a_{\bm k}^\dagger a_{{\bm k}} - 1)
~.
\ee

\section{The QSL spin-spin correlation function}
\label{app:QSL_spin_spin}

Let us consider the imaginary-time-ordered correlation function~\cite{Giuliani_and_Vignale}, as defined in Eq.~(\ref{eq:spin_spin_corr_summed_main}):
\be \label{eq:QSL_T_ord_corr_1}
Q^{\gamma}({\bm r}_i,{\bm r}_{i'},\tau) &=& -\langle {\cal T} s_{i}^\gamma(\tau) s_{i'}^\gamma \rangle
=\langle {\cal T} e^{{\cal H}_{\rm m}^{\rm eff} \tau} b_i^\gamma c_{i} e^{-{\cal H}_{\rm m}^{\rm eff}  \tau} b_{i'}^\gamma c_{i'} \rangle
~.
\ee
We now commute the operators between the exponentials, gaining a minus sign. Next, we commute $b_i^\gamma$ with $e^{-{\cal H}_{\rm m}^{\rm eff}  \tau}$. Hence, Eq.~(\ref{eq:QSL_T_ord_corr_1}) becomes
\be \label{eq:QSL_T_ord_corr_2}
Q^{\gamma}({\bm r}_i,{\bm r}_{i'},\tau) &=& 
i\langle {\cal T} e^{{\cal H}_{\rm m}^{\rm eff} \tau} c_{i} e^{-({\cal H}_{\rm m}^{\rm eff} + V^{(1)}_{i,\gamma} + V^{(2)}_{i,\gamma}) \tau} c_{i'} (ib_i^\gamma b_{i'}^\gamma ) \rangle
~,
\ee
where~\cite{Knolle_thesis}
\be
\label{eq:app_V_1_def}
&&
V^{(1)}_{\ell,\eta} = -2 iJ \sum_{\langle \ell,i \rangle_\eta} u_{i\ell}^\eta c_{i} c_\ell 
~,
\\
\label{eq:app_V_2_def}
&&
V^{(2)}_{\ell,\eta} = -2 i g \sum_{\langle \langle \ell, k\rangle \rangle} \varepsilon_{\eta\gamma'\gamma''} u_{\ell j}^{\eta} D_{j} u_{jk}^{\gamma''} c_{\ell} c_{k}
-2 i g \sum_{\langle \langle i, k\rangle \rangle} \varepsilon_{\gamma\eta\gamma''} u_{i\ell}^{\gamma} D_{\ell} u_{\ell k}^{\gamma''} c_{i} c_{k}
~.
\ee
Here, the sum in Eq.~(\ref{eq:app_V_1_def}) is therefore restricted to all sites $i$ that are nearest neighbors of $\ell$ in the direction $\eta$. Conversely, the first sum in Eq.~(\ref{eq:app_V_2_def}) is restricted to all next-nearest neighbors $k$ of $\ell$, such that the intermediate site $j$ is in the direction $\eta$. Finally, the last term in~(\ref{eq:app_V_2_def}) is summed over all next-nearest neighbors $i$ and $k$ such that the intermediate site is $\ell$.
We now observe that, since all the $u_{ij}^\gamma$ commute with the Kitaev hamiltonian, the density matrix factorizes into a product of $c$- and $b$-density matrices. Hence,
\be \label{eq:QSL_T_ord_corr_3}
Q^{\gamma}({\bm r}_i,{\bm r}_{i'},\tau) &=& 
i \langle {\cal T} c_{i}(\tau) c_{i'} S_{i,\gamma}(\tau) \rangle (i\delta_{ii'} + u_{ii'}^\gamma \delta_{\langle i,i'\rangle_\gamma})
~,
\ee
where $\delta_{ij}$ and $\delta_{\langle i,j\rangle_\gamma}$ constrain $i$ and $j$ to either coincide or to be nearest-neighbors along the direction $\gamma$, respectively. Here we introduced~\cite{Knolle_thesis}
\be
S_{i,\gamma}(\tau) &\equiv& e^{{\cal H}_{\rm m}^{\rm eff} \tau}e^{-({\cal H}_{\rm m}^{\rm eff} + V^{(1)}_{i,\gamma} + V^{(2)}_{i,\gamma}) \tau} 
={\cal T} \exp\left(- \int_0^\tau d\tau' \big[V^{(1)}_{i,\gamma}(\tau') + V^{(2)}_{i,\gamma}(\tau')\big] \right)
~.
\ee
As usual, the time-evolution of Majorana particles is generated by ${\cal H}_{\rm m}^{\rm eff}$. 
To continue the calculation, we now fix the unit cell such that it includes the sites $i$ and $j$, the nearest neighbor of $i$ in the direction $\gamma$ ($j=i'$ if $i'\neq i$). Thus, we rewrite Eq.~(\ref{eq:QSL_T_ord_corr_3}) as
\be \label{eq:QSL_T_ord_corr_6}
Q^{\gamma}({\bm r}_i,{\bm r}_{i'},\tau) &=& 
i\delta({\bm r} + {\bm \delta}_\gamma^\alpha - {\bm r}' - {\bm \delta}_\gamma^{\alpha'}) (i\delta_{\alpha\alpha'} + i \sigma^y_{\alpha\alpha'}) \langle {\cal T} c_{{\bm r},\alpha}(\tau) c_{{\bm r}, \alpha'} S_{{\bm r},\gamma}^{\alpha}(\tau) \rangle
~,
\ee
where we defined ${\bm r}_i \equiv {\bm r}+{\bm \delta}_\gamma^\alpha$ and ${\bm r}_{i'} \equiv {\bm r}+{\bm \delta}_\gamma^{\alpha'}$ as the site positions for later convenience. In these expressions, ${\bm r}$ and ${\bm r}'$ are the positions of the unit cells that contain the two sites, while $\alpha,\alpha' = A,B$ denote their type. Here ${\bm \delta}_\gamma^A = {\bm 0}$ and ${\bm \delta}_\gamma^B = {\bm d}_\gamma^A$ are the position of sites of types $A$ and $B$, respectively, in a given unit cell. 

Using the definition in Eq.~(\ref{eq:c_f_op_def}), we rewrite Eq.~(\ref{eq:QSL_T_ord_corr_6}) as~\cite{Knolle_thesis}
\be \label{eq:QSL_T_ord_corr_6_rew}
Q^{\gamma}({\bm r}_i,{\bm r}_{i'},\tau)  &\simeq& 
-\delta({\bm r} + {\bm \delta}_\gamma^\alpha - {\bm r}' - {\bm \delta}_\gamma^{\alpha'}) (\sigma^z_{\alpha\alpha'} + i \sigma^y_{\alpha\alpha'}) 
\frac{ \langle {\cal T} \big[f_{\bm r}(\tau) + \eta_{\alpha} f_{\bm r}^\dagger(\tau)\big] (f_{\bm r} + \eta_{\alpha'} f_{\bm r}^\dagger) S_{{\bm r},\gamma}^\alpha(\tau) \rangle }{ \langle {\cal T} S_{{\bm r},\gamma}^\alpha(\tau) \rangle } \langle {\cal T} S_{{\bm r},\gamma}^\alpha(\tau) \rangle 
~,
\nn
\ee
where
\be
S_{{\bm r},\gamma}^\alpha(\tau) =
{\cal T} \exp\left(- \int_0^\tau d\tau' \big[V^{(1)}_{{\bm r},\gamma,\alpha}(\tau') + V^{(2)}_{{\bm r},\gamma,\alpha}(\tau')\big] \right)
~.
\ee
Using that all $u_{ij}^\gamma=-1$ and $D_j=1$, we find
\be
V^{(1)}_{{\bm r},\gamma,\alpha} &=& 
2 iJ c_{{\bm r},\alpha} c_{{\bm r},{\bar \alpha}} = - 2 J (2 f_{\bm r}^\dagger f_{\bm r} - 1)
~.
\ee
It is similarly possible to express $V^{(2)}_{{\bm r},\gamma,\alpha} \equiv V^{(2)}_{i,\gamma}$ in terms of $f_{\bm r}$ and $f_{\bm r}^\dagger$. The final expression is quite lengthy and will not be reported here: in what follows we will in fact neglect $V^{(2)}_{{\bm r},\gamma,\alpha}$ since this is proportional to the coupling constant $g$, taken to be much smaller than $J$. 
Therefore,
\be
S_{{\bm r},\gamma}^\alpha(\tau) \to S_{{\bm r}}(\tau) \simeq
{\cal T} \exp\left(2 J \int_0^\tau d\tau' \big[2 f_{\bm r}^\dagger(\tau') f_{\bm r}(\tau') - 1\big] \right) 
~.
\ee

The following step consists in performing the so-called ``adiabatic approximation''~\cite{Knolle_thesis,Knolle_prl_2014,Knolle_prb_2015} in Eq.~(\ref{eq:QSL_T_ord_corr_6_rew}), whereby we extend the time-ordered exponential up to $\tau = \beta = (k_{\rm B} T)^{-1}$ (the upper limit of the imaginary-time interval) in the fraction on its right-hand side. We therefore assume that the perturbation is not switched off abruptly at $\tau'=\tau$, but extends up to the end of the imaginary-time interval ($=\beta$). When we replace $S_{{\bm r}}(\tau) \to S_{{\bm r}}(\beta)$, the fraction becomes the usual definition of the imaginary-time-ordered Green's function. 
It is also possible to prove that $\langle {\cal T} f_{\bm r}(\tau) f_{\bm r} \rangle = \langle {\cal T} f^\dagger_{\bm r}(\tau) f^\dagger_{\bm r} \rangle =  0$ because of the symmetry properties of $\xi_{\bm k}$, $\Delta_{\bm k}$ and $\kappa_{\bm k}$. Since the $S$-matrix does not introduce anomalous couplings, also the dressed anomalous imaginary-time-ordered Green's functions are zero. Hence, we get
\be
\label{eq:QSL_T_ord_corr_7}
Q^{\gamma}({\bm r}_i,{\bm r}_{i'},\tau)  &=& 
-\delta({\bm r} + {\bm \delta}_\gamma^\alpha - {\bm r}' - {\bm \delta}_\gamma^{\alpha'}) 
\frac{ \langle {\cal T} \big[f_{\bm r}(\tau) f_{\bm r}^\dagger - \eta_{\alpha} \eta_{\alpha'}  f_{\bm r} f_{\bm r}^\dagger(\tau) \big] S_{{\bm r}}(\beta) \rangle }{ \langle {\cal T} S_{{\bm r}}(\beta) \rangle } \langle {\cal T} S_{{\bm r}}(\tau) \rangle 
~.
\ee
We now define the {\it connected} Green's function~\cite{Knolle_thesis}
\be \label{eq:Q_c_Q_L_def}
&&
Q_{{\rm c}}^{\gamma}({\bm r},\tau,\tau') \equiv - \frac{ \langle {\cal T} f_{\bm r}(\tau) f_{\bm r}^\dagger(\tau') S_{{\bm r}}(\beta) \rangle }{ \langle {\cal T} S_{{\bm r}}(\beta) \rangle } 
~,
\ee
which we use to rewrite Eq.~(\ref{eq:QSL_T_ord_corr_7}) as
\be 
\label{eq:QSL_T_ord_corr_7b}
Q^{\gamma}({\bm r}_i,{\bm r}_{i'},\tau)  &=& 
\big[Q_{{\rm c}}^{\gamma}({\bm r},\tau,0) - \eta_\alpha \eta_{\alpha'} Q_{{\rm c}}^{\gamma}({\bm r},0,\tau)\big]
\langle {\cal T} S_{{\bm r}}(\tau) \rangle 
\delta({\bm r} + {\bm \delta}_\gamma^\alpha - {\bm r}' - {\bm \delta}_\gamma^{\alpha'}) 
~.
\ee
$Q_{{\rm c}}^{\gamma}({\bm r},\tau,\tau')$ can be calculated by resumming the entire RPA-like series of Feynman diagrams~\cite{Giuliani_and_Vignale} in the impurity potential $V_{\rm imp}({\bm r}) = 4 J f_{\bm r}^\dagger f_{\bm r}$, which is in this case exact. The resummation gives~\cite{Knolle_thesis,Giuliani_and_Vignale}
\be
\label{eq:Q_c_Dyson}
Q_{{\rm c}}^{\gamma}({\bm r},\tau,\tau') = Q_{{\rm c},0}^{\gamma}({\bm r},\tau,\tau') - 4 J \int_0^\beta d\tau'' Q_{{\rm c},0}^{\gamma}({\bm r},\tau, \tau'') Q_{{\rm c}}^{\gamma}({\bm r},\tau'',\tau')
~,
\ee
where $Q_{{\rm c},0}^{\gamma}({\bm r},\tau, \tau') \equiv - \langle {\cal T} f_{\bm r}(\tau) f_{\bm r}^\dagger(\tau') \rangle$.
The minus sign in front of the integral in Eq.~(\ref{eq:Q_c_Dyson}) is due to the sign in the definition~(\ref{eq:Q_c_Q_L_def}).

Coming now to the term $\langle {\cal T} S_{{\bm r}}(\tau) \rangle$, we rewrite it as
\be
\langle {\cal T} S_{{\bm r}}(\tau) \rangle  =  \sum_n P_n\, {}_0\langle n | e^{{\cal H}_{\rm m} \tau} e^{-({\cal H}_{\rm m} -2 i J c_{{\bm r},A} c_{{\bm r},B}) \tau} | n \rangle_0 
~,
\ee
where $P_n$ is the occupation factor of the eigenstate $| n \rangle_0$. The subscript ``$0$'' in $| n \rangle_0$ denotes that it is an eigenstate of the bare Hamiltonian ${\cal H}_{\rm m}^{\rm eff}$. Inserting the resolution of the identity in terms of the eigenstates $| m \rangle_J$ of the Hamiltonian ${\cal H}_{{\rm m},J}^{\rm eff} = {\cal H}_{\rm m}^{\rm eff} - 2 i J c_{{\bm r},A} c_{{\bm r},B}$, we get
\be
\langle {\cal T} S_{{\bm r}}(\tau,0) \rangle  = \sum_{n,m} P_n\, \big| {}_J\langle m | n \rangle_0 \big|^2 e^{(E_n^{(0)}- E_m^{(J)})\tau}
~.
\ee
In the limit of zero temperature, $P_n \simeq 0$ for all states but the ground state of ${\cal H}_{\rm m}^{\rm eff}$ (essentially the state with no spinons). We will therefore assume that only such state is occupied. Furthermore, for sufficiently large times the exponential factor is dominated by the state with the minimum $E_m^{(J)}$ (note that $E_m^{(J)}> E_n^{(0)}$, since the latter is the ground state), which we denote with $m=0$. Hence, we approximate
\be \label{eq:TS_final}
\langle {\cal T} S_{{\bm r}}(\tau,0) \rangle  \simeq \big| {}_J\langle 0 | 0 \rangle_0 \big|^2 e^{-\Delta_F\tau}
~.
\ee
We stress that $| n \rangle_0$ and $| m \rangle_J$ describe only $c$-Majorana particles (or, equivalently, the $f$-fermions): the Hilbert-space sector has been assumed from the very beginning to be the zero-flux one. In particular, $|0\rangle_0$ and $|0\rangle_J$ are the ground states of the Hamiltonians ${\cal H}_{\rm m}^{\rm eff}$ and ${\cal H}_{{\rm m},J}^{\rm eff}$, respectively. These are written in terms of only the $c$-particles: all $u_{ij}^\gamma$ have been set equal to $-1$. Note, however, that ${\cal H}_{{\rm m},J}^{\rm eff}$ can also be viewed as the Hamiltonian of a Kitaev model in which one bond eigenvalue has been flipped, and therefore two fluxes have been introduced into neighboring plaquettes. The wavefunction $| 0 \rangle_J$ has therefore the same form of the ground state of the two-flux sector, which has an energy $\Delta_F$ above the zero-flux one. We wish to stress that $|0\rangle_0$ and $|0\rangle_J$ are {\it not} orthogonal, as one could naively expect, since they describe only $c$-Majorana particles and belong to the same flux sector (the true zero- and two-flux ground states, which describe both $c$- and $b$-Majorana particles, are orthogonal). In Eq.~(\ref{eq:TS_final}), $\big| {}_J\langle 0 | 0 \rangle_0 \big|^2$ is just a numerical factor and will be neglected in what follows. Hence, Eq.~(\ref{eq:QSL_T_ord_corr_7b}) becomes
\be
\label{eq:QSL_T_ord_corr_8}
Q^{\gamma}({\bm r}_i,{\bm r}_{i'},\tau) &\simeq& 
\big[Q_{{\rm c}}^{\gamma}({\bm r},\tau,0) - \eta_\alpha \eta_{\alpha'} Q_{{\rm c}}^{\gamma}({\bm r},0,\tau)\big]
e^{-\Delta_F\tau}
\delta({\bm r} + {\bm \delta}_\gamma^\alpha - {\bm r}' - {\bm \delta}_\gamma^{\alpha'}) 
~.
\ee

\section{The function $Q_{{\rm c},0}^{\gamma}$}
\label{eq:non_int_Q}
We now study the properties of the function $Q_{{\rm c},0}^{\gamma}({\bm r},\tau, \tau')$. We rewrite it explicitly as
\be
Q_{{\rm c},0}^{\gamma}({\bm r},\tau,\tau') 
&=&
-\theta(\tau-\tau')\langle f_{\bm r}(\tau) f_{\bm r}^\dagger(\tau') \rangle + \theta(\tau'-\tau)\langle f_{\bm r}^\dagger(\tau') f_{\bm r}(\tau) \rangle 
~.
\ee
Hence, for $\tau>\tau'$,~\cite{Knolle_thesis}
\be
Q_{{\rm c},0}^{\gamma}({\bm r},\tau,\tau')\Big|_{\tau>\tau'} &=& - \frac{1}{N} \sum_{{\bm q},{\bm q}'} e^{i({\bm q} - {\bm q}')\cdot{\bm r}} \langle f_{\bm q}(\tau) f_{{\bm q}'}^\dagger(\tau') \rangle
\nn
&=&
- \frac{1}{N} \sum_{{\bm q},{\bm q}'} e^{i({\bm q} - {\bm q}')\cdot{\bm r}} 
\langle 
\big(u_{\bm q} a_{\bm q} e^{-2\varepsilon_{\bm q}\tau}+ v_{\bm q} a_{-{\bm q}}^\dagger e^{2\varepsilon_{-{\bm q}}\tau} \big) 
\big(u_{{\bm q}'} a^\dagger_{{\bm q}'} e^{2\varepsilon_{{\bm q}'}\tau'} + v_{{\bm q}'} a_{-{\bm q}'} e^{-2\varepsilon_{-{\bm q}'}\tau'} \big) 
\rangle
\nn
&=&
- \frac{1}{N} \sum_{{\bm q}} \big(
|u_{\bm q}|^2 e^{-2\varepsilon_{\bm q}(\tau-\tau')} \langle a_{\bm q} a^\dagger_{{\bm q}} \rangle
+
|v_{\bm q}|^2 e^{2\varepsilon_{-{\bm q}}(\tau-\tau')} \langle a_{-{\bm q}}^\dagger a_{-{\bm q}} \rangle
\big)
~.
\ee
Therefore, it is independent of the coordinate ${\bm r}$ and of $\gamma$. 
Similarly, for $\tau'>\tau$ we get 
\be
Q_{{\rm c},0}^{\gamma}({\bm r},\tau,\tau')\Big|_{\tau'>\tau} &=& \frac{1}{N} \sum_{{\bm q}} \big(
|u_{\bm q}|^2 e^{-2\varepsilon_{\bm q}(\tau-\tau')} \langle a^\dagger_{{\bm q}} a_{\bm q} \rangle
+
|v_{\bm q}|^2 e^{2\varepsilon_{-{\bm q}}(\tau-\tau')} \langle a_{-{\bm q}} a_{-{\bm q}}^\dagger \rangle
\big)
~.
\ee
Putting everything together [ignoring from now on the dependence of $Q_{{\rm c},0}^{\gamma}({\bm r},\tau,\tau')$ on ${\bm r}$] and taking the Fourier transform we find
\be \label{eq:Q_c_iomega}
Q_{{\rm c},0}^{\gamma}(i\omega_m) 
&=&
\frac{1}{N} \sum_{{\bm q}} \left(
\frac{|u_{\bm q}|^2}{i\omega_m-2\varepsilon_{\bm q}}
+
\frac{|v_{\bm q}|^2}{i\omega_m+2\varepsilon_{\bm q}}
\right)
~.
\ee
Analytically continuing $i\omega_m \to \omega+i 0^+$ and taking the imaginary part we then get~\cite{Knolle_thesis}
\be \label{eq:ImQ_c_retarded}
\Im m\big[Q_{{\rm c},0}^{\gamma}(\omega)\big] &=& 
-\frac{\pi}{N}\sum_{{\bm q}} \big[
|u_{\bm q}|^2 \delta(\omega-2\varepsilon_{\bm q})
+
|v_{\bm q}|^2\delta(\omega+2\varepsilon_{\bm q})
\big]
~.
\ee
This function can be easily calculated numerically (the procedure is equivalent to the calculation of the density-of-states of graphene).
The real part of $Q_{{\rm c},0}^{\gamma}(\omega)$ is obtained via a Kramers-Kronig transform~\cite{Giuliani_and_Vignale}:
\be
\Re e\big[Q_{{\rm c},0}^{\gamma}(\omega)\big] = {\cal P}\int_{-\infty}^{\infty} \frac{d\omega'}{\pi} \frac{\Im m\big[Q_{{\rm c},0}^{\gamma}(\omega)\big]}{\omega'-\omega}
~,
\ee
where ${\cal P}$ denotes the principal value.

\section{The QSL spin-spin correlation function in momentum and frequency space}
\label{sect:QSL_spin_spin_momentum_frequency}
We now take the Fourier transform of Eq.~(\ref{eq:QSL_T_ord_corr_8}) in imaginary time~\cite{Giuliani_and_Vignale} and get
\be
\label{eq:QSL_T_ord_corr_8_FT}
Q^{\gamma}({\bm r}_i,{\bm r}_{i'},i\omega_n) &=& \int_0^{\beta} d\tau e^{i\omega_n\tau} Q^{\gamma}({\bm r}_i,{\bm r}_{i'},\tau) 
\nn
&=& 
-\frac{1}{\beta} \sum_{\omega_{n'}} 
\big[Q_{{\rm c}}^{\gamma}({\bm r},i \omega_{n'}) - \eta_\alpha \eta_{\alpha'} Q_{{\rm c}}^{\gamma}({\bm r},- i \omega_{n'})\big]
\frac{e^{-\beta \Delta_F} - 1}{i \omega_{n'} - i \omega_{n} + \Delta_F}
\delta({\bm r} + {\bm \delta}_\gamma^\alpha - {\bm r}' - {\bm \delta}_\gamma^{\alpha'}) 
~.
\ee
Here $\omega_n$ and $\omega_{n'}$ are fermionic Matsubara frequencies~\cite{Giuliani_and_Vignale}. To perform the sum, we rewrite it as a contour integral over the poles~\cite{Mahan_book,Bruus_Flensberg} of $n_{\rm F}(z)$, {\it i.e.}
\be \label{eq:app_Q_contour}
Q^{\gamma}({\bm r}_i,{\bm r}_{i'},i\omega_n) = \delta({\bm r} + {\bm \delta}_\gamma^\alpha - {\bm r}' - {\bm \delta}_\gamma^{\alpha'})  \oint \frac{dz}{2\pi i} n_{\rm F}(z)  \big[Q_{{\rm c}}^{\gamma}({\bm r},z) - \eta_\alpha \eta_{\alpha'} Q_{{\rm c}}^{\gamma}({\bm r},- z)\big]
\frac{e^{-\beta \Delta_F} - 1}{z - i \omega_{n} + \Delta_F}
~.
\ee
Taking the analytical continuation $i\omega_n\to \omega+i0^+$ in Eq.~(\ref{eq:app_Q_contour}) and its imaginary part, one finds
\be\label{eq:ImQ_semifinal}
\Im m\big[Q^{\gamma}({\bm r}_i,{\bm r}_{i'},\omega)\big] &=& 
-(1-e^{-\beta\Delta_F}) \delta({\bm r} + {\bm \delta}_\gamma^\alpha - {\bm r}' - {\bm \delta}_\gamma^{\alpha'}) 
\big[ n_{\rm F}(\omega - \Delta_F) + n_{\rm B}(-\Delta_F) \big] 
\nn
&\times&
\Big\{
\Im m\big[Q_{{\rm c}}^{\gamma}({\bm r},\omega - \Delta_F)\big] + \eta_\alpha \eta_{\alpha'} \Im m\big[Q_{{\rm c}}^{\gamma}({\bm r},\Delta_F- \omega)\big]
\Big\}
~.
\ee
With a similar procedure, Eq.~(\ref{eq:Q_c_Dyson}) gives
\be
\label{eq:Q_c_Dyson_FTtime_sol_Im}
\Im m\big[Q_{{\rm c}}^{\gamma}({\bm r},\omega)\big] = \frac{\Im m\big[Q_{{\rm c},0}^{\gamma}(\omega)\big]}{\big|1 + 4 J Q_{{\rm c},0}^{\gamma}(\omega)\big|^2}
~.
\ee
Here we used that, as proven in App.~\ref{eq:non_int_Q}, $Q_{{\rm c},0}^{\gamma}({\bm r},i\omega_m)$ is independent of the coordinate ${\bm r}$. Since, at is evident from Eq.~(\ref{eq:Q_c_Dyson_FTtime_sol_Im}), $Q_{{\rm c}}^{\gamma}({\bm r},\omega)$ is also independent of ${\bm r}$, hereafter we will neglect its dependence on such variable.
Therefore, Eq.~(\ref{eq:ImQ_semifinal}) becomes
\be\label{eq:ImQ_final}
\Im m\big[Q^{\gamma}({\bm r}_i,{\bm r}_{i'},\omega)\big] &=& 
-(1-e^{-\beta\Delta_F}) \delta({\bm r} + {\bm \delta}_\gamma^\alpha - {\bm r}' - {\bm \delta}_\gamma^{\alpha'}) 
\big[ n_{\rm F}(\omega - \Delta_F) + n_{\rm B}(-\Delta_F) \big] 
\nn
&\times&
\left\{
\frac{\Im m\big[Q_{{\rm c},0}^{\gamma}(\omega-\Delta_F)\big]}{\big|1 + 4 J Q_{{\rm c},0}^{\gamma}(\omega-\Delta_F)\big|^2} + \eta_\alpha \eta_{\alpha'} \frac{\Im m\big[Q_{{\rm c},0}^{\gamma}(\Delta_F-\omega)\big]}{\big|1 + 4 J Q_{{\rm c},0}^{\gamma}(\Delta_F-\omega)\big|^2}
\right\}
~.
\nn
\ee
Finally, we take the Fourier transform over ${\bm r}-{\bm r}'$ in Eq.~(\ref{eq:ImQ_final}) and sum over $\alpha$ and $\alpha'$ to get
\be
\label{eq:ImQ_final_FT}
\Im m\big[Q^{\gamma}({\bm q},\omega)\big] &=& 
-2(1-e^{-\beta\Delta_F})
\big[ n_{\rm F}(\omega - \Delta_F) + n_{\rm B}(-\Delta_F) \big] 
\Bigg\{
\frac{\Im m\big[Q_{{\rm c},0}^{\gamma}(\omega-\Delta_F)\big]}{\big|1 + 4 J Q_{{\rm c},0}^{\gamma}(\omega-\Delta_F)\big|^2} 
\big[1 + \cos({\bm q}\cdot {\bm d}_\gamma^A)\big]
\nn
&+& 
\frac{\Im m\big[Q_{{\rm c},0}^{\gamma}(\Delta_F-\omega)\big]}{\big|1 + 4 J Q_{{\rm c},0}^{\gamma}(\Delta_F-\omega)\big|^2}
\big[1 - \cos({\bm q}\cdot {\bm d}_\gamma^A)\big]
\Bigg\}
~.
\ee
In the zero-temperature limit $\beta\to \infty$, the term
\be
n_{\rm F}(\omega - \Delta_F) + n_{\rm B}(-\Delta_F) \to -\Theta(\omega-\Delta_F)
~,
\ee
and therefore
\begin{equation}
\Im m\big[Q^{\gamma}({\bm q},\omega)\big] =
2 \Theta(\omega-\Delta_F)
\Bigg\{
\frac{\Im m\big[Q_{{\rm c}}^{0,\gamma}(\omega-\Delta_F)\big]}{\big|1 + 4 J Q_{{\rm c}}^{0,\gamma}(\omega-\Delta_F)\big|^2} 
\big[1 + \cos({\bm q}\cdot {\bm d}^A_\gamma)\big]
+
\frac{\Im m\big[Q_{{\rm c}}^{0,\gamma}(\Delta_F-\omega)\big]}{\big|1 + 4 J Q_{{\rm c}}^{0,\gamma}(\Delta_F-\omega)\big|^2}
\big[1 - \cos({\bm q}\cdot {\bm d}^A_\gamma)\big]
\Bigg\}
~.
\end{equation}

\end{widetext}

\end{document}